\RequirePackage{lineno}
\documentclass[%
10pt,
aps,
prd,
twocolumn,
superscriptaddress,
showpacs,
altaffilletter,
nofootinbib,
 amsmath,
 amssymb,
]{revtex4-2}

\usepackage{xcolor}
\usepackage{xspace}
\modulolinenumbers[5]
\usepackage{siunitx}
\usepackage{subfigure}
\usepackage{verbatim}
\usepackage{rotating}
\usepackage{bm}
\usepackage{ulem}
\begin{document}

\newcommand{\authornote}[1]{{%
  \let\thempfn\relax
  \footnotetext[0]{$\diamond${ }#1}
}}

\bibliographystyle{elsarticle-num}

\newcommand*\diff{\mathop{}\!\mathrm{d}}
\newcommand*\Diff[1]{\mathop{}\!\mathrm{d^#1}}
\newcommand{\pb}{PbF$_2$}
\newcommand{\mtca}{$\mu$TCA}
\newcommand{\fix}[1]{{\color{red}{#1}}}
\newcommand{\needref}{[{\color{red}{ref}}]}
\newcommand{\gm}{\ensuremath{g-2}\xspace}
\newcommand{\gmtwo}{\ensuremath{g-2}\xspace}


\renewcommand{\wp}{\ensuremath{\omega_{p}}\xspace}
\renewcommand{\amu}{\ensuremath{a_{\mu}}\xspace}
\newcommand{\oa}{\ensuremath{\omega_{a}^{m}}\xspace}
\newcommand{\oai}{\ensuremath{\omega_{a}}\xspace}

\newcommand{\op}{\ensuremath{\omega_{p}}\xspace}
\renewcommand{\ns}[1]{\SI{#1}{ns}\xspace}
\newcommand{\mus}[1]{\SI{#1}{\micro\second}\xspace}
\newcommand{\mum}[1]{\SI{#1}{\micro m}\xspace}
\newcommand{\runone}{Run-1\xspace}
\newcommand{\runtwo}{Run-2\xspace}
\newcommand{\runthree}{Run-3\xspace}
\newcommand{\runfour}{Run-4\xspace}
\newcommand{\runonea}{Run-1a\xspace}
\newcommand{\runoneb}{Run-1b\xspace}
\newcommand{\runonec}{Run-1c\xspace}
\newcommand{\runoned}{Run-1d\xspace}
\newcommand{\precession}{precession-run1}
\newcommand{\field}{field-run1}
\newcommand{\BD}{BD-run1}
\newcommand{\PRL}{PRL-run1}

\title{Measurement of the anomalous precession frequency of the muon in the Fermilab Muon \gm experiment}
%
\affiliation{Argonne National Laboratory, Lemont, IL, USA}
\affiliation{Boston University, Boston, MA, USA}
\affiliation{Brookhaven National Laboratory, Upton, NY, USA}
\affiliation{Budker Institute of Nuclear Physics, Novosibirsk, Russia}
\affiliation{Center for Axion and Precision Physics (CAPP) / Institute for Basic Science (IBS), Daejeon, Republic of Korea}
\affiliation{Cornell University, Ithaca, NY, USA}
\affiliation{Fermi National Accelerator Laboratory, Batavia, IL, USA}
\affiliation{INFN Gruppo Collegato di Udine, Sezione di Trieste, Udine, Italy}
\affiliation{INFN, Laboratori Nazionali di Frascati, Frascati, Italy}
\affiliation{INFN, Sezione di Napoli, Napoli, Italy}
\affiliation{INFN, Sezione di Pisa, Pisa, Italy}
\affiliation{INFN, Sezione di Roma Tor Vergata, Roma, Italy}
\affiliation{INFN, Sezione di Trieste, Trieste, Italy}
\affiliation{Istituto Nazionale di Ottica - Consiglio Nazionale delle Ricerche, Pisa, Italy}
\affiliation{Department of Physics and Astronomy, James Madison University, Harrisonburg, VA, USA}
\affiliation{Institute of Physics and Cluster of Excellence PRISMA+, Johannes Gutenberg University Mainz, Mainz, Germany}
\affiliation{Joint Institute for Nuclear Research, Dubna, Russia}
\affiliation{Department of Physics, Korea Advanced Institute of Science and Technology (KAIST), Daejeon, Republic of Korea}
\affiliation{Lancaster University, Lancaster, United Kingdom}
\affiliation{Michigan State University, East Lansing, MI, USA}
\affiliation{North Central College, Naperville, IL, USA}
\affiliation{Northern Illinois University, DeKalb, IL, USA}
\affiliation{Northwestern University, Evanston, IL, USA}
\affiliation{Regis University, Denver, CO, USA}
\affiliation{Scuola Normale Superiore, Pisa, Italy}
\affiliation{School of Physics and Astronomy, Shanghai Jiao Tong University, Shanghai, China}
\affiliation{Tsung-Dao Lee Institute, Shanghai Jiao Tong University, Shanghai, China}
\affiliation{Institut für Kern - und Teilchenphysik, Technische Universit\"at Dresden, Dresden, Germany}
\affiliation{Universit\`a del Molise, Campobasso, Italy}
\affiliation{Universit\`a di Cassino e del Lazio Meridionale, Cassino, Italy}
\affiliation{Universit\`a di Napoli, Napoli, Italy}
\affiliation{Universit\`a di Pisa, Pisa, Italy}
\affiliation{Universit\`a di Roma Tor Vergata, Rome, Italy}
\affiliation{Universit\`a di Trieste, Trieste, Italy}
\affiliation{Universit\`a di Udine, Udine, Italy}
\affiliation{Department of Physics and Astronomy, University College London, London, United Kingdom}
\affiliation{University of Illinois at Urbana-Champaign, Urbana, IL, USA}
\affiliation{University of Kentucky, Lexington, KY, USA}
\affiliation{University of Liverpool, Liverpool, United Kingdom}
\affiliation{Department of Physics and Astronomy, University of Manchester, Manchester, United Kingdom}
\affiliation{Department of Physics, University of Massachusetts, Amherst, MA, USA}
\affiliation{University of Michigan, Ann Arbor, MI, USA}
\affiliation{University of Mississippi, University, MS, USA}
\affiliation{University of Oxford, Oxford, United Kingdom}
\affiliation{University of Rijeka, Rijeka, Croatia}
\affiliation{Department of Physics, University of Texas at Austin, Austin, TX, USA}
\affiliation{University of Virginia, Charlottesville, VA, USA}
\affiliation{University of Washington, Seattle, WA, USA}
\author{T.~Albahri}  \affiliation{University of Liverpool, Liverpool, United Kingdom}
\author{A.~Anastasi} \thanks{Deceased} \affiliation{INFN, Sezione di Pisa, Pisa, Italy}
\author{A.~Anisenkov} \altaffiliation[Also at ]{Novosibirsk State University}  \affiliation{Budker Institute of Nuclear Physics, Novosibirsk, Russia}
\author{K.~Badgley}  \affiliation{Fermi National Accelerator Laboratory, Batavia, IL, USA}
\author{S.~Bae{\ss}ler} \altaffiliation[Also at ]{Oak Ridge National Laboratory}  \affiliation{University of Virginia, Charlottesville, VA, USA}
\author{I.~Bailey} \altaffiliation[Also at ]{The Cockcroft Institute of Accelerator Science and Technology}  \affiliation{Lancaster University, Lancaster, United Kingdom}
\author{V.~A.~Baranov}  \affiliation{Joint Institute for Nuclear Research, Dubna, Russia}
\author{E.~Barlas-Yucel}  \affiliation{University of Illinois at Urbana-Champaign, Urbana, IL, USA}
\author{T.~Barrett}  \affiliation{Cornell University, Ithaca, NY, USA}
\author{A.~Basti}  \affiliation{INFN, Sezione di Pisa, Pisa, Italy}\affiliation{Universit\`a di Pisa, Pisa, Italy}
% ...
\collaboration{The Muon \gmtwo Collaboration} \noaffiliation
\vskip 0.25cm

%
%

\begin{abstract}
The Muon \gm Experiment at Fermi National Accelerator Laboratory
(FNAL) 
has measured the muon anomalous
precession frequency \oa
to an uncertainty of 434 parts per billion (ppb), statistical, and
56 ppb, systematic,  
with data collected
in four storage ring configurations  
during its first physics run in 2018.  When combined with 
a precision measurement of the magnetic field of 
the experiment's muon storage ring, the precession frequency
measurement determines a muon magnetic anomaly
of $a_{\mu}({\rm FNAL}) = 116\,592\,040\, (54) \times 10^{-11}$ (0.46
ppm).
This article describes the multiple techniques 
employed in the reconstruction, analysis and fitting of the data to
measure the precession frequency.  It also  
presents the averaging of the results from the eleven separate
determinations of \oa, and the systematic uncertainties  
on the result.
\end{abstract}


\maketitle
\tableofcontents
\section{Introduction}
\label{sec:intro}

Reference~\cite{\PRL} reports a new measurement of the 
muon magnetic anomaly
 $a_{\mu}=(g_{\mu}-2)/2$ made by our Muon \gm
Collaboration based on its \runone data at Fermi National Accelerator
Laboratory (FNAL).  That initial physics run
occurred over a period of 15 weeks in Spring 2018. We find 
\begin{equation*}
a_{\mu}({\rm FNAL}) = 116\,592\,040\, (54) \times 10^{-11}.
\end{equation*}
where the total uncertainty includes the dominant statistical
uncertainty combined with combinations from the precession rate
systematic, magnetic systematic, and beam-dynamics systematic
uncertainties.   This combined uncertainty corresponds 
to a $0.46$ parts per million (ppm) measurement.  

Three companion papers to that letter describe in detail the key
inputs to this result.  Reference~\cite{\field} presents the
detailed analysis of the precision measurement of the magnetic field
within our storage ring.  Reference~\cite{\BD} details the small
corrections to our anomalous moment measurement from effects
associated with the dynamics of the stored muon beam.  This paper
presents the data reconstruction, analysis, and systematic uncertainty
evaluation for the determination of the average muon spin precession frequency
within the precision magnetic field of our storage ring.    The letter brings
the results from these three papers together, combining the corrected
muon precession frequency with the precision field measurement to obtain
the \amu result given above.

\subsection{Status of \gm of the muon}
The measurement of the muon magnetic anomaly performed by the E821
experiment at the Brookhaven National Laboratory (BNL)
\cite{Bennett:2006fi} of $\amu =
116\,592\,092(63)$\footnote{Updated to reflect recent CODATA values of
  external inputs} has shown an excess with respect to  
the Standard Model (SM) prediction by over
3.5 standard deviations.  Since the publication  
of the final E821 result, the evaluation of the SM prediction has
undergone significant scrutiny.  The quantum electrodynamics (QED) 
contributions to  
\gm, calculated to order
$(\alpha/\pi)^{5}$~\cite{Aoyama:2012wk,Aoyama:2019ryr}, agree well
with precise measurement  
of \gm for the electron~\cite{Hanneke:2008tm}.  Recent discrepancies
in the measurement of the fine structure  
constant~\cite{Morel2020,Parker191} do not significantly affect the
muon \gm prediction.  Electroweak corrections include the  
complete two-loop evaluation, hadronic 
effects, and the leading log 3-loop  
contributions~\cite{PhysRevD.67.073006,PhysRevD.73.119901,PhysRevD.88.053005}.
The dominant theoretical uncertainties  
arise in the QCD hadronic vacuum polarization (HVP) and hadronic
light-by-light (HLxL) corrections, which   
the Muon \gm Theory Initiative~\cite{G2Theory} has recently reviewed thoroughly.
The review, covering dispersive, lattice and modeling methods,
arrived at a consensus~\cite{Aoyama:2020ynm} for the hadronic
contributions and their uncertainties, and predicts 
$a_{\mu}^{SM} = 116\,591\,810(43)\times
10^{-11}$~\cite{Aoyama:2012wk,Aoyama:2019ryr,Czarnecki:2002nt,Gnendiger:2013pva,Davier:2017zfy,Keshavarzi:2018mgv,Colangelo:2018mtw,Hoferichter:2019gzf,Davier:2019can,Keshavarzi:2019abf,Kurz:2014wya,Melnikov:2003xd,Masjuan:2017tvw,Colangelo:2017fiz,Hoferichter:2018kwz,Gerardin:2019vio,Bijnens:2019ghy,Colangelo:2019uex,Blum:2019ugy,Colangelo:2014qya}.   
Comparison with the E821 result 
yields a difference of $(279\pm76)\times 10^{-11}$, which remains over the 
3.5 standard deviation level.   
In order to confirm, or refute, that discrepancy, Experiment
E989~\cite{Grange:2015fou} was constructed at Fermi National Laboratory. 

\subsection{Principles of the experiment}
\label{subsec:principles-experiment}

The Fermilab E989 (Muon \gm) experiment follows a sequence of
polarized muon beam storage experiments pioneered at CERN and
BNL.
In particular, it uses an experimental approach  based 
on the muon anomalous precession within a storage ring
with a highly uniform and precisely known magnetic field.
This approach was pioneered in the 
 CERN experiment \cite{Bailey:1978mn} and 
refined with muon, rather than with pion, injection
by the E821 experiment
at BNL \cite{Bennett:2006fi}.

The technique is based on the convergence of three
fundamental effects: the relative precession rates of the muon spin
and momentum within a uniform magnetic field, parity violation in
muon decay, and the Lorentz boost of the muon decay
products between the muon
rest frame and the lab frame.  When a muon orbits horizontally within
the uniform vertical magnetic field of a perfect storage ring, its
momentum vector precesses at the cyclotron frequency $\vec{\omega}_{c}
= -q\vec{B}/m\gamma$.  For a relativistic muon polarized in the
horizontal plane, the Larmor precession, combined with Thomas
precession, yields a total spin precession frequency of 
\begin{equation*}
\vec{\omega}_{s} = -g_{\mu}\frac{q\vec{B}}{2m} - (1-\gamma)\frac{q\vec{B}}{m\gamma}.
\end{equation*}
The relative precession frequency of the spin with respect to the
momentum, denoted hereafter as the anomalous precession frequency
\oai, is therefore 
\begin{equation}
\vec{\omega}_{a} = \vec{\omega}_{s} - \vec{\omega}_{c} = -\left(\frac{g_{\mu}-2}{2}\right)\frac{q\vec{B}}{m} = -a_{\mu}\frac{q\vec{B}}{m}.
\label{eq:ideal_wa}
\end{equation}
A measurement of the anomalous precession frequency, coupled with
precise knowledge of the storage ring magnetic field, therefore
provides a direct probe of the anomalous magnetic moment. 

Parity violation within the weak decay of the muon provides the means
for such a direct measurement of the anomalous precession frequency:
the highest energy positrons from muon decay are emitted, within its
rest frame, in a direction strongly correlated with the muon spin
direction.  
When coupled with the Lorentz boost, this
spin-energy correlation results in a modulation of the positron energy
spectrum in the laboratory frame:  the stiffest spectrum occurs when
the spin and muon momentum directions are aligned, and the softest
occurs when they are anti-aligned.  This modulation occurs at the rate
of the anomalous precession frequency.   

As a result of the energy modulation, the number of positrons above a
given energy threshold $E_{\rm th}$
from muon decay within this ideal stored beam varies with time as
\begin{equation}
N(t) = N_{0}e^{-t/\gamma\tau_{\mu}}\left(1 + A(E_{\rm th})\cos(\oai t
  + \phi_{0})\right).
\label{eq:5param}
\end{equation}
The parameter $N_{0}$ represents the initial beam intensity,
$\gamma\tau_{\mu}$ the lifetime of the boosted muon, and $\phi_{0}$ the average
initial angle of the muon spins relative to the beam direction.  The
asymmetry parameter $A(E_{\rm th})$, which governs the amplitude of
the rate oscillation about the average exponential for muon decay,
depends on the threshold energy: the energy-spin correlation weakens
as the positron energy decreases.  In fact, since the total decay rate
must fall as a pure exponential, the asymmetry, evaluated for the
lowest energy positrons, changes sign.  
The choice of energy threshold then 
requires balancing the increased muon statistics with the dilution of
the
average asymmetry, and the optimal choice varies with the method used
to extract the anomalous precession frequency (see
Sec.~\ref{sec:fitting}). 
Details of the statistical power of the \oai determination are
described in~\cite{Bennett:2006fi} where it is shown that,
for the optimal method, the variance of the measured precession
frequency \oai scales as  
\begin{equation}
\sigma^2 \propto \frac{1}
{N\langle A^2\rangle_{E_{\rm th}}} \, .
\end{equation}

While a vertical magnetic field provides the horizontal confinement
necessary to store a muon beam, 
storage of the beam for any significant period requires additional
vertical focusing.  A pulsed electrocstatic quadrupole (ESQ) system, comprising
four discrete sections symmetrically spaced about the muon storage
ring and covering 43\% of its circumference, provides this focusing.
Allowing for the presence of such an electric field $\vec{E}$, as well
as for muon beam motion that is not strictly perpendicular to the
magnetic field, the anomalous precession frequency of
Eq.~\ref{eq:ideal_wa} becomes~\footnote{We are ignoring the possibility
of the existence of 
a muon {\it electric dipole moment} which would contribute with
additional terms.}~\cite{\BD} 
\begin{eqnarray}
\vec{\omega}_{a} & = & -\frac{q}{m}\left[ a_{\mu}\vec{B} -
   a_{\mu} \left(\frac{\gamma}{\gamma+1}\right) 
(\vec{\beta}\cdot\vec{B})\vec{\beta}\right. \nonumber \\
                 &   & \left. - \left(a_{\mu} -
\frac{1}{\gamma^{2}-1}\right)\frac{\vec{\beta}\times\vec{E}}{c}\right].
\label{eq:realistic_wa}
\end{eqnarray}
The $\vec{\beta} \cdot \vec{B}$ term accounts for a possible component
of the muon velocity parallel to the magnetic field. 
The last term, which corresponds to the additional magnetic field
component that the muon experiences in its rest frame from $\vec{E}$,
vanishes for a muon with momentum $p_{0} = 3.094$~GeV/c, or $\gamma \sim
29.3$. This experiment has been designed to accept and store a beam of
muons with a narrow momentum spread (0.15\%) about $p_{0}$.  The
corrections to $a_{\mu}$ arising from both vertical beam motion and
the residual electric field correction are discussed in detail in
Ref.~\cite{\BD}. 
Due to these and to other effects detailed in~\cite{\PRL}, the
measured precession frequency needs to be corrected in order to
obtain the quantity \oai required to evaluate \amu. 
This paper describes the procedure followed to obtain the observed
precession frequency \oa. After the corrections to bring this
observed frequency to the ideal \oai above, combination with
the precision field measurements detailed in Reference~\cite{\field}
allow determination of \amu.

Muons stored at this momentum possess a boosted lifetime of
$\gamma\tau_{\mu} \approx\,$\mus{64.4}.  This lifetime limits the
practical storage time of the beam: almost all of the muons have
decayed away after \mus{700}. We therefore need many muon beam
``fills'', cycles of muon beam injection and storage, which occur at
a rate of 16 fills every 1.4~s for E989. In each fill, a muon bunch
of time width  120~ns, to be compared with  
a cyclotron period  $T_{c} = \SI{149.2}{ns}$, 
is injected within the 7.112 m radius
ring, with its 1.45 T field.

The muons within the storage ring undergo betatron oscillations --
stable oscillations about the equilibrium orbit -- with
characteristics that depend on the strength of the ESQ electric
field.  The system is weak-focusing and
properly characterized by the field index $n$ for a continuous
ESQ given by  
\begin{equation}
n = \frac{R_{0}}{v B_{0}}\frac{\partial E_{y}}{\partial y},
\end{equation}
where $R_{0}$ is the equilibrium orbit radius, $v$ is the muon
velocity, $B_{0}$ is the magnetic field, and $E_{y}$ is the effective vertical
quadrupole field component.
The horizontal ($x$) and vertical ($y$) tunes -- the
number of betatron oscillations per cyclotron revolution -- are
related to the field index by $\nu_{x} \approx \sqrt{1-n}$ and
$\nu_{y} \approx \sqrt{n}$, respectively.  These tunes introduce two
key oscillation frequencies into the experiment, 
\begin{align}
f_{x} &\approx f_{c}\sqrt{1-n} \\
f_{y} &\approx f_{c}\sqrt{n}.
\end{align}
with $f_c = \omega_c / 2\pi$.
The radial and vertical betatron motion of the muons within the beam is strongly
coherent when the beam is first injected into the storage ring.  
The lattice
chromaticity, due to the $\sim 0.15\%$ momentum
spread of the stored muon beam, 
and the ESQ non-linearities, related to higher
order multipoles, cause this motion to decohere.

The finite acceptance of the detector system couples with the beam
motion resulting from coherent betatron oscillations (CBO) to
introduce  additional time modulation into the rate of detected
positrons and into the shape of the positron energy spectrum.  As
Sec.~\ref{sec:fitting} and Ref.~\cite{\field} discuss in detail,
these CBO effects introduce a time variation into the effective
asymmetry $A(E_{\rm th})$ and phase $\phi_{0}$ terms in
Eq.~\ref{eq:ideal_wa}.  Radial motion of the beam (within the horizontal
plane) introduces particularly strong oscillations at multiples of the
frequency $f_{CBO} = f_{C} - f_{x}$.  Accurate modeling of the
time dependence of our data requires incorporation of both the
horizontal and vertical effects. The betatron oscillations
 do not, though, couple strongly
to the anomalous precession frequency \oa as long as they are
stable while the muons are stored. 

Table~\ref{tab:frequencies}, in the Appendix, summarizes the nominal
frequencies that characterize the \gm storage ring for the two
values of the field index employed during \runone.

The remainder of this article proceeds as follows.  After a summary of the instrumentation
relevant for the precession frequency analysis in Section~\ref{sec:instrumentation}, Section~\ref{sec:histogram}
presents the analysis strategies behind the determination
of the precession frequencies, followed by the data reconstruction strategies
employed to enable those strategies in Section~\ref{sec:positron_recon}.  Section~\ref{sec:data_corrections} 
outlines the two major corrections applied to
the data: the gain corrections input to the reconstruction and the pileup correction needed before
fitting.  Section~\ref{sec:fitting} then presents the data model, the fit, the fit results and the stability of 
the fit results.  After a discussion of the systematic uncertainties affecting the precession
measurement in Section~\ref{sec:systematics}, the article concludes with a discussion of the averaging procedure to combine
the results from the different analysis efforts in Section~\ref{sec:combination}, followed by the 
summary of results in Section~\ref{sec:discussion}.
\section{Instrumentation overview}
\label{sec:instrumentation}

\begin{figure}[tb]
\centering
\includegraphics[width=0.99\linewidth]{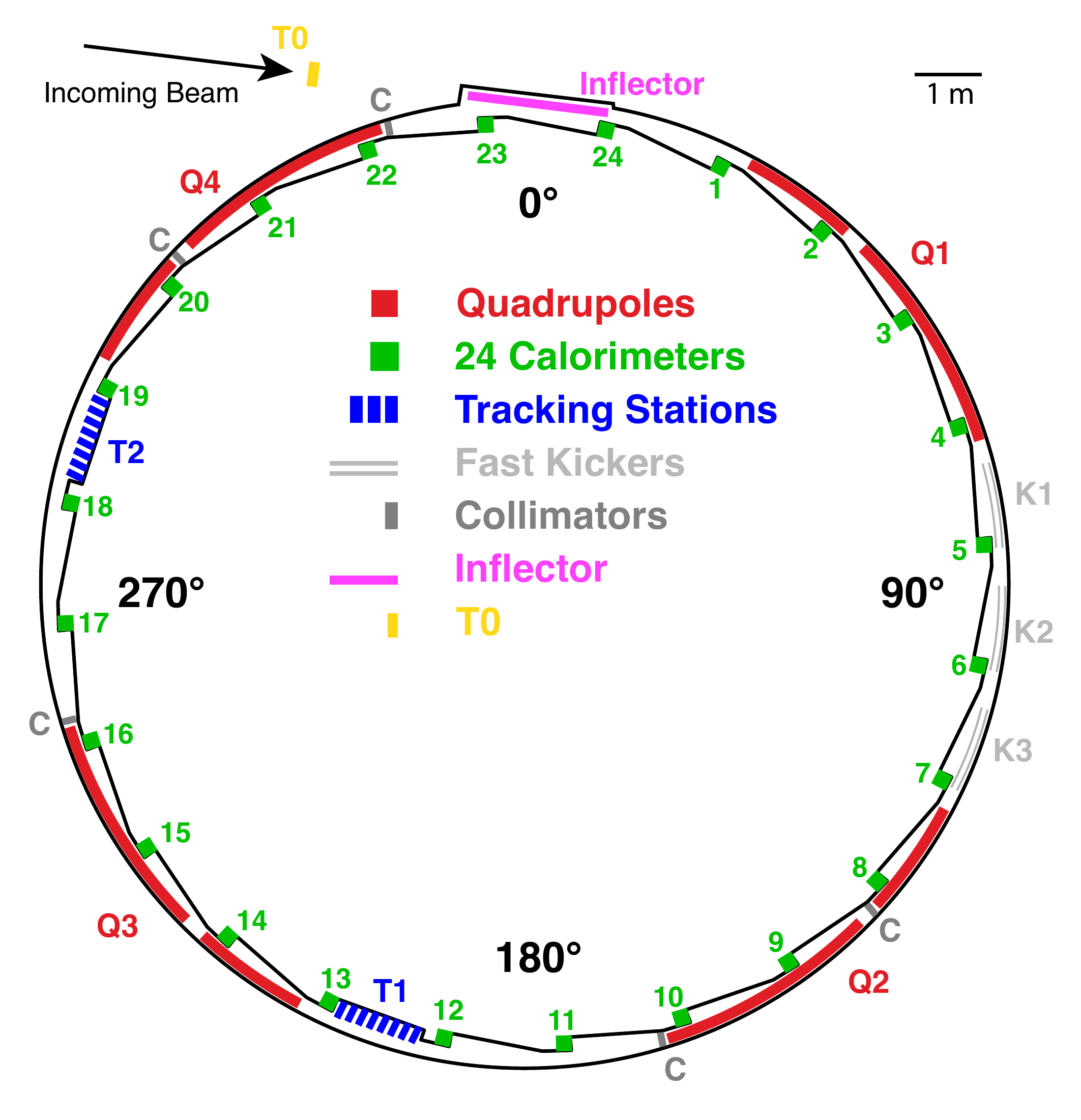}
\caption{Schematic of the Muon \gm storage ring and instrumentation showing
  the elements directly involved in the muon precession analysis.
  Key elements include the suite of 24 electromagnetic calorimeters (green or medium grey),
  the straw tracker system (dark blue or near back), the ESQs
  (red or dark grey), a fast kicker system (light blue or light grey), and the beam entrance (T0)
  detectors (yellow or very light grey).}
\label{fig:layout}
\end{figure}

The primary system for measurement of the positron energy and time
distribution consists of a suite of 24 small electromagnetic
calorimeters distributed around
 the interior of the storage ring and
positioned behind a scallop in the vacuum chamber to minimize the
material traversed by the daughter positrons,
as
shown in Fig.~\ref{fig:layout}.  The positrons from muon decay,
have  
momenta too small to be stored in the ring and drift inwards in the
magnetic field towards the calorimeters. 
At any given time, a
single calorimeter will 
detect positrons emitted
from muons over only a small range of spin precession phases.  The highest
energy positrons can travel a significant fraction of an orbit before
encountering a calorimeter.  Softer positrons travel smaller
distances, so have been produced later in a muon precession
cycle.\footnote{A full spin procession cycle corresponds to roughly 30
  cyclotron periods.}   
As a result, the phase of the muon when it decayed varies over the energy
range of accepted daughter positrons.
The phase difference over this range does not significantly dilute the precession signal.

Each calorimeter station, described in detail
elsewhere~\cite{Kaspar:2016ofv,Khaw:2019yzq,Anastasi:2016luh}, consists of a 9 column by 6 row
array of PbF$_{2}$ crystals instrumented with silicon photomultiplier
(SiPM) photodetectors.  Digitization of the output from each of the
$24\times 54$ channels occurs continuously over an entire fill at a
rate of approximately 800 Megasamples per second (MSPS).
This scheme eliminates dead time 
and potential rate-dependence.
 A beam-arrival signal from the Fermilab accelerator
complex triggers the digitization process for a fill.  The master
digitization clock for the experiment is completely independent of the
accelerator clocks that determine the beam-arrival timing.  Blinding
of the precise
digitization rate at the hardware level avoids the potential for
unconscious bias in the data analysis.  During data analysis, an
additional level of blinding occurs in software, as described in
Sec.~\ref{subsec:analysis-blinding}. 

The blinded clock for digitization derives from a master 40 MHz
precision clock, in turn driven by a GPS-stablized 10 MHz rubidium
clock source.  To achieve the hardware-level blinding, two Fermilab
staff (independent of the collaboration) detune the 40~MHz clock to a
frequency in the range 39,997\,kHz to 39,999\,kHz. 
Correction for the blinding offset occurred as the last stage of the
analysis, after completion of all systematic bias evaluations and cross
checks, and following the decision to unblind and publish. 
We mix a second
blinded clock with the master clock to  monitor the  clock system
stability without revealing the blinding offset. The monitoring of the
resulting blind frequency difference utilizes a second GPS-stabilized
reference clock that is completely independent of the master
clock and its GPS stabilitization. 

The set of complete waveforms obtained from a fill then pass to the
frontend processors of the data acquisition (DAQ)
system~\cite{Gohn:2016shi} for data reduction, which proceeds as follows.  Each calorimeter
has a dedicated frontend processor and GPU that perform the data reduction
necessary to keep the stored data volume manageable.  The
DAQ system prepares two data streams for  offline analysis.
The first ``event-based'' data stream corresponds to identification of
particle activity within the detector.  Whenever a waveform sample for
any crystal in a calorimeter exceeds a $\sim 50$~MeV threshold, the
DAQ system extracts a  time window of approximately \mus{40},
depending on the pulse width, surrounding that sample from
all crystals in that calorimeter for offline analysis.   
The second data stream provides a continuous sampling of the waveforms for each
fill that allows an ``integrated energy'' approach
(Sec.~\ref{sec:fitting}) to the precession frequency determination.
To achieve a manageable data output rate, the DAQ system combines the
raw crystal waveform samples into contiguous 75~ns windows over a
range of $-$\mus{6} to $+$\mus{231} relative to the muon beam arrival
time for the \runone data presented here.  The system also allows
summing of a configurable number of consecutive fills, but that was
not utilized for this dataset.   

For the measurement of \oa, time stability relative to the start of the
fill drives the design of the detector as well as the data
reconstruction algorithms.  Suppose, for example, the gain of the SiPM
photodetectors drift in a fashion correlated with time since muon
injection (referred to as ``time into the fill'').  Without
correction, the true positron energy distribution above a fixed
threshold in an analysis would shift.  Because of the energy-precession
phase correlation discussed above, such a shift would effectively
introduce a time dependence into the  phase $\phi_0$ in the precession
term in the decay rate (Eq.~\ref{eq:5param}).  With $\phi_0 \rightarrow
\phi(t)$, the extracted precession phase $\omega_{a}$ would be
directly biased~\footnote{While the CBO motion noted above introduces
  an oscillatory behavior into the phase, this effect averages to
  zero.}.  A laser-based  system~\cite{Anastasi:2019lxf} provides
monitoring and assessment of such gain variations in each of the 1296
crystals.  The system sweeps a set of laser pulses over the time into
the fill on a subset of data and directly measures the beam-correlated
gain variations.  This system also provides a common pre-beam
pulse, for each fill,
that allows time synchronization of all of the digitizer channels
and it is used to monitor time stability across the fill.

Reconstruction effects that are sensitive to particle flux, and thus
can vary early to late, can also introduce an effective $\phi(t)$ and a
possible bias to \oa. These effects, such as random overlap of
different positron showers in a calorimeter (pileup), 
will be noted in later
sections of the paper. 

Several other subsystems indicated in Fig.~\ref{fig:layout} play a
role in the analysis of the spin precession data.  The T0 counter,
located at the beam entrance to the storage ring, provides a
measurement of the beam arrival time, which is used as the
reference start time for the spin precession measurements.  The signal
from this counter is digitized within the same system as the
calorimeters and also receives the common laser time synchronization
pulse.  A fast kicker system~\cite{Grange:2015fou} places the injected
beam onto a trajectory that allows stable storage.  The amplitude of
the momentum kick affects the amplitude of the CBO
 that
must be modeled in the data.  Finally, two stations of straw
trackers~\cite{Grange:2015fou} allow the measurement of effects arising from 
the dynamics of the stored beam that
 affect analysis of the data.

\subsection{\runone data subsets}
\label{subsec:subsets}

Over the course of the \runone dataset, the pulsed high voltage systems
(fast kicker and electrostatic ESQs) operated at several
different 
set points as we commissioned them and tuned for optimal running
conditions.  These systems play significant roles in determining the
beam  
dynamics, such as the amplitude and frequency of the CBO, which in turn can modulate the positron rate. We
therefore 
determine \oa during each operating condition individually.  
Table~\ref{tab:subsets}
summarizes the key characteristics of these four data subsets. 

\begin{table}[tb]
\caption{Summary of the \runone data subsets. The positron statistics
  correspond to those with energy greater than 1.7~GeV after a time of
  \mus{30} into a fill, according to the selection criteria
  described in section \ref{subsec:DataSelection}.}
\begin{center}
\begin{tabular}{ccccc}
\hline\hline
\runone  &  Tune & Kicker & Fills      & Positrons  \\
Subset &  (n)  &  (kV)  & ($10^{4}$) & ($10^{9}) $  \\
\hline
 1a    & 0.108 & 130    &  151       & 0.92 \\
 1b    & 0.120 & 137    &  196       & 1.28 \\
 1c    & 0.120 & 130    &  333       & 1.98 \\
 1d    & 0.107 & 125    &  733       & 4.00 \\
 \hline\hline
\end{tabular}
\end{center}
\label{tab:subsets}
\end{table}%

During this physics run, two of the 32 high voltage resistors for the
ESQs became damaged.  While the 
ESQs still operated, the
resulting 
change in resistance altered the RC time constant for some ESQ
plates and increased the time required to reach operating voltages. 
As a result, some of the
voltages varied at the beginning of the time window used
for the determination of \oa.  This variation introduced a time dependence
into the CBO-related frequencies, which could be measured directly and
incorporated into the \oa analyses (see Sec.~\ref{sec:fitting}).
The variation also introduced a time dependence to the beam width.
Because the average muon precession phase varies across the transverse
beam storage volume (due to positron acceptance effects), this change
of width introduced a time-dependent drift to the average precession
phase $\phi(t)$.  Such a phase
drift shifts the observed precession frequency and must be corrected.   
Reference~\cite{\BD} discusses the determination of the beam
storage related corrections to \oa for these four subsets in detail. 
\section{Analysis techniques}
\label{sec:histogram}

By pursuing multiple independent analyses of the muon spin precession
data, we obtain powerful cross checks on the value of the precession
frequency \oa determined from the data.  For the \runone
results described here, six analysis efforts have been developed, 
each utilizing a unique
mix of reconstruction, analysis and independent data-driven
corrections to determine \oa.  
These approaches
have varying sensitivities to potential systematic effects, as well as
varying statistical sensitivities.  This section summarizes the four
general analysis approaches that have been used to determine
\oa from the \runone data, as well as the common selection
criteria.  The six efforts draw from these four techniques to arrive
at a total of eleven determinations of \oa for each data subset.
The following sections provide the details of data
reconstruction, data correction and fitting. 

Reference~\cite{Bennett:2007zzb} provides a detailed mathematical
analysis of the statistical sensitivity for each of the approaches
described here. 

\subsection{Data selection}
\label{subsec:DataSelection}

The data selection criteria applied in all analyses include fill-level
discriminants that ensure that all critical subsystems, such as the
electrostatic quadrupoles (ESQs), the fast kickers and all the calorimeter
channels were operating in a standard, stable condition.  The criteria
identify and eliminate, for example, time intervals surrounding
sparking in the ESQ system.  Additional criteria ensured
stable, uniform conditions for delivery of the beam to the storage
ring, as well as stable magnetic field conditions. 

All analysis methods select reconstructed positron candidates
(Sec.~\ref{subsec:local_fitting})
or integrated energy samples
(Sec.~\ref{subsec:global_fitting}) that are at least \mus{30} into the
fill after beam injection. 
Prior to \mus{30},
programatic variation of the ESQ plate voltages moves the beam edges into
collimators to reduce the population of muons at the boundaries of
phase space accepted by the storage ring~\cite{\BD}.  This
procedure helps to minimize beam loss during the period over which we
observe the muon spin precession.  By \mus{30}, the ESQ
plates stabilize at their nominal value.   This start time choice also
reduces other effects,
like event pileup (Sec.~\ref{subsec:multi_positron_pileup}),
related to high detector rates at injection time
that could potentially bias \oa, yet strikes a reasonable
balance with statistical losses.   

For the \runoned subset, we shift the analysis starting time to
\mus{50} into the muon fill because of effects related to the 
damaged ESQ high voltage resistors.  Ref.~\cite{\BD}
discusses these 
effects and their corrections in detail. 

In all analyses, the precise start time of the fit
 corresponds to a node in the
anomalous precession cycle, which minimizes the sensitivity to
time-dependent effects like a gain change correlated with time into
the fill. 
The end time of the fit is at $T \simeq\,$\mus{650}, 
corresponding to approximately 10 muon 
lifetimes  at $p_{0}=3.094$ GeV/c. 

\subsection{Event-based methods}
\label{subsec:event-methods}

Within the event-based approach, an analysis selects candidate decay
positron events reconstructed with energies above an optimal
threshold, and bins them in time relative to beam injection.  The
different methods correspond to different positron weighting schemes.
These methods reflect the physical process
described in Sec.~\ref{subsec:principles-experiment}, in which the
positron rate asymmetry 
grows with increasing energy threshold because of the increasing
correlation between decay positron direction and muon spin.  With unit
weighting per positron ($w(E) = 1$), this method maps directly onto
the rate prediction of Eq.~\ref{eq:5param}, though with additional
effects from positron acceptance and beam dynamics.  Alternatively,
weighting each positron by the effective decay asymmetry at its energy ($w =
A(E)$) provides the optimal statistical
sensitivity~\cite{Bennett:2007zzb}.  Four of the analysis efforts for
\runone use both the  threshold method, with unit weighting, and the
asymmetry-weighted method. Each team extracts the asymmetry
function $A(E)$ directly from the data by binning the data in positron
energy $E$ and fitting the time distribution in each bin (see
Sec.~\ref{sec:fitting} for a discussion of the fitting method).

\begin{figure}[tb]
\centering
\includegraphics[width=.95\linewidth]{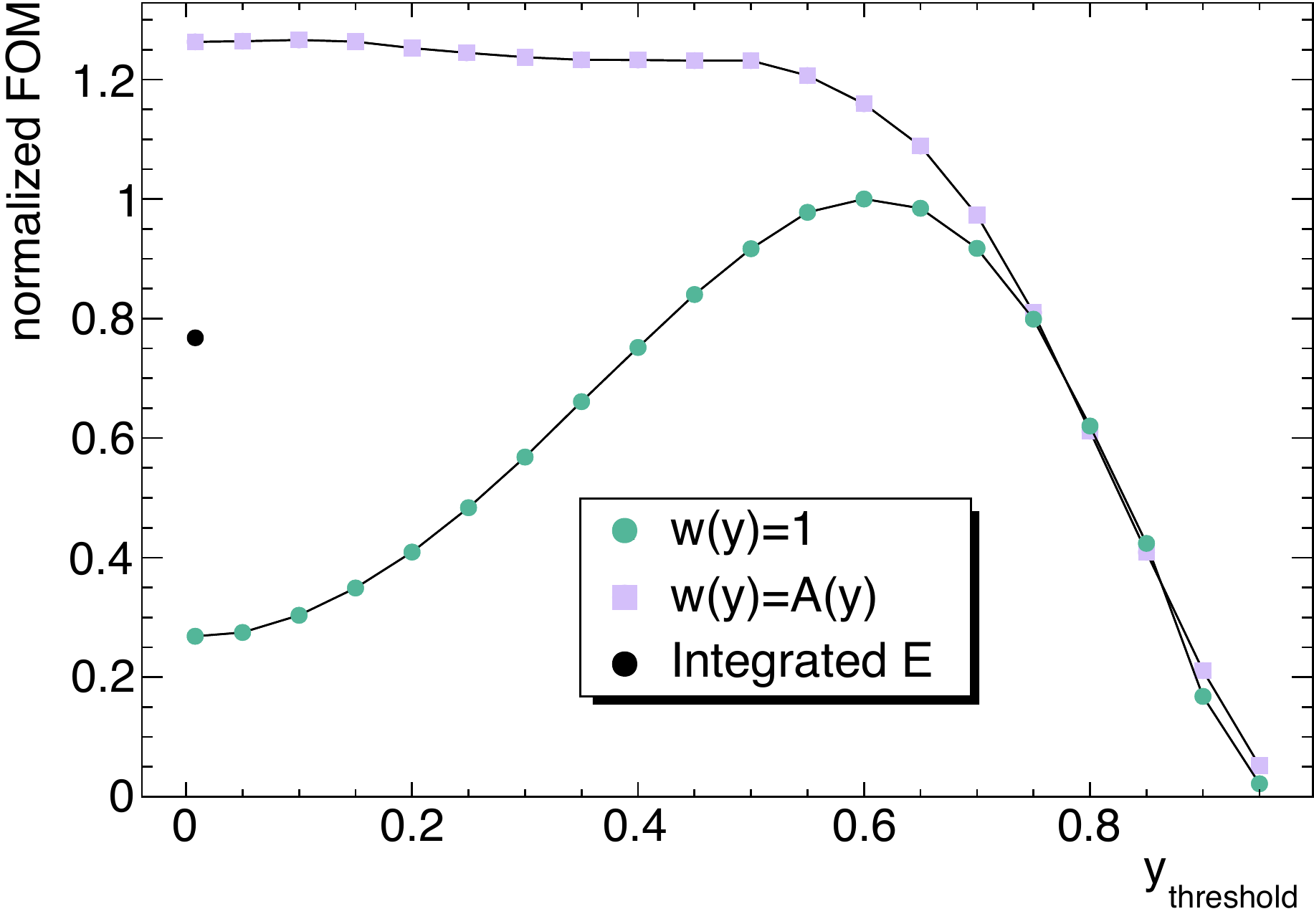}
\caption{The statistical figures of merit $N\bar{A}^{2}$ 
  calculated using a simple Monte Carlo simulation for the
  threshold method ($w(y) = 1$) and $NA_{\rm rms}^{2}$ for the
  asymmetry-weighted method ($w(y) = A(y)$) as a function of threshold
  energy. The simulation
  included basic detector acceptance.  The normalized energy
  $y = E_{e^{+}}/E_{\rm max}$, where $E_{\rm max} \approx 3.1$~GeV is the
  maximum allowed positron energy in the laboratory frame from muon decay.
  The isolated black point
  indicates the corresponding figure of merit for the integrated
  energy method in case of no energy threshold
  (Sec.~\ref{subsec:q_method}).} 
\label{fig:FOM}
\end{figure}

The inverse of the
\oa variance scales as $N\bar{A}^{2}$ for the threshold
method, where $N$ represents the total positron statistics above
threshold and $\bar{A}$ the average asymmetry, and as $NA_{\rm
  rms}^{2}$ for the asymmetry-weighted method, where $A_{\rm rms}$ is
the root mean square asymmetry above threshold.  Figure~\ref{fig:FOM}
illustrates the behavior of these two statistical figures of merit
(FOM)
from a simple Monte Carlo simulation that includes basic detector
acceptance effects but assumes perfect knowledge of the absolute
energy scale. For the threshold method, the lower energy positrons
dilute the asymmetry to an extent that overwhelms the statistical
gains, causing the overall sensitivity to drop off.  For the
asymmetry-weighted method, the asymmetry weighting itself minimizes
the dilution, and, in principle, it allows using positrons of all
energies, 
including those of negative asymmetry.  

\begin{figure}[tb]
\centering
\includegraphics[width=.95\linewidth]{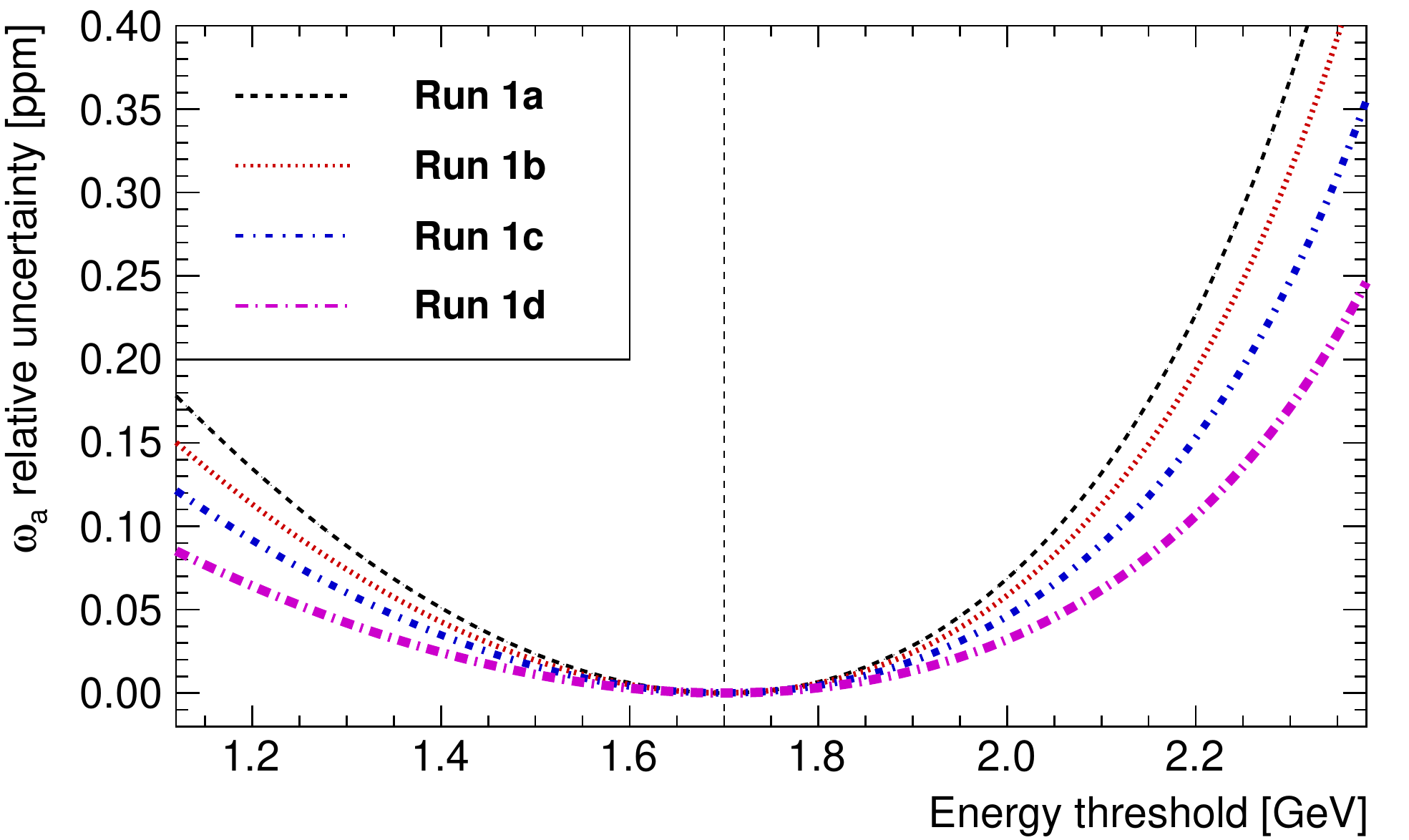}
\caption{Relative uncertainty on \oa 
 versus energy threshold for the four \runone datasets 
  determined from a simple five-parameter fit ({\it cf.}
  Eq.~\ref{eq:5param}) to
  data with varying threshold. The different curvature is due to the
  different statistics among the datasets.} 
\label{fig:threshSweep}
\end{figure}

In practice, acceptance, detector effects and uncertainties in the absolute energy
scale all affect the optimal choice of energy threshold.  For the
threshold method, a sweep over a range of threshold energies
determines the optimal threshold from the data itself.  At each trial
threshold energy, a fit to the time-binned data with the ideal
functional form of Eq.~\ref{eq:5param} provides the \oa precision
estimate. Figure~\ref{fig:threshSweep} shows a representative sweep.  The
optimal threshold occurs near 1.7~GeV for the threshold method.  For
the asymmetry-weighted method, a 1.0 GeV threshold choice balances
detector noise mitigation with the marginal statistical gain from a
lower threshold.   

\subsection{Integrated energy method}
\label{subsec:q_method}
The integrated energy method extracts the anomalous precession
frequency from the calorimeter data with a very different strategy.
Rather than using disjoint time windows with discrete positron events,
this method examines a continuous total energy sum in the calorimeters
from a combination of many muon fills.  
An energy versus time histogram is then formed from this data.
This method uses different raw data and analysis procedures, thus
inheriting different systematic sensitivities and providing
complementary statistics.
In particular, contributions from pulse pileup events and
the initially bunched muon beam, both key issues in controlling
systematic effects, require very different handling. As such, the
integrated energy method, although statistically less powerful,
remains valuable in demonstrating the robustness of the extraction of
the anomalous frequency.

\subsection{Ratio method}
\label{subsec:r_method}

The ratio method, described in detail in
Ref.~\cite{phdthesis:2020Kinnaird}, provides a way of processing the
data to remove the exponential decay and reduce any slowly
or smoothly varying effects in the data, such as muon losses. This
method can be combined with any of the event-based or integrated
energy approaches.  For the \runone results presented here, we have
applied this technique to a threshold method analysis.  Elimination of
these slowly varying effects  shifts the relative importance of
different systematic sensitivities compared to the event-based
analyses. 

\begin{figure}[tb]
\centering
\includegraphics[width=.98\linewidth]{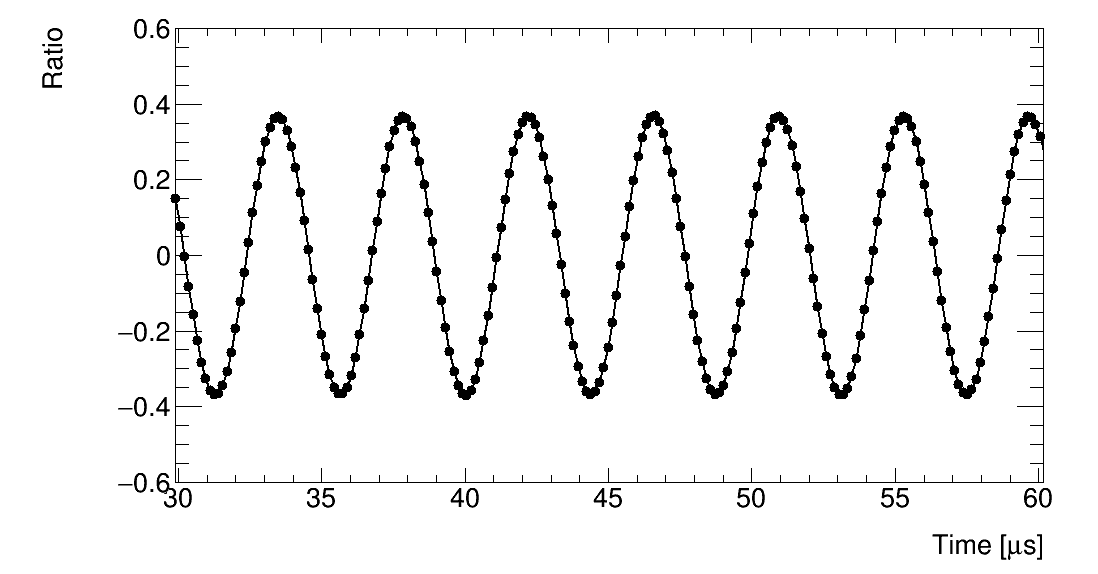}
\caption{The ratio $r(t)$ (see text) obtained from the \runoned data
  subset. The ratio preserves the amplitude and the frequency of the
  \gm oscillation, while eliminating the exponential behavior and
  reducing other slow and smooth terms.} 
\label{fig:Ratio}
\end{figure}

To eliminate the slow variations, this method randomly divides the
positron candidates into four subsets.  When time binning the data,
the times for one subset receive a shift forward by $T_{a}/2$, where
$T_{a}$ is the anomalous precession period~\footnote{$T_a$ is known at
  the ppm level from previous experiments, a precision which is more
  than sufficient for the Ratio method}, those in a second subset 
receive a shift backwards by $T_{a}/2$, while those in the other two 
remain unchanged. In terms of
the number of events $n(t)$ collected in the bin at time $t$, the
rebinning process yields the four binned functions  
    \begin{align}
        u_{+}(t) &= \frac{1}{4} n(t+T_{a}/2), \\
        u_{-}(t) &= \frac{1}{4} n(t-T_{a}/2), \\
        v_{1}(t) &= \frac{1}{4} n(t), \\
        v_{2}(t) &= \frac{1}{4} n(t),
    \end{align}
Forming the sum and difference ratio
\begin{equation} \label{eq:ratio}
r(t) = \frac{[u_{+}(t)-v_{1}(t)]+[u_{-}(t)-v_{2}(t)]}
{[u_{+}(t)+v_{1}(t)]+[u_{-}(t)+v_{2}(t)]},
\end{equation}
suppresses the exponential decay term and other slowly varying
effects.  Re-expressing the yields $n(t)$ in terms of the rate
function in Eq.~\ref{eq:5param} and expanding in $(T_A/\gamma\tau_\mu)$
the functional
form of the ratio becomes
\begin{multline} \label{eq:ratioApprox}
r(t) = 
A\cos{(\oa t + \phi)} - \frac{1}{16}\left(
  \frac{T_{a}}{\gamma\tau_{\mu}}\right)^{2} + \\
{\cal O}\left((T_{a}/(4\gamma\tau_{\mu}))^{4}\right),
\end{multline}
which illustrates the suppression of the
lifetime. Figure~\ref{fig:Ratio} presents the ratio function obtained
from the \runoned data subset. 

Reweighting the four rebinned subsets according to 
\begin{multline}
u_{+}(t):u_{-}(t):v_{1}(t):v_{2}(t) =\\
e^{T_{a}/2\gamma\tau_{\mu}}:e^{-T_{a}/2\gamma\tau_{\mu}}:1:1
\end{multline}
eliminates the last two terms in
Eq.~\ref{eq:ratioApprox} and a  simple sinusoidal description
of the ratio time series
becomes exact in the absence of beam-related
effects. Those effects, such as betatron
oscillations and  muon loss,
do not cancel exactly in the ratio, therefore this 
analysis approach 
utilizes the full functional form of $r(t)$ described in
Sec.~\ref{sec:fitting}. 

All bins in the $u$ and $v$ functions for \runone contain sufficient
statistics to allow standard Gaussian error estimation and
propagation.  With the lifetime correction factors incorporated into
the definition of the $u$ functions, the expression for the
statistical uncertainty on the $r(t)$ binned ratios becomes 
\begin{equation}
\sigma_{r}^{2}(t) = \frac{1-r^{2}(t)}{u_{+}(t)+u_{-}(t)+v_{1}(t)+v_{2}(t)}.
\end{equation}
This provides a statistical uncertainty that is comparable to the
  event-based methods.

\subsection{Finite beam length}
\label{subsec:beam_bunch_structure}

\begin{figure}[tb]
    \centering
    \includegraphics[width=.45\textwidth]{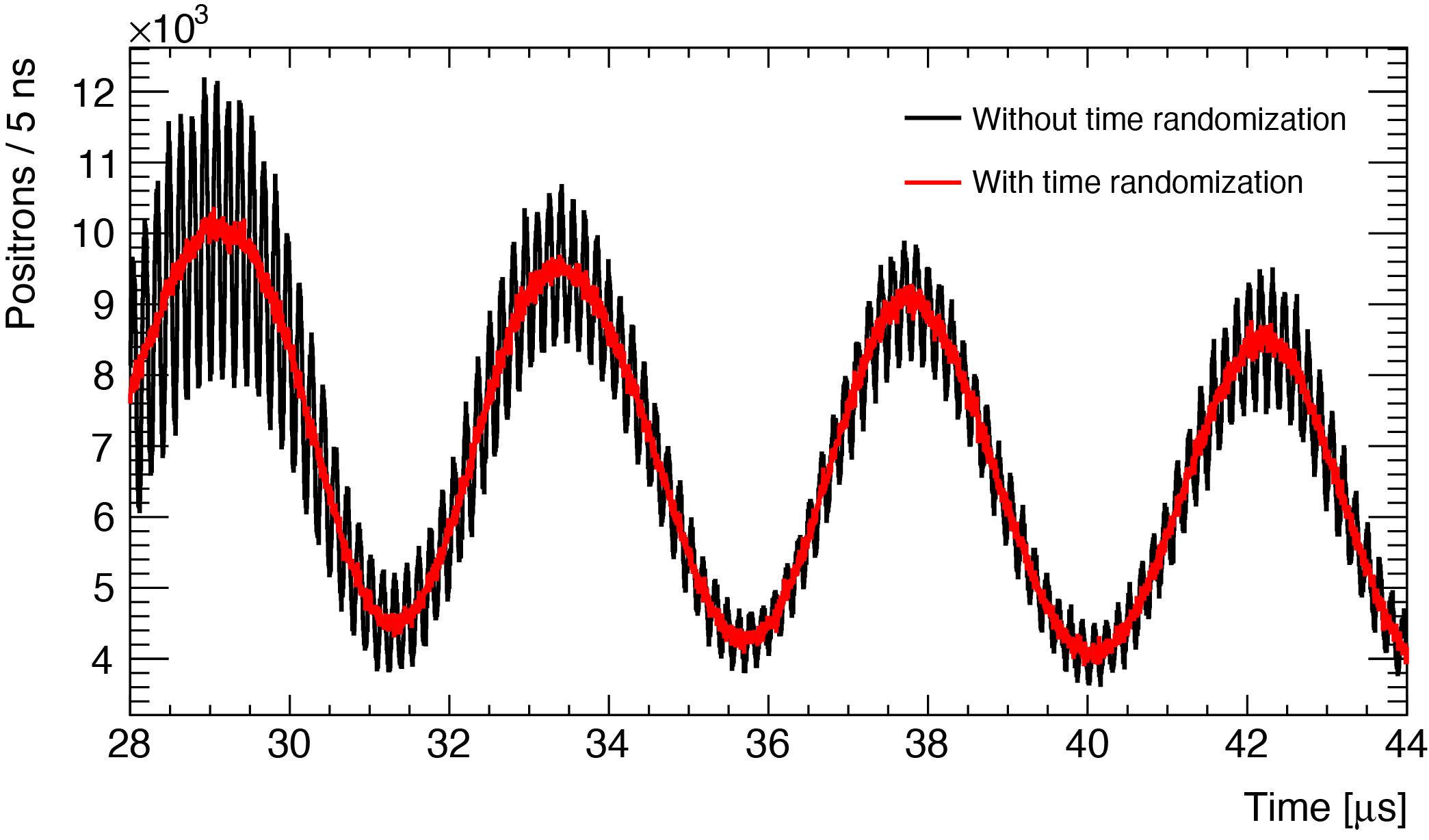}
    \caption{The positron intensity variation in one calorimeter as a
      function of time.  Unrandomized data (black) clearly show a
      variation at the $\SI{149.2}{ns}$ cyclotron periodicity on top
      of the slower (\mus{4.365}) \oa variation from the residual
      beam bunching.  Time-randomization of the data (red or grey) suppresses
      this variation, and binning in the cyclotron period suppresses further
      to a negligible level. Data are from a subset of \runone.} 
    \label{fig:fastrotation}
\end{figure}

At injection time, the 120~ns long beam does not spread evenly along the
storage ring.  As a result, the initial positron intensity at
individual calorimeter stations oscillates at the cyclotron frequency
($T_{c} = \SI{149.2}{ns}$).
The beam, however, debunches because higher momentum muons orbit at
larger radii, and therefore with longer periods, than lower momentum
muons.  After \mus{5}, the leading edge of the beam first laps the
trailing edge.  By the analysis start time of \mus{30} (approximately
two hundred orbits), the muon beam populates the ring almost
uniformly.  Figure~\ref{fig:fastrotation} shows the positron intensity
variation in one calorimeter from the residual beam bunching. 

Combining the positron data in widths of the average $T_{c}$ largely
filters out this effect, leaving only a small residual sinusoidal
trend in \oa as a function of calorimeter position.  Because of the
varying phase of this signal around the ring, summing data 
from all calorimeters
almost completely eliminates the residual effects.  As
Fig.~\ref{fig:fastrotation} also shows, randomizing the measured
positron arrival times uniformly over the interval $\pm T_{c}/2$ while
binning eliminates this effect, even at the calorimeter level.  All
event-based \oa analysis approaches for \runone employ this
randomization procedure. 

\section{Data reconstruction}
\label{sec:positron_recon}

The two raw data paths from the DAQ system, 
event and integrated energy based approaches as discussed in
Section~\ref{sec:instrumentation}, 
require distinct reconstruction algorithms. 
For the event-based analyses, the data reconstruction stage transforms
the raw waveform data in each saved time window into positron
candidates with quantities such as positron hit energies and times. We
have independently developed two methods for this positron
reconstruction: {\it local-fitting} and {\it global-fitting}. Both
fitting approaches utilize pulse templates, empirical descriptions of
each individual SiPM's response to positron showers and laser pulses,
to extract times and energies from digitizer waveforms. We construct
the template for each channel using the data, and each template includes
the well-defined oscillatory behavior for that channel after the main pulse,
which results from imperfections in the pole zero subtraction in the SiPM
readout electronics.
The physics
objects resulting from the two methods will necessarily differ
somewhat because of diverging decisions made during the respective
algorithm and software development processes. These differences
between reconstruction procedures aid in characterizing and
understanding each approach. Applying multiple reconstructions to the
same raw data helps verify correctness 
of the reconstruction and provides an important check
on systematic effects.   

For the integrated energy analysis, the reconstruction involves
careful combination of the contiguous waveforms over all crystals and
all muon fills to obtain a final integrated waveform that preserves a
good signal to noise ratio. 

\subsection{Local-fitting approach}
\label{subsec:local_fitting}

\begin{figure}[htbp]
\centering
\includegraphics[width=0.95\linewidth]{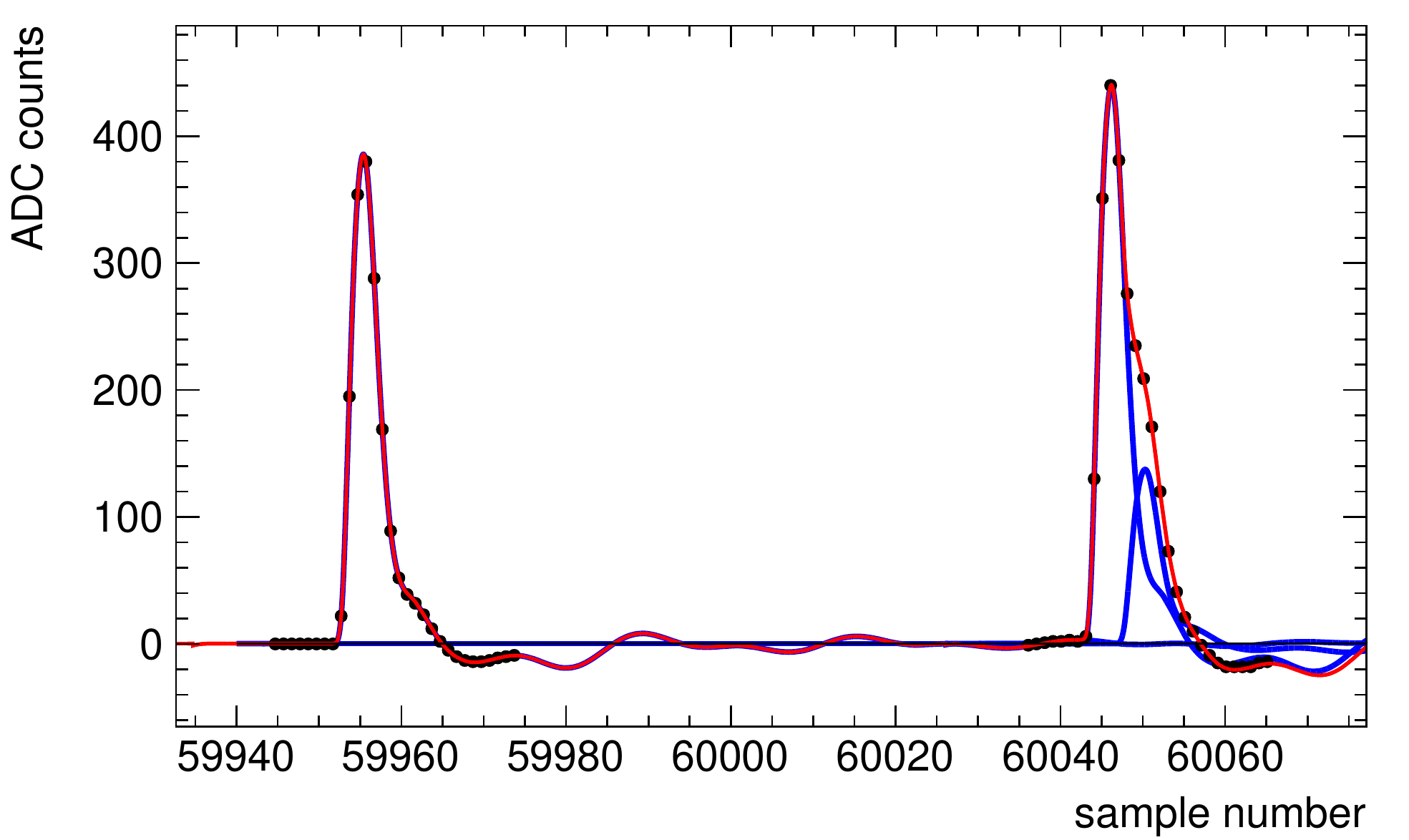}
\caption{Example of a template fit selected from \runone data. The black
  points are digitizer sample values and the smooth curves are fit
  results. Each ``sample number'' corresponds to 1.25 ns.
This figure shows a chain fit containing waveforms from two
  separate time windows and three pulses. The baseline perturbations
  from the first pulse persist into the second time window, in which
  two pulses separated by 5\,ns were identified.} 
\label{fig:templateFitResults}
\end{figure}

The local approach fits pulses with an amplitude over a configurable
threshold in each crystal
independently. References~\cite{Fienberg:2014kka, Khaw:2019yzq}
describe the template pulse fitting algorithm utilized in this step
in detail.  Should two or more pulses occur within the length of
the pulse template (250\,ns), the algorithm refits them simultaneously,
using the results of the initial fits as starting parameters, to
remove effects due to the tail of the first pulse overlapping with the
second one.  This fitting algorithm correctly
handles scenarios in which multiple pulses spread over two or more
distinct time windows from the DAQ system, as shown in
Fig.~\ref{fig:templateFitResults}. 
The individual pulses receive relative energy and timing alignment
corrections determined from studies of the minimum-ionizing-particle
(MIP) signal from muons passing through the
calorimeters~\cite{Khaw:2020}.  
Timing of all calorimeter channels gets aligned to the muon beam
arrival time through a synchronization (sync) pulse generated by the
laser system. 
All calorimeter channels and the T0 detector receive this common
sync pulse.  The difference from the sync pulse time for the
calorimeter channels' sample times compared to the beam 
arrival time
in the T0 detector provides the aligned time into fill for all channels.
Sec.~\ref{subsec:gain_fluctuation} discusses the application of gain
corrections on various timescales.  The location of the optimal
\oa threshold in each calorimeter (see
Fig.~\ref{fig:threshSweep}) then sets the absolute energy scale. 

The final step of reconstruction involves the clustering of pulses
from individual channels into a candidate positron with an estimate of
the total energy of the incident positron. The clustering combines all
pulses in a calorimeter station within a tunable artificial dead time
window into one candidate.  We have used windows of both 3\,ns
and 5\,ns for the \runone analyses. During clustering, the impact
position of the positron is also inferred using a center-of-gravity
method with logarithmic weights~\cite{Awes:1992yp,
  phdthesis:2020Sweigart}. For more details about the  local
reconstruction approach, please refer to Sec.4 of
Ref.~\cite{phdthesis:2019Fienberg}.  While not used for the \runone
analysis, 
spatial clustering can be added
to the time-based one.

\subsection{Global-fitting approach}
\label{subsec:global_fitting}

In the global-fitting approach, the algorithm simultaneously fits
clusters of pulse waveforms from multiple crystals in a given time
window from the DAQ.
This approach inherently imposes spatial separation between positrons that hit a
calorimeter close in time, reducing the size of the pileup correction
discussed in Sec.~\ref{subsec:pu_empirical_approach}. In particular,
each positron with an energy over a threshold of 60 analog-to-digital
counts (ADC), corresponding to approximately 50 MeV, 
above noise is identified with a $3 \times 3$ cluster of
crystals.  After applying a time correction to each crystal similar to
that described in Sec.~\ref{subsec:local_fitting}, the clusters
identified in the time window are fit by minimizing a $\chi^2$
described in Section~\ref{subsec:par_extract}.  
Because the SiPM pulse shape for a crystal does not depend on
the pulse magnitude~\cite{Khaw:2019yzq}, we can model each trace by a
crystal-dependent template that scales with energy and translates with
time. The pulse magnitude for each crystal pulse  floats independently
in the fit. The algorithm constrains the templates for each crystal to
peak at a shared time. Clusters that share one or more crystals must
be separated by at least 1.25\,ns; otherwise, they will be merged into
one larger cluster.  When a pulse template extends across multiple
time windows, the algorithm refits all identified clusters within
these windows simultaneously.  Relative energy corrections determined
using the MIP energy peak from muons adjust the pulse amplitude for
each crystal in the cluster. 
An \oa energy threshold scan
determines the absolute energy, similarly to the
  local-fitting approach (Fig.~\ref{fig:threshSweep}).
A refined version of the
center-of-gravity method with logarithmic energy weights provides an
estimate of the position of each cluster.  For more details about this
reconstruction approach, refer to Ch.~4 in
Ref.~\cite{phdthesis:2020Sweigart}.

\subsection{Integrated energy waveform}
\label{subsec:qmethodrecon}

\begin{figure}[tb]
\centering
\includegraphics[width=\linewidth]{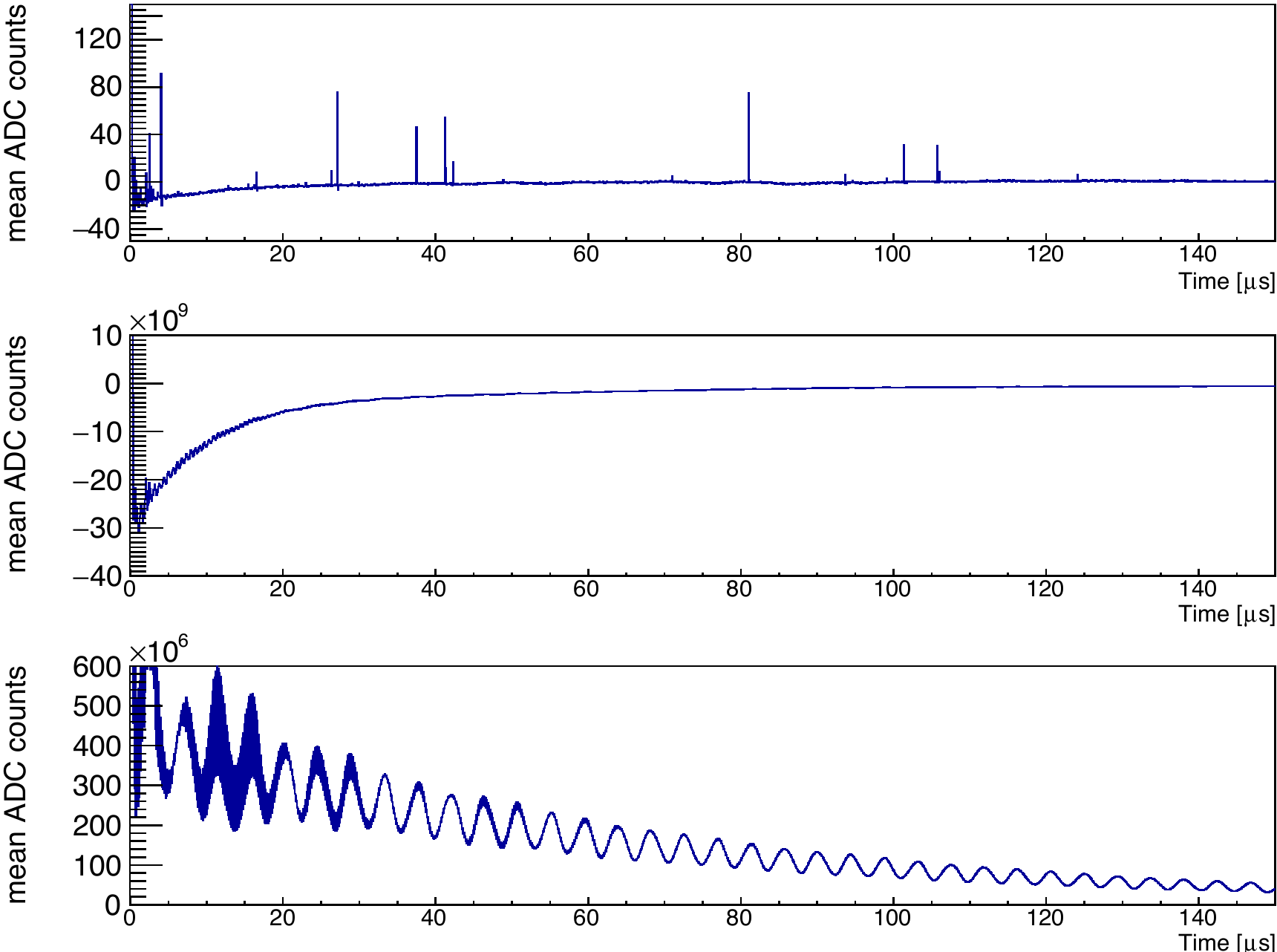}
\caption{Representative single-fill rebinned waveforms for a high
  rate crystal in Calorimeter 12 (top), 
the corresponding waveform sum over all crystals and all fills in
  \runonec for Calorimeter 12 
(middle), and the above threshold
  integrated energy waveform (bottom).  
The vertical axis of time-decimated ADC counts  is the mean value of
the 60 raw ADC samples of each time-decimated bin. 
The beam injection and
  pedestal recovery signals appear clearly for both the single-fill
  and summed time distributions.  While individual positron pulses
  appear clearly in the single-fill distributions, the pedestal
  structure overwhelms their contribution in the summed distribution. 
\label{fig:qHist}}
\end{figure}

As discussed in Sec.~\ref{sec:instrumentation}, 1296 contiguous,
time-rebinned, crystal-by-crystal waveforms comprise the integrated
energy dataset. These waveforms span a time period of $-$\mus{6}$\,< t
< +$\mus{231} relative to the beam arrival time with $75$\,ns wide
bins. The reduced time range and increased time binning were chosen to
limit the rate and volume of the integrated energy data.
Ideally, a simple sum of the waveforms over the 54 crystals from a
calorimeter would yield the integrated energy waveform for that
calorimeter.  As Fig.~\ref{fig:qHist} illustrates, while positron
pulses appear clearly in single-fill waveforms, the O($100\,$ns)
pedestal recovery structure overwhelms the positron precession signal
in the waveform over all fills in a dataset.
 We have therefore developed a threshold integration method to
separate the integrated time distribution from the pedestal variation. 

\begin{figure}[tb]
\centering
\includegraphics[width=0.95\linewidth]{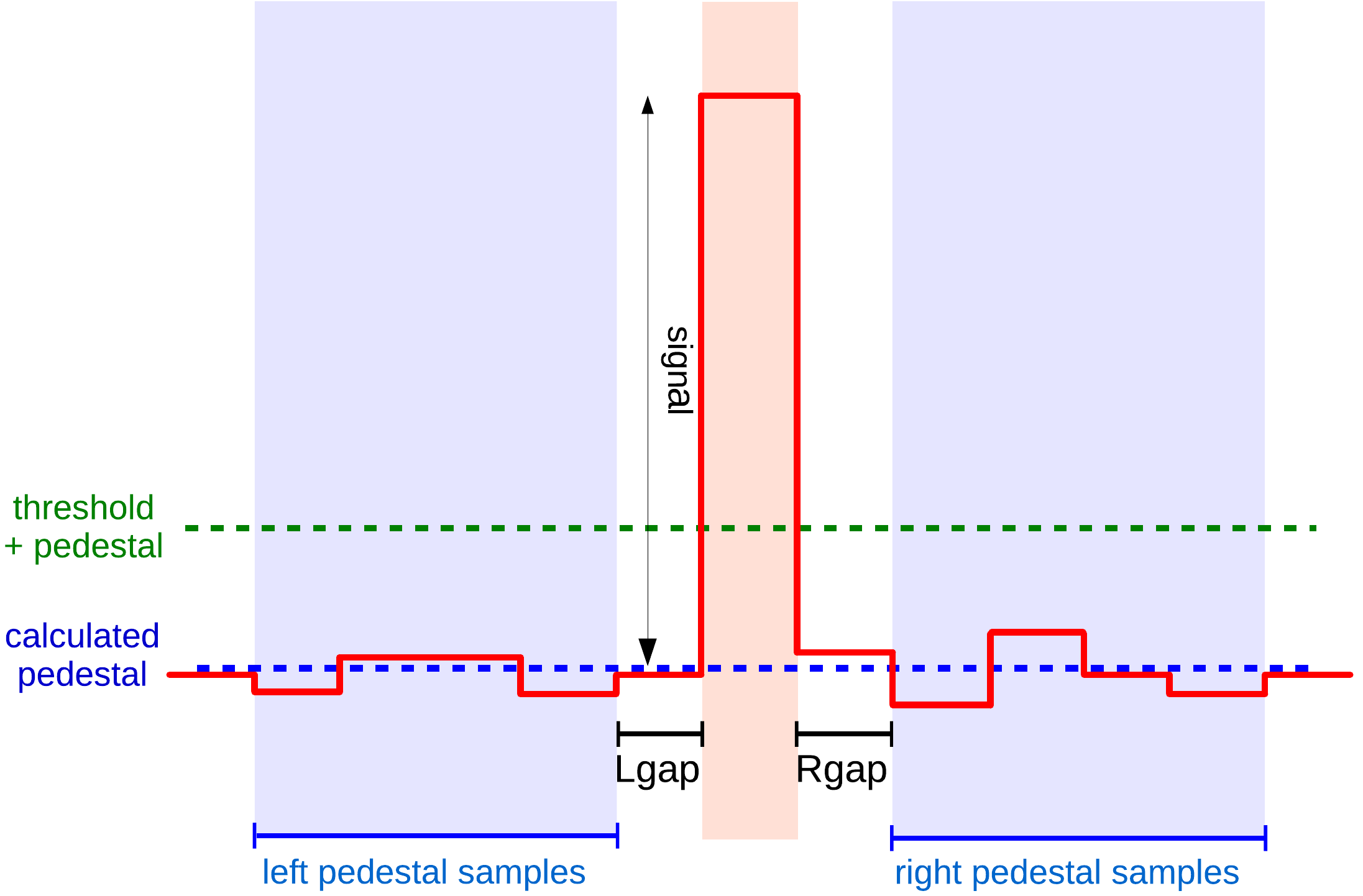}
\caption{Diagram illustrating the pedestal calculation algorithm  and
  the application of the threshold for the threshold integrated energy
  waveforms. The mean of the below-threshold
  samples in the left / right pedestal
  window provides the pedestal estimate. $L_{\rm{gap}}$ /
  $R_{\rm{gap}}$ are adjustable gaps between the time bin and the left
  / right pedestal windows. 
\label{fig:thresholdQ}}
\end{figure}

Fig.~\ref{fig:thresholdQ} depicts the threshold integration
method. For each fill-level crystal waveform from a calorimeter, a
rolling pedestal algorithm provides a pedestal estimate at each time
bin.  After gain correction (see Sec.~\ref{subsec:gain_fluctuation}),
any pedestal-subtracted energy that exceeds a pre-defined threshold
setting is added to the threshold integrated energy waveform $E(t)$ for that
calorimeter.   

In the \runone analysis, the mean value of the below-threshold
 ADC samples in equal-sized
time windows to the left and right of each time bin
provides the pedestal estimate.  To avoid biases from pulse
undershoot and ringing in the estimate, the algorithm introduces a gap
between the pedestal windows and the time bin.  The threshold setting,
pedestal window size, and gap size are all adjustable parameters
common to all 
crystals. The nominal settings in processing \runone
data correspond to a threshold setting of $\sim$300\,MeV, left and
right pedestal windows of 300\,ns, and left and right gap sizes of
75\,ns.  

While the event based methods use time randomization to ameliorate the
residual effects of the finite beam length (see
Sec.~\ref{subsec:beam_bunch_structure}), correction of the integrated
energy waveform requires a different approach.  Combining the above
waveforms pairwise into $T_{b} = 150$\,ns wide bins, which is close
to the cylcotron period $T_{c} = \SI{149.2}{ns}$, would suppress these
effects. However, an aliased modulation at a frequency $f_{\rm alias}
= 1/T_{c} - 1/T_{b}$ would persist.  We instead employ a smoothing
algorithm to combine the 75\,ns binned waveform $\{E_{i}^{75}\}$ into the
150\,ns binned waveform $\{E_{i}^{150}\}$ via 
\begin{equation}
E^{150}_{i} = \frac{1}{4} E^{75}_{2i-1} + \frac{1}{2} E^{75}_{2i} + \frac{1}{4} E^{75}_{2i+1}.
\end{equation}
where $i$ refers to the bin number of the 150\,ns wide binned data.
This approach eliminates both the fundamental and the aliased
modulations.  While the procedure introduces bin-by-bin correlations,
these can be accommodated straightforwardly in subsequent fitting
procedures. 

The associated uncertainties for the above-threshold, integrated
energy, histogram bins  
were computed using Poisson statistics. Given a bin energy 
$E = \sum_j E_j$, obtained 
by summing recorded positron energies $E_j$, the associated bin
uncertainty is $\sigma =  (  \sum_j E^2_j )^{1/2}$.  
Small corrections arise from
effects of positron pileup and the division of a positron's energy
between two adjacent time bins. Such effects are
order 10$^{-2}$ on the normalized $\chi^2$. 

\section{Data corrections}
\label{sec:data_corrections}

A number of time-dependent effects require application of corrections 
to the reconstructed data to avoid bias in \oa.
These effects include gain variations on a
number of timescales, 
pileup effects in the
calorimeters, and
the loss of beam muons through mechanisms other than decay.

\subsection{Detector gain fluctuation and time synchronization}
\label{subsec:gain_fluctuation}

The energy scale of each calorimeter channel can vary with external
factors such as temperature and hit rate.  
These effects occur over  different timescales: hours or days for
temperature-related effects, and microseconds or tens of microseconds for
effects related to muon rate. 
A laser calibration system~\cite{Anastasi:2019lxf} provides the ability to 
correct for these effects.  The system operates in
different modes to provide correction functions at different time
scales: {\it long-term correction} for daily effects, {\it in-fill
  gain correction} for the tens of microseconds scale, {\it short term
  gain correction} for hits which are tens of nanoseconds apart. 

The above-mentioned 
effects
affect the physics output in different
ways. In particular, any variation of the calorimeter response between
the beginning and the end of a fill,
if uncorrected,
 results in a early-to-late energy threshold variation
and thus in a potential shift of \oa as
mentioned in Sec.~\ref{sec:instrumentation}. 

The E989 systematic uncertainty goal related to detector gain variation is
20\,ppb, which requires control of systematic gain changes over the
\mus{700} long muon fills better than 0.5 per
mille (see Fig. 16.5 in~\cite{Grange:2015fou}).    
The long-term corrections,
which do not couple as directly to the determination of \oa, do not
require as strict a control.

\begin{figure}[tb]
\includegraphics[width=1.0\linewidth]{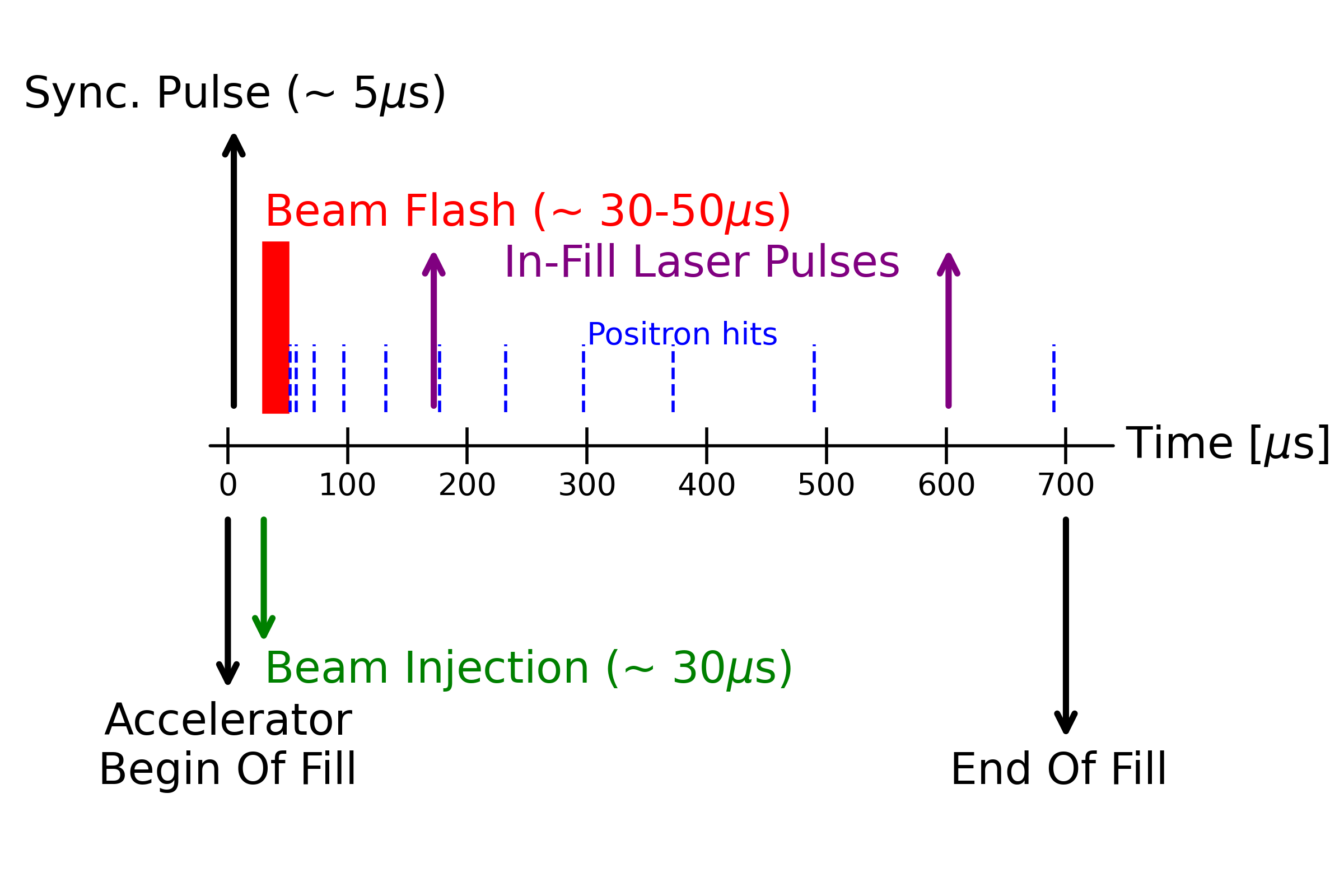}
\caption{Timing
of the sync pulse and representative in-fill  pulses provided
by the laser system.}
\label{fig:IFL}
\end{figure}

Reference~\cite{Anastasi:2019lxf} provides details of the laser system. 
Briefly, a programmable Laser Control Board triggers a
pattern of laser pulses which illuminate, during
standard data taking, the
calorimeter crystals through quartz fibers coupled to the crystal
face. The amount of emitted light approximately corresponds to an
energy release of 1 GeV.
Figure~\ref{fig:IFL} shows a schematic of this pattern, which
includes a reference signal issued before injection that provides
precise time synchronization, and a set of pulses during a
fraction
 of the muon fills that accurately measure the detector
response as a function of rate.  
An additional set of pulses between fills (not shown)
provide the long term calibration.
The fills with laser pulses are not to be used for the
  analysis, as the laser itself modifies the detector response.
  Therefore only a fraction of approximately 10\% of the muon fills
  include the laser pulses.

\begin{figure}[tb]
\includegraphics[width=0.95\linewidth]{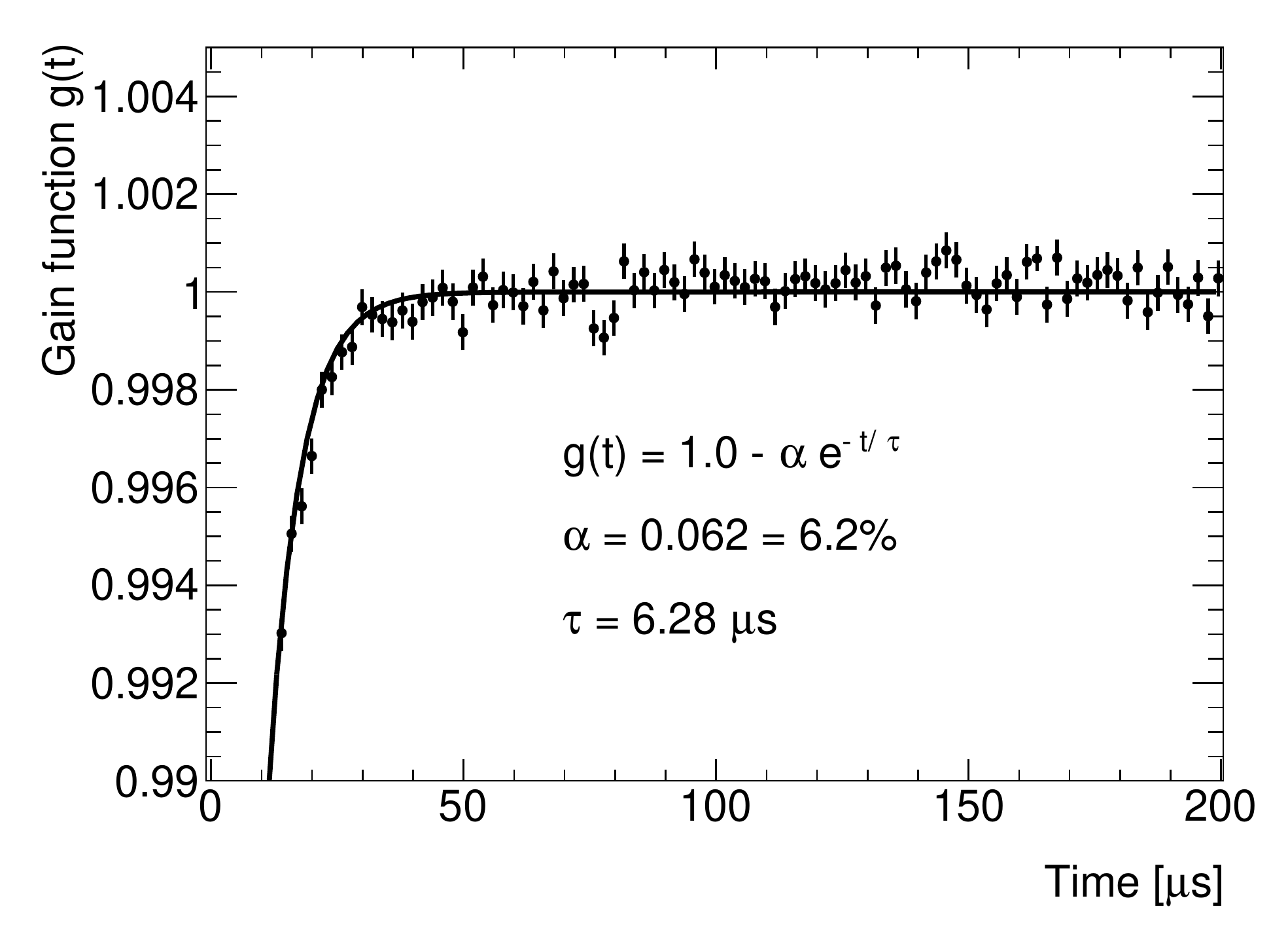}
\caption{The gain function describes the gain drop in the photodetection 
  system for a representative calorimeter channel due to the flash of muons and beam positrons
  at injection. It is
  expanded to show the behavior of the gain after $t=30\,\mu s$, the starting
  time of the \oa fit (see section \ref{subsec:DataSelection}).
\label{fig:IFG}}
\end{figure}

Figure~\ref{fig:IFG} shows a representative gain curve for a
single crystal, as measured by the laser calibration system, during
the first \mus{200} after muon injection.   
The initial gain sag, clearly visible at the time of injection, results
from SiPM charge depletion that occurs when the initial flash of particles,
accompanying the storable muon beam at injection, strikes the calorimeters.

A model for the gain function based on an exponential decay
returning asymptotically to unity, 
with average amplitude of approximately  6\% and time constant of order
\mus{6}, adequately describes the calorimeter response to laser data.
Thus  \mus{30}
after injection, the start time of the \oa fit, 
the gain correction is at the per mille level 
and it rapidly decreases to zero. 
While small, this correction is not negligible 
and its effect on \oa  is discussed in Sec.~\ref{subsec:gainsys}.

\begin{figure}[tb]
\includegraphics[width=0.95\linewidth]{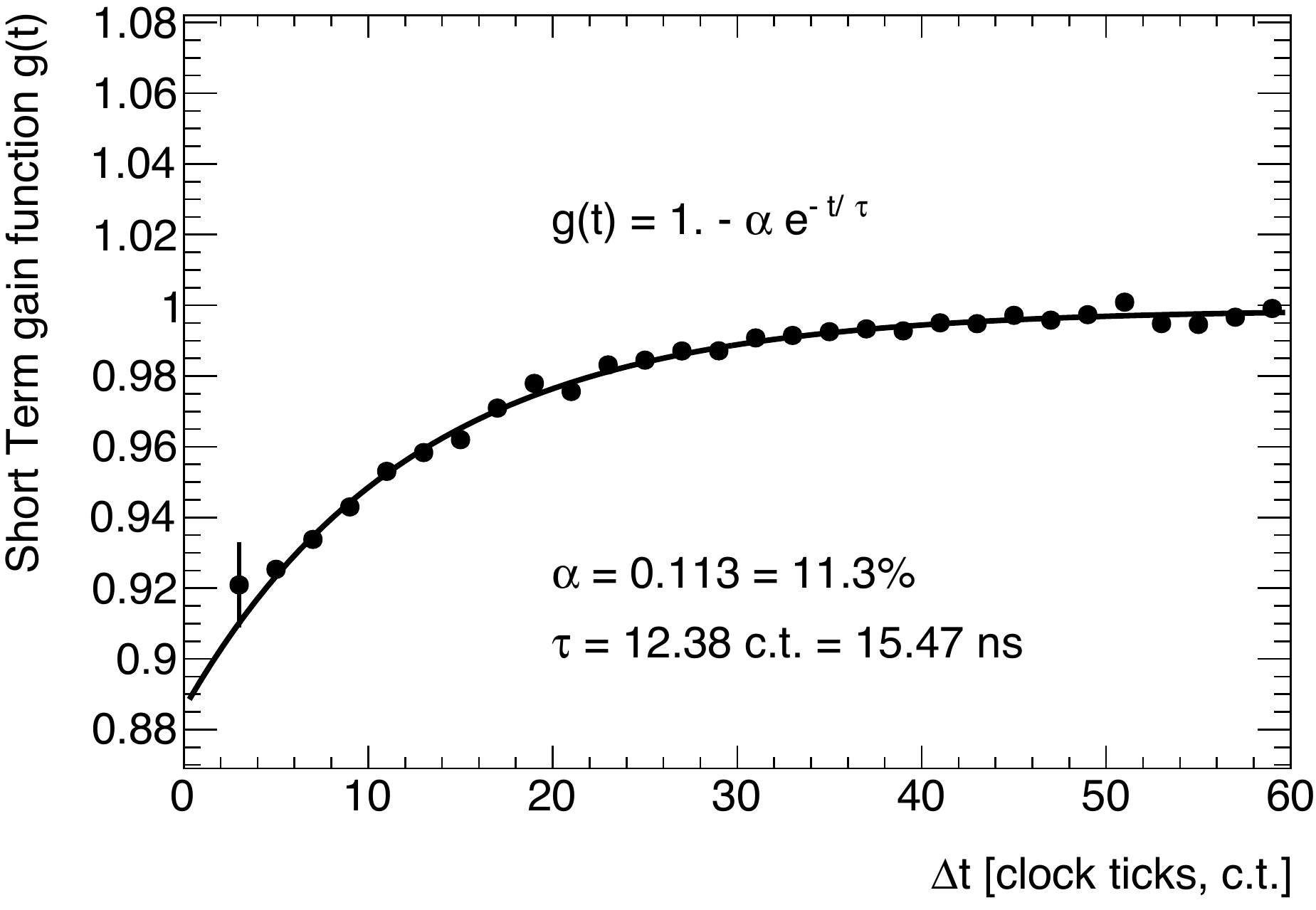}
\caption{Gain correction function for consecutive hits in the same
  crystal as a function of their time separation expressed in {\it
    clock ticks}. The clock sampling frequency is 800 MHz and 1 clock
  tick corresponds to  1.25 ns.}   
\label{fig:STDP}
\end{figure}

When two positrons hit the same crystal
within a few tens of nanoseconds, the finite recovery time of the SiPM
and amplifier can reduce the gain experienced by the second particle.  
We map this short term gain correction as a function of energy and time
by redirecting the laser light so that two lasers can pulse
a set of crystals with programmable delay and intensity.
Figure~\ref{fig:STDP} shows the gain drop for a typical channel.
The amplitude varies linearly with the energy of
the first particle with an average slope 
of 5\%/GeV, while the
exponential recovery time has an average value of $\ns{15}$.

While the short time correction can be readily applied to the event-based
analysis, in which single positron clusters are selected,
the integrated energy method requires a different approach.
A second in-fill gain correction is determined 
which combines the gain drop effects due both to the initial muon
flash and to the hit of consecutive positrons, providing an average
combined correction.

\subsection{Multi-positron pileup}
\label{subsec:multi_positron_pileup}


The positron reconstruction approaches described in
Sec.~\ref{sec:positron_recon} cannot resolve multiple positrons that
strike a calorimeter 
sufficiently close in time or space. Event-based analyses must account
for such pileup 
by statistically subtracting a constructed pileup
spectrum. Without this correction, the unresolved pileup could bias
the fitted \oa in 
Sec.~\ref{sec:fitting} by as much as $\mathcal{O}(100~\text{ppb})$.
The integrated energy approach, by design, has no inherent pileup
bias,
in the limit of zero energy threshold,
because it looks only at total energy and does not need to associate
energy contributions to individual positrons. 
This subsection presents three different approaches developed to
correct for the pileup contamination present in the spectrum of
reconstructed positrons. 

\subsubsection{Shadow window approach}
\label{subsec:Shadow_window_approach}

The shadow window approach described here builds on and refines the original
algorithm developed for the BNL E821
experiment \cite{Bennett:2006fi}.  Reference \cite{phdthesis:2020Kinnaird}
provides further details on the algorithm and attendant modifications
of the statistical uncertainties of the positron data. 

The algorithm assumes that the probability of observing a pileup
positron (doublet)  
equals that for observing two individual positrons
(singlets) that are separated in time by an amount much smaller than the
cyclotron period. The shadow window method searches in 
a fixed time window
after a given positron (the trigger) for 
a second trailing positron (the shadow). A time offset 
$T_{G}$, also called shadow gap time,
from the trigger and a shadow 
window width $T_{D}$ define the search window.  

When the shadow window contains a positron, the trigger (T) and shadow (S)
positrons are combined into a shadow doublet with energy and time
         \begin{gather}
            E_{\text{doublet}} = C \cdot (E_{\rm T} + E_{\rm S}), \label{eq:Edoublet} \\
            t_{\text{doublet}} = \frac{t_{\rm T} \cdot E_{\rm T} + (t_{\rm S}-T_{G}) \cdot E_{\rm S}}{E_{\rm T} + E_{\rm S}} + \frac{T_{G}}{2}. \label{eq:tdoublet}
        \end{gather}
The constant $C$ in the energy sum corrects for a response difference of the
calorimetry for true pileup compared to the resolved positrons. The
\runone analyses employing the shadow window approach use the nominal
value $C=1$. The energy-weighted time of the two 
singlets provides the time for the doublet, with a shift of $T_{G}/2$
that accounts for the muon flux variation 
 across that gap time.   

Application of this procedure to all time-ordered positron candidates
within each fill provides a 
data driven statistical estimate of the pileup contamination.  Pileup
distorts the data time distribution by adding the doublets while
removing the individual positron contributions.  Therefore the
difference                                  
\begin{equation} \label{eq:ShadowPileup}
P(E,t) = D(E,t) -  S_{T}(E,t) - S_{S}(E,t),
\end{equation}
where $D(E,t)$ is the distribution of doublets, and $S_{T}(E,t)$ and
$S_{S}(E,t)$ 
are the distribution of trigger and shadow singlets respectively,
provides the correction to be subtracted from the reconstructed time
series. The single positrons used to build up the doublet
enter in $S_{T}(E,t)$ and $S_{S}(E,t)$ shifting  their time 
to $t=t_D$.

For the \runone analyses that employ the shadow window method, the
shadow window width 
$T_{D}$ is tuned depending on the specific analysis
artificial dead time parameters, with a value typically close to
5\,ns. The shadow gap time $T_{G}$, typically near 10\,ns, has been
tested for values ranging from 10~ns up to the beam cyclotron period
of $\sim 150$\,ns.

\subsubsection{Empirical approach}
\label{subsec:pu_empirical_approach}

The shadow window approach is based on models for how the reconstruction in Sec.~\ref{subsec:local_fitting} would treat two positron hits close in time or space. 
To avoid such modeling challenges, we have developed a more empirical approach where the multiple pulses are superimposed at the waveform level \cite{phdthesis:2020Sweigart}.  
The use of the reconstruction directly on the combined waveforms eliminates the need for modeling behavior of the global reconstruction (Sec.~\ref{subsec:global_fitting}).

This algorithm first identifies pairs of reconstructed clusters that
spatially overlap and fall within $149.2 \pm 5.0$ ns of each other,
corresponding to a cyclotron period.   
For each pair, the raw time windows
 are corrected for gain effects, such as
the short term 
effect (Sec.~\ref{subsec:gain_fluctuation}), and
superimposed. 
The reconstruction algorithm is then run on this
combined time window
(Sec.~\ref{subsec:global_fitting}) and
in case a single cluster is identified, it populates the
energy-time distribution $\rho_{1+2}(E,t)$, while the original
clusters  populate
$\rho_1(E,t)$ and $\rho_2(E,t)$.  The difference 
\begin{equation}
\label{eq:empirical-pileup-1}
\delta \rho_\text{pileup} (E, t) = \frac{\rho_{1+2} (E, t) - \rho_1 (E, t) - \rho_2 (E, t)}{2}
\end{equation}
provides the pileup spectrum correction, with 
the factor of $1/2$ correcting for combinatorics.  
Subtracting $\delta
\rho_\text{pileup} (E, t)$ from the reconstructed spectrum
statistically corrects it for pileup. 

Because pileup contaminates the sample of single clusters themselves,
the pileup spectrum in Eq.~\ref{eq:empirical-pileup-1} requires a
correction for higher-order pileup.   
In particular, each of the two singlets is contaminated by the two-positron pileup rate, so the next order correction can be determined by extending the above procedure
to include triplets of reconstructed clusters.  The above superposition and reconstruction procedure of different combinations of three raw waveforms produces four energy-time distributions, one for the triple combination and one for each of the three pairings.
The combination
\begin{align}
\delta &\rho_\text{correction} (E, t) \nonumber \\
= &- \left[ \rho_{1+2+3} (E, t) - \rho_{1} (E, t) - \rho_{2} (E, t) - \rho_{3} (E, t) \right] / 2 \nonumber \\
&+ \left[ \rho_{1+2\phantom{+3}} (E, t) - \rho_{1} (E, t) - \rho_{2} (E, t) \right] / 2 \nonumber \\
&+ \left[ \rho_{2+3\phantom{+1}} (E, t) - \rho_{2} (E, t) - \rho_{3} (E, t) \right] / 2 \nonumber \\
&+ \left[ \rho_{1+3\phantom{+2}} (E, t) - \rho_{1} (E, t) - \rho_{3} (E, t) \right]
\end{align}
gives the correction to be added to Eq.~\ref{eq:empirical-pileup-1}
(for details see~\cite{phdthesis:2020Sweigart}).  The
indices on the energy-time distributions indicate the time order of
original cluster candidates when the corresponding waveforms are
superimposed.  For \runone, no corrections beyond this order are
necessary to sufficiently correct the reconstructed spectra for
pileup. 

As in the case of the shadow window approach, the determination of \oa from the pileup-subtracted time series uses an exact calculation of the bin uncertainties~\cite{phdthesis:2020Sweigart}.  

Overall, this empirical approach provides an excellent description of the pileup events present in the reconstructed data.  This method is also robust against modifications to the reconstruction algorithm.  In addition, it avoids the need for simulation to characterize, for example, the possible dependencies of $C$ in Eq.~\ref{eq:Edoublet}.  Ref.~\cite{phdthesis:2020Sweigart} provides further detail about the procedures and characterization for this approach.

\subsubsection{Probability density function approach}


\begin{figure*}[tb]
\centering
\includegraphics[width=.48\linewidth]{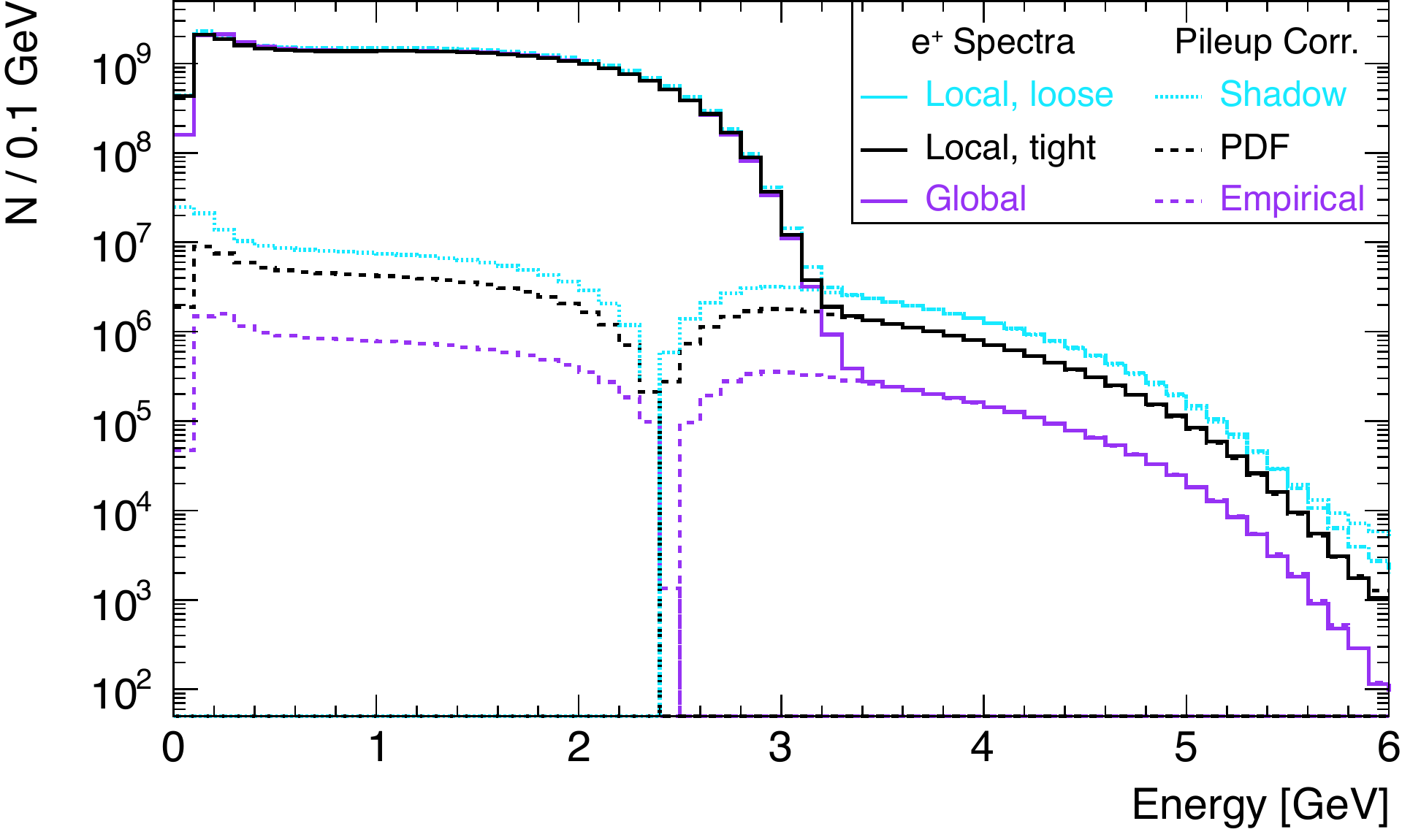}
\includegraphics[width=.48\linewidth]{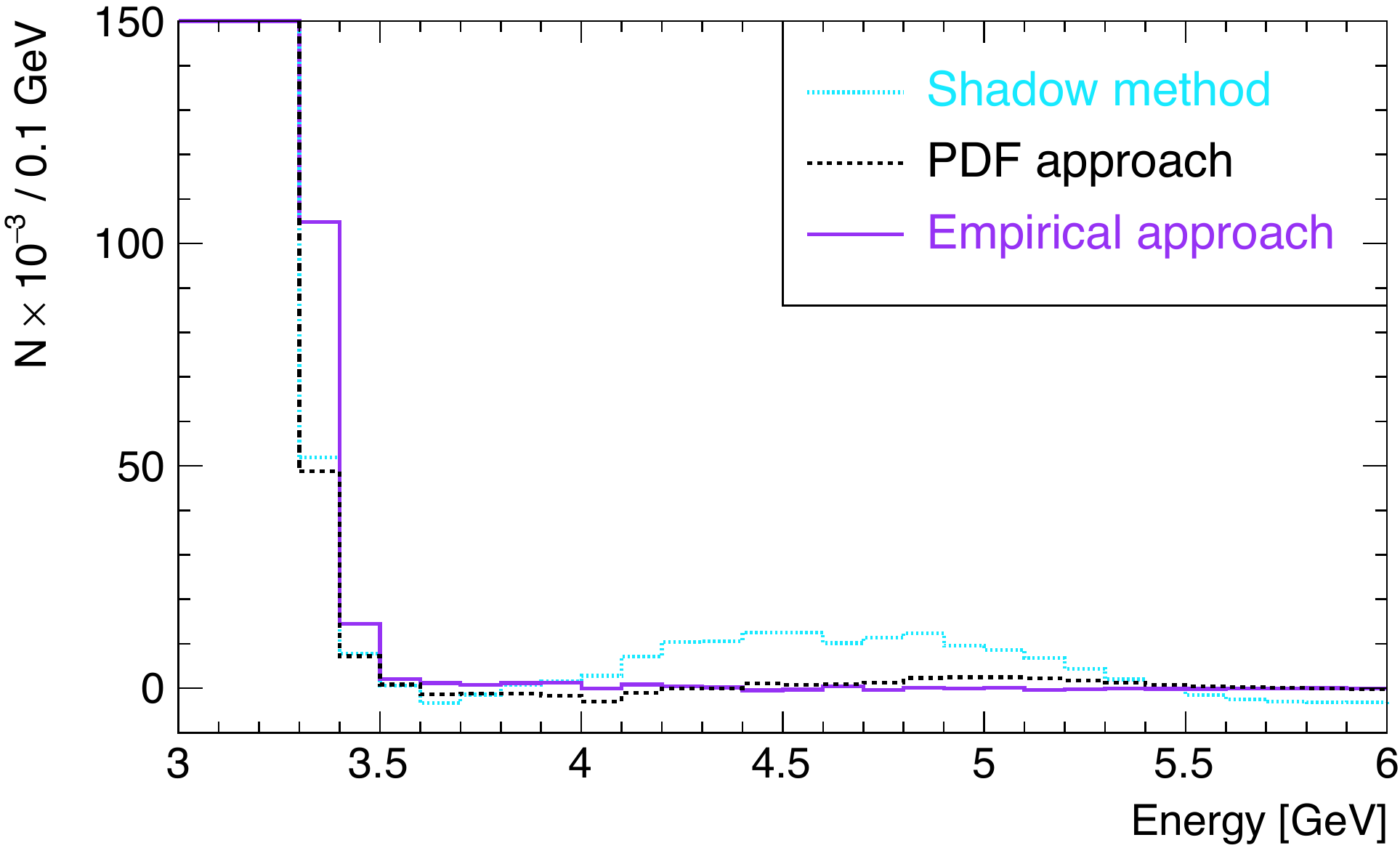}
\caption{Measured energy spectrum, summed over all calorimeters,
  along with the total pileup correction. 
Left:  number of
  positrons per energy interval.  The three solid curves
  correspond to the uncorrected spectra from the different clustering procedures, the global-fitting
  approach (purple or dark gray), local-fitting with tight clustering cuts (black), and
  local-fitting with loose clustering cuts (blue or light gray). The dashed lines
  correspond to the associated pileup correction evaluated with three different
  methods: empirical approach (purple or dark gray), probability density
  function approach (black), shadow method approach (blue). The correction curves show the
  absolute value of the pileup contribution, which has a negative sign (events
  to be subtracted from energy spectrum) above $E \sim 2.4$ GeV and a positive sign (events
  to be added to the energy spectrum) below $E \sim 2.4$ GeV.
Right: The corrected spectra obtained from the (signed) difference between solid and
  dashed lines for the three methods. Few residual
  events remain in the unphysical region above the endpoint of the
  spectrum which, due to detector resolution, extends up to 
  $E \sim 3.4$ GeV.}
\label{fig:eDepPileup}
\end{figure*} 

Unlike the previous approaches, where the pileup spectrum is  created
by ``combining" two clusters or waveforms, the {\it probability
  density function approach} constructs the pileup spectrum by
considering the energy-time distribution of an entire dataset. 

Let $\rho(E,t)$ represent the ideal calorimeter hit distribution that
would be measured by a detector with perfect resolution in time and
space and  $\delta\rho_{pu,d}(E,t)$ the double pileup
perturbation. The sum $\rho_{pu,d}(E,t) = \rho(E,t) +
\delta\rho_{pu,d}(E,t)$ 
describes the effect of two-particle pileup $\rho_{pu}(E,t)$.  
A leading-order estimate of $\delta\rho_{pu,d}(E,t)$ yields~\cite{phdthesis:2019Fienberg}
%
%
%
\begin{equation}
    \begin{array}{l}
\delta \rho_{pu,d}(E, t) = \\
r(t) \cdot \Delta t \bigg[\rho_{d+}(E, t) - 2\rho(E,t)\int\rho(E_2,t) \diff E_2 \bigg]~,
\end{array}
\label{eqn:drhodoub}
\end{equation}
with the double pulse sum term defined as
\begin{equation}
    \rho_{d+}(E, t) \equiv \int \rho(E-E_2, t)\cdot\rho(E_2,t) \diff E_2.
\end{equation}
The parameters $\Delta t$ and $r(t)$ represent the detector reconstruction dead time and the overall hit rate as a function of time, respectively. The first term in Eq.~\ref{eqn:drhodoub} corresponds to the false counts measured when two positron showers are mistaken for one, and the second term corresponds to the two true positron showers that are lost. The former will in principle be affected by nonlinearities in the treatment of unresolved pulse pairs by the reconstruction. These nonlinearities are not included in the pileup correction approach described here. 

Eq.~\ref{eqn:drhodoub} describes the contamination of the measured
energy spectrum from double pileup in terms of the uncontaminated
spectrum $\rho(E,t)$.  By iterative application of the expression
starting with the measured
hit spectrum, which is itself
contaminated by pileup, Eq.~\ref{eqn:drhodoub} can also generate the
pileup correction.
Because the relative double pileup contamination appears at the order
$r(t)\cdot\Delta t$, even with a conservative detector reconstruction
dead time and no spatial cluster separation employed in the
reconstruction, $r(t)\cdot\Delta t$ distorts the term in brackets by at most 1\% to 2\%. 
Thus, use of the pileup contaminated hit spectrum, instead of the ideal
one, to generate the expected double pileup contamination 
distorts the correction by order $r^2(t) \Delta t^2$, or $10^{-4}$.  Repeating this procedure using the spectrum $\rho_{c}(E_2,t)$ obtained from the first correction estimate yields a final spectrum also correct to order $r^2(t) \Delta t^2$.
These key observations motivate this pileup correction method.  
One can also determine the expected contamination from triple pileup,
which appears at order $r(t)^2\Delta t^2$.

The treatment of double pileup shown above assumes that all pulse
pairs within the detector reconstruction dead time of one another will
yield a false count at the summed energy and the loss of a count at each
of the two constituent pulse energies. This assumption is not valid when three pulses 
all fall within the
reconstruction  dead time. In this case, one expects a
loss of three true counts and a gain of one false count. A simple application of the double
pileup treatment, however, would count three pulse pairs and thus
erroneously remove six true counts and add three false counts. A
triple pileup correction must then account both for the
reconstruction's treatment of groups of three unresolved pulses and
for the error in the double pileup correction that occurs at the order
of triple pileup. Ref.~\cite{phdthesis:2019Fienberg} shows  that the correction %
%
%
\begin{equation}
    \begin{array}{l}
    \delta \rho_{pu, t}(E,t) =\\
     r(t)^2 \Delta t^2 \bigg[ \int\rho(E-E_d)\cdot\rho_{d+}(E_d, t) \diff E_d\\
  - 3 \rho_{d+}(E,t)\cdot\int \rho(E_3,t) \diff E_3 \\
    + 3 \rho(E, t) \cdot \left(\int\rho(E_2,t)\diff E_2\right)^2\bigg]
\end{array}
    \label{eqn:drhotripsimpl}
\end{equation}
removes the triple pileup perturbation.
The bias in the triple
pileup correction from use of the pileup-contaminated spectrum,  
rather than the true one, is of order $r(t)^3 \cdot \Delta t^3$, or $10^{-6}$. 

Ref.~\cite{phdthesis:2019Fienberg} provides the details of the implementation of this method.  As done for the other two methods, each final bin uncertainty of the corrected spectrum includes the contribution from this procedure.  

Figure~\ref{fig:eDepPileup} summarizes, for the three methods, the
initial pileup contribution (left) and the residual contamination 
above the positron end point (right)  after pileup subtraction.


\subsubsection{Pileup and the threshold integrated energy analysis}
Conceptually, a threshold-free integrated energy
analysis is free from distortion by pileup of positrons
in space and time. The integrated energy
correctly receives the energy contribution
from all positrons -- whether proximate or not.

However, a threshold-based integrated energy
analysis can suffer pileup distortions.
Therefore, an algorithm was developed for calculating
the pedestal and applying the threshold that mitigated
such distortions.
 
To understand the algorithm it is important to note
that pileup pulses may occur either on the trigger sample
or in the pedestal window.  A pileup pulse on the
trigger sample will increase the corresponding, pedestal-subtracted,
ADC value. A pileup pulse in the pedestal window
will decrease the corresponding, pedestal-subtracted, ADC value.

By requiring both the trigger sample to be above the energy threshold
and the pedestal samples to be below the energy threshold,
the effects of pileup are mitigated. To understand this mitigation
it is important to note the four categories of pulse pileup:
an above-threshold pulse on the trigger sample,
an above-threshold pulse in the pedestal window,
a below-threshold pulse on the trigger sample, and a
below-threshold pulse in the pedestal window.

\begin{enumerate}

\item
An above-threshold pileup on the trigger sample is properly handled as
the correct energy of the 
two above-threshold pulses on the trigger sample is recorded

\item
An above-threshold pileup in the pedestal window is properly handled
as the correct energy of 
the single above-threshold pulse on the trigger sample is recorded due
to rejection of the 
above-threshold pileup pulse in the pedestal window,

\item
A below-threshold pileup on the trigger sample causes an overestimate
of the correct energy of 
the single above-threshold pulse on the trigger sample. 

\item
A below-threshold pileup on the pedestal
window causes an underestimate of the correct energy of the single
above-threshold pulse on 
the trigger sample. However, overall, the energy overestimation from
below-threshold, trigger 
sample pileup and energy underestimation from below-threshold,
pedestal window pileup, 
statistically cancel.
 
\end{enumerate}

Extensive studies with Monte Carlo simulations show that the residual
contribution from  higher-order pileup has negligible effect on \oa.
\section{Determination of \oa}
\label{sec:fitting}

\begin{figure*}[tb]
\centering
\includegraphics[width=.45\linewidth]{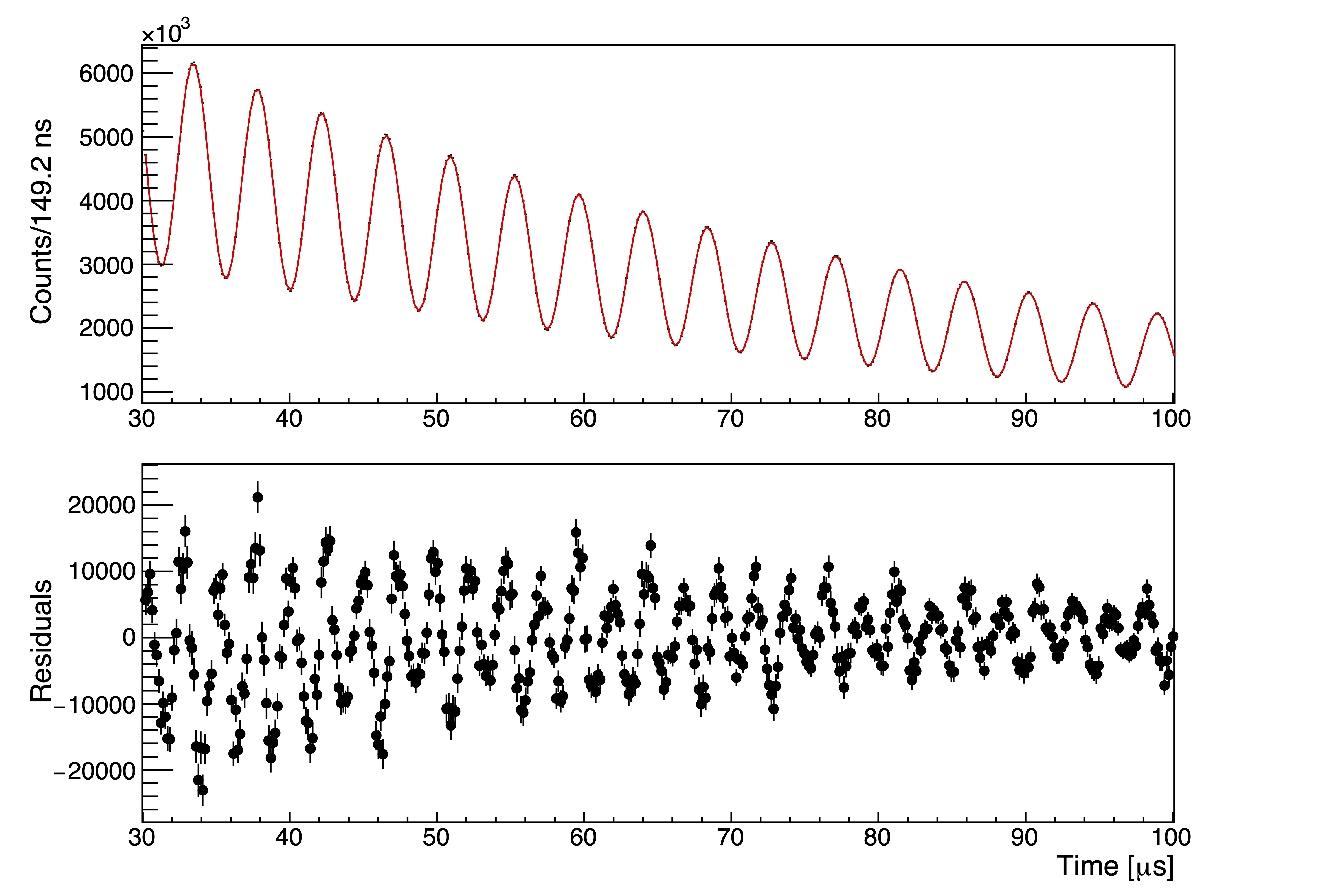}
\includegraphics[width=.48\linewidth]{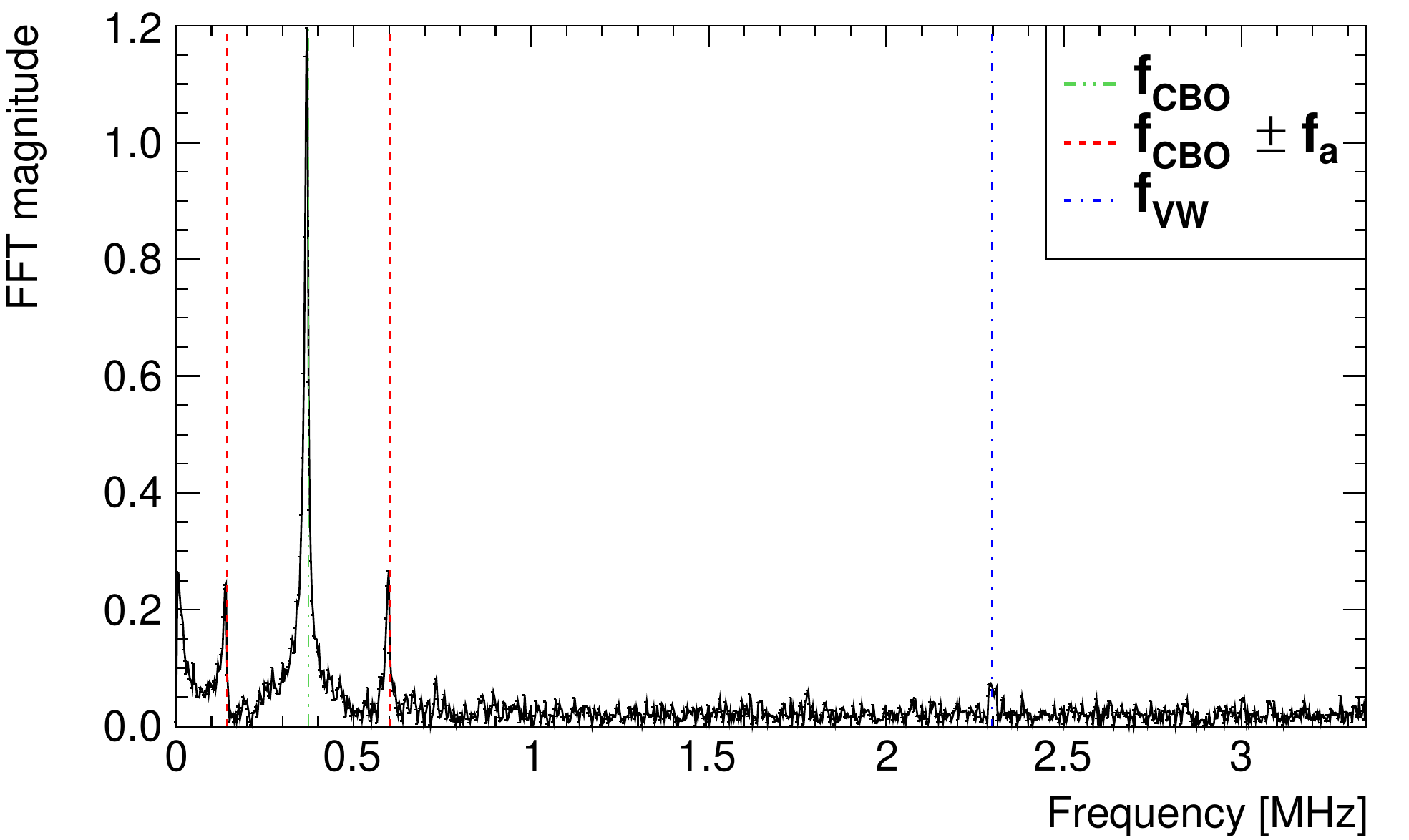}
\caption{
The results of a five-parameter fit based on Eq.~\ref{eq:5param} to
the time series from the unit-weighted event analysis of the \runonea
data. Left: the fit at early times (top) and the fit residuals
(bottom), showing beam effects that the simple five-parameter
function does not describe.  Right: Fourier transform of the fit
residuals showing the peaks at the expected beam oscillation
frequencies.  These distributions emphasize the need to incorporate
the effects related to the beam dynamics into the fit model, as discussed
in the text.} 

\label{fig:fiveParamTMeth}

\end{figure*}
An unbiased determination of \oa requires a physically motivated functional form that describes the positron time series detected by the calorimeters.  This section discusses the dynamical effects included in our fitting model, and the fits to determine the anomalous precession frequency.

Figure~\ref{fig:fiveParamTMeth} shows the function and residuals for
the first \mus{70} of a five-parameter fit (Eq.~\ref{eq:5param}) 
to the time series from
the unit-weighted event analysis of the \runonea data.  As discussed
earlier, the fit starts from $t \simeq $\mus{30} after muon
injection. The figure also shows the Fast Fourier Transform (FFT) of
the residual distribution, which illustrates that the five-parameter
model does not adequately capture all dynamics present in the data.
In particular, the FFT shows several peaks that arise mainly due to
beam dynamics. 

Coherent betatron oscillation (CBO) of the beam produces the
predominant oscillation frequency at $f_{\rm CBO}\simeq 0.372$ MHz
present in the residuals (see Sec.~\ref{sec:intro}).  Two side
frequencies are also evident at $f_{\rm CBO} \pm f_a$, 
where $f_a = \oa/2\pi$ is the
anomalous precession frequency. The vertical beam
oscillations occur at higher frequencies of $f_{\rm VW}\simeq 2.297$
MHz, while the peak at low frequencies indicates the presence of
effects, such as muon loss, that evolve slowly over the course of a
muon fill.  The data used in this fit have had the corrections for
pileup and gain perturbations applied.  Without those corrections, the
peak at low frequency would be considerably higher. 


\subsection{Muon loss}
\label{subsec:muloss}

\begin{figure*}[tb]
\centering
\includegraphics[width=0.48\textwidth]{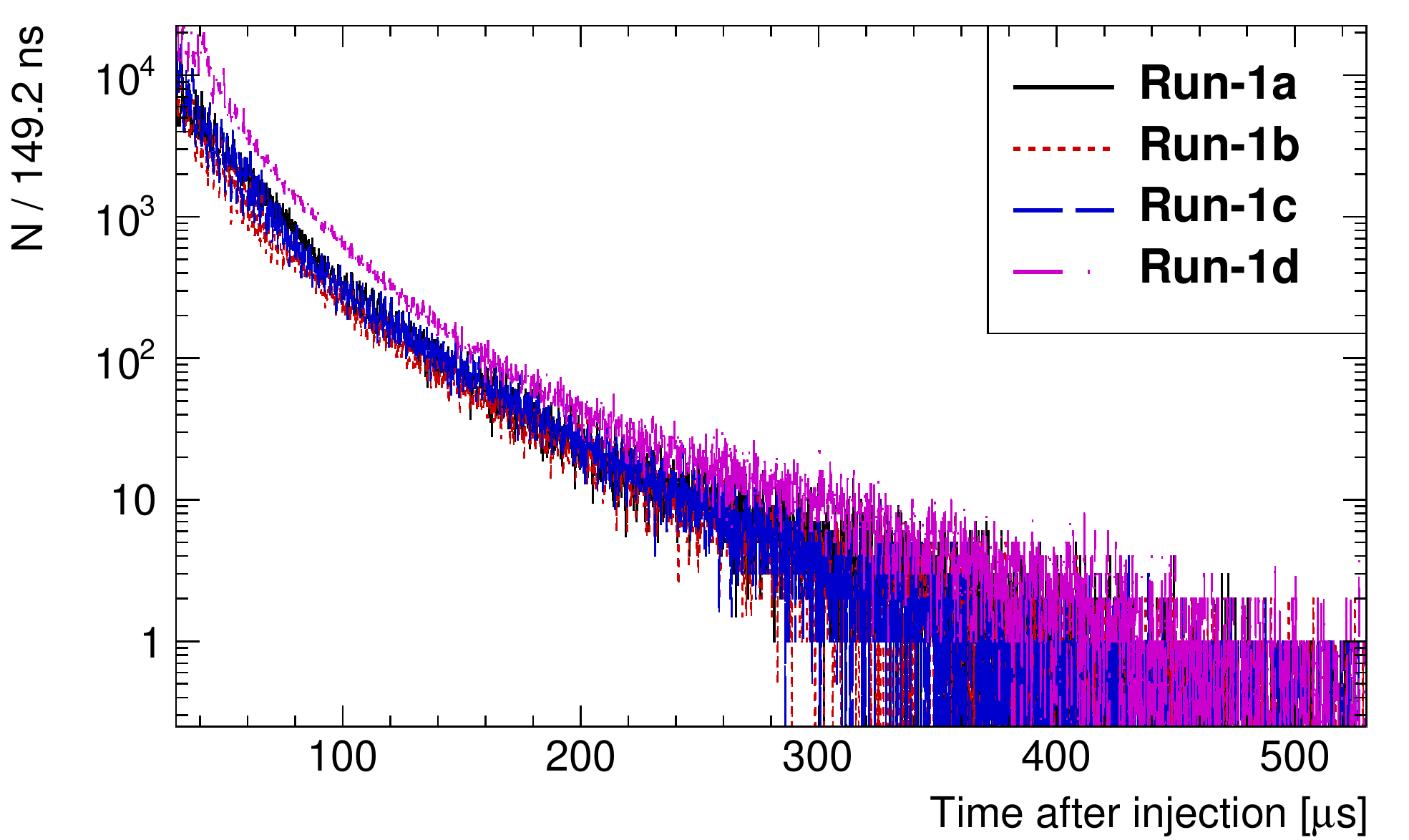}
\hfill
\includegraphics[width=0.48\textwidth]{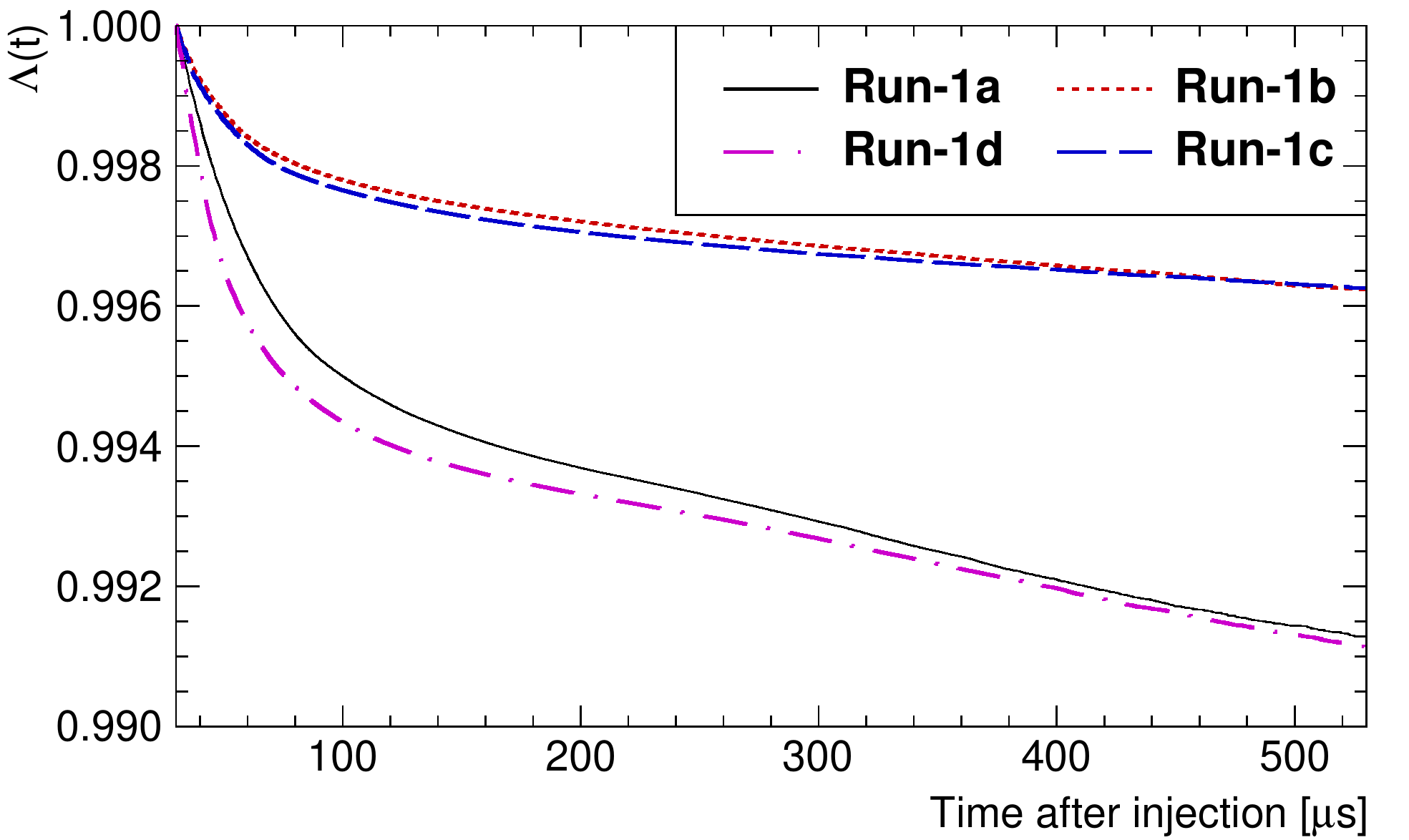}
\caption{Muon losses measured for the \runonea (black), \runoneb (red or dark grey), \runonec (blue or medium grey) and \runoned (magenta or light grey) data subsets.
  Left: The number of triple coincidences, 
  as measured by the selection criteria described in the
  text.  Right: The correction function $\Lambda(t)$. The value of
  $\Lambda(t)$ is set to 1 at the fit start time, which is
  approximately \mus{30}. The two upper curves correspond to
  the \runoneb and \runonec datasets, while the two lower curves correspond to
  \runonea and \runoned (see text).} 
\label{fig:LMF}
\end{figure*}

Not all muons remain stored throughout their lifetime in the storage
ring; a fraction of them exit the storage ring after striking
collimators or other obstacles.  The resulting energy loss, which
shifts the energy of a muon below the storage ring momentum acceptance
range ($\pm$0.15\% of 3.1\,GeV/c), dominates the beam loss mechanisms.
A loss of muons leads to a time dependence of the normalization factor
$N$ in the decay time spectrum of Eq.~\ref{eq:5param} and requires
correction. 

A fraction of these {\it lost muons} will pass through one or more
calorimeters, depositing  in each an energy typical of a MIP of 
about $170$\,MeV.  The lost muons
passing through multiple calorimeters have a time of flight between
successive calorimeters of 6.15\,ns.  These two characteristics allow
identification of lost muons and a measurement of the loss rate up
to an overall acceptance
factor~\cite{phdthesis:2020Sweigart,phdthesis:2019Fienberg}. 
As a balance between statistics and accidental contamination, we
require that the lost muon candidates cross at least three calorimeters.
The remaining, minimal amount of accidental contamination in the
triple coincidence sample can be corrected for on average by
searching for coincidences in nearby time-of-flight windows.
Figure~\ref{fig:LMF} (left) shows the corrected time spectrum of lost
muons for each dataset taken during \runone. 

\begin{figure}[tb]
\includegraphics[width=0.95\linewidth]{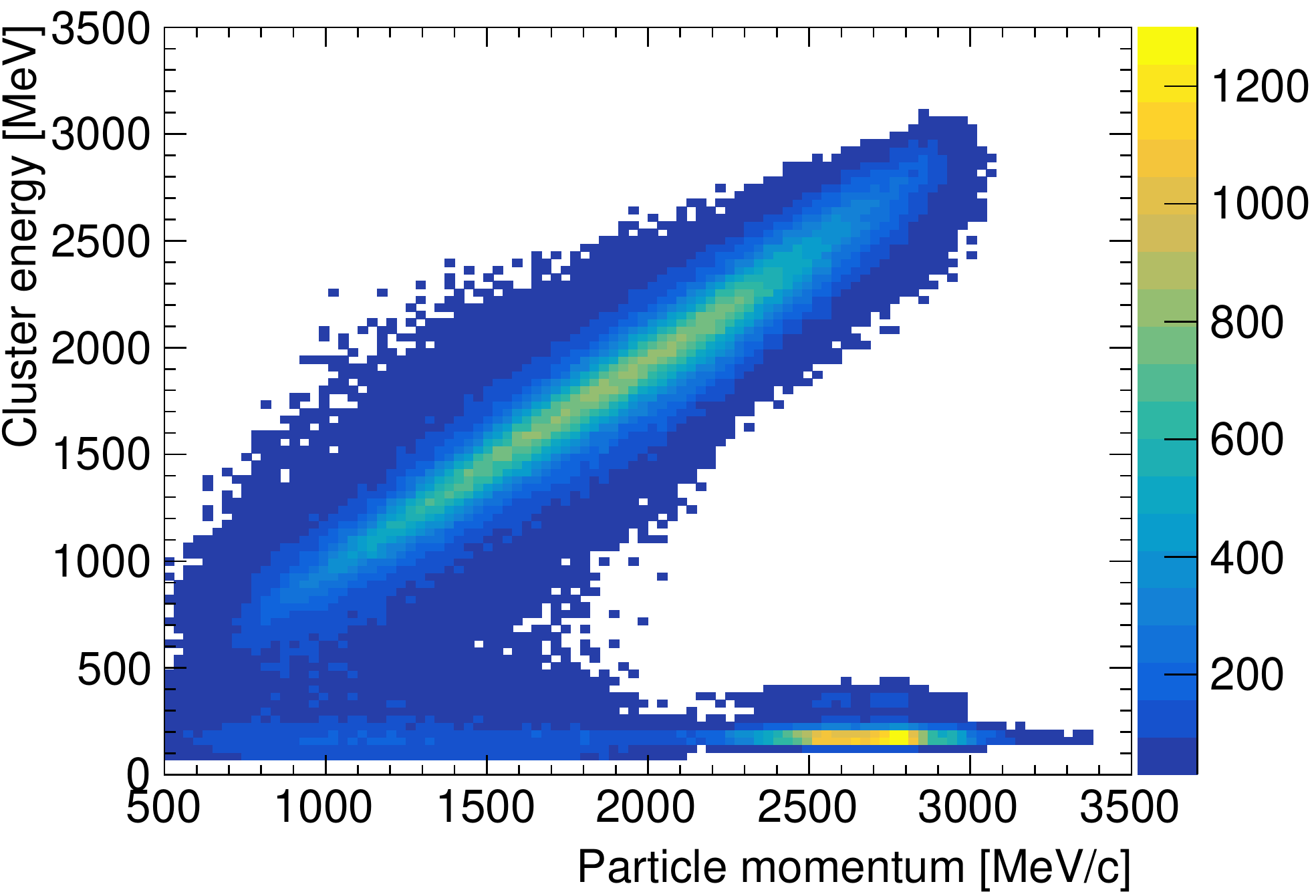}
\caption{Cluster energy versus particle momentum for tracks crossing a
  tracker station and hitting the following calorimeters.  The
  positron line, along the diagonal, and the muon peak, with deposited
  energy of 170 MeV and particle momentum slightly lower than 3 GeV/c,
  are clearly visible. 
Scattered muons can also have a lower momentum, while still depositing
the energy of a Minimum Ionizing Particle. 
Note that the tracker geometrical acceptance
  decreases below $p=1$ GeV/c due to the particle curvature. } 
\label{fig:EvsP}
\end{figure}

For the two calorimeters that each sit behind a tracking station, 
muons can
be easily identified by comparing the momentum ($p$) and the energy ($E$) measured
by the two detectors, as shown in Fig.~\ref{fig:EvsP}.  
Thus, as an alternative method to the one described above, lost muon
candidates can be selected with the following approach. First, we
apply a cut on the $E/p$ ratio of the detected particles. We then build
a likelihood function based on the measurements made by the two
calorimeters. This function includes information regarding the
deposited energy, position distribution, and time of flight with
respect to temporally adjacent calorimeters. 
This
likelihood function allows selection of muons in all 24
calorimeters, providing a muon loss spectrum that is totally
compatible with the one identified by the method described above. 

The presence of the muon loss spectrum $L(t)$ modifies the simple exponential decay by introducing a multiplicative correction function:
\begin{equation}
\Lambda(t) = 1 - K_{\rm loss} \int_{0}^{t} e^{t'/\gamma\tau} L(t') dt'.
\label{eq:LM}
\end{equation}
Reference~\cite{phdthesis:2019Fienberg} presents a derivation of this 
correction function.   The normalization parameter
$K_{\rm loss}$, related to the calorimeter geometrical acceptance
and to the selection efficiency, is determined by the \oa fit.
 Figure \ref{fig:LMF} shows the typical
distortion of the simple exponential introduced by these lost muons:
the effect is concentrated in the first tens of microseconds and the total loss rate,
integrated over the fill, varies between $3-4$ per mille for datasets 1b and 1c,
in which the ESQs operated at a high tune value $n=0.120$ 
(see Tab.~\ref{tab:subsets}), and
$7-8$ per mille for datasets 1a and 1d, for which
$n=0.108$. 

Reference~\cite{\BD} discusses the small correction to \oa that can
result if the lost muon sample has a different phase content than the
muon decay sample used in the fits.

\subsection{Beam dynamics and detector acceptance-based fit model}
\label{subsec:fit_model}


Four fundamental frequencies, first introduced in
Sec.~\ref{sec:intro}, can fully describe the dynamics of the muon
ensemble: the anomalous precession frequency, $f_a = \oa/(2\pi)$;
the cyclotron frequency, $f_c$; the horizontal betatron frequency,
$f_x$; and the vertical betatron frequency, $f_y$.  Together with
their harmonics and admixtures, these frequencies account for each
frequency observed in Fig.~\ref{fig:fiveParamTMeth}.
Ref.~\cite{phdthesis:2020Sweigart} provides a physical description of these
frequency combinations. 
The fitting model %
\begin{eqnarray}
F(t) = N_{0} \cdot N_{x} (t) \cdot N_{y} (t) \cdot \Lambda(t) \cdot
e^{-t/\gamma\tau_{\mu}} \cdot \nonumber \\ 
\left[1 + A_{0} \cdot A_{x}(t) \cdot \cos{(\oa t + \phi_{0}
    \cdot \phi_{x}(t))} \right]
\label{eq:fit_function}
\end{eqnarray}
modifies the basic rate model of Eq.~\ref{eq:5param} to incorporate
the effects of detector acceptance and beam dynamics.  The parameter
$N_{0}$ is the overall normalization, $\Lambda(t)$ is the muon loss
correction given in Sec.~\ref{subsec:muloss}, $A_{0}$ is the decay
asymmetry, and $\phi_{0}$ is the initial average spin precession phase. The terms $N_x$,
$N_y$, $A_x$, and $\phi_x$ describe the interplay between 
calorimeter acceptance
and beam dynamics that affect the overall rate, the average asymmetry
and the average phase.  These functions are defined as 
\begin{alignat}{7}
&N_x &(t) ={}& 1 &{}+{}& e^{-1 t/\tau_\text{CBO}} & &A_{N,x,1,1} && \cos (1 \omega_\text{CBO}&  t + \phi_{N,x,1,1} &)& \nonumber \\
& & & &{}+{}& e^{-2 t/\tau_\text{CBO}} & &A_{N,x,2,2} && \cos (2 \omega_\text{CBO}&  t + \phi_{N,x,2,2} &)& , \label{eq:nxt} \\
&N_y &(t) ={}& 1 &{}+{}& e^{-1 t/\tau_y} & &A_{N,y,1,1} && \cos (1 \omega_y&  t + \phi_{N,y,1,1} &)&  \nonumber \\
& & & &{}+{}& e^{-2 t/\tau_y} & &A_{N,y,2,2} && \cos (1
\omega_\text{VW}&  t + \phi_{N,y,2,2} &)&  ,  \label{eq:nyt} \\
&A_x &(t) ={}& 1 &{}+{}& e^{-1 t/\tau_\text{CBO}} & &A_{A,x,1,1} && \cos (1 \omega_\text{CBO}& t + \phi_{A,x,1,1} &)& ,\label{eq:axt} \\
&\phi_x &(t) ={}& 1 &{}+{}& e^{-1 t/\tau_\text{CBO}} & &A_{\phi,x,1,1} && \cos (1 \omega_\text{CBO}& t + \phi_{\phi,x,1,1} &)& .
\label{eq:phixt}
\end{alignat}
For the case of $N_{x}(t)$ in Eq.~\ref{eq:nxt}, the parameters of the 
form $A_{N,x,i,j}$ and $\phi_{N,x,i,j}$
correspond to the effect of the $i^{\rm th}$ moment of the radial
($x$) beam distribution at the $j^{\rm th}$ multiple of the
fundamental frequency (for $N_{x}(t)$, $\omega_{\rm CBO}$)  on the rate
normalization $N$~\cite{phdthesis:2020Sweigart}.  Analogous parameters
in Eqs.~\ref{eq:nyt}--\ref{eq:phixt} 
model the modulation of the average asymmetry $A$ and phase $\phi$, as
well as the effect of moments of the vertical ($y$) beam distribution.
Some analysis groups employ small variations of the higher order terms of the
beam dynamics modeling in their fitting function compared to the model
presented here, providing a valuable cross check.  Other model variations include
an additive rather than multiplicative correction to the phase term.
Those terms couple very weakly to \oa
with the statistics of the \runone datasets, and these model variations
have negligible effect. 

%

The damaged high voltage resistors for the electrostatic ESQs in
\runone (see Sec.~\ref{subsec:subsets}) add one further modeling
requirement by necessitating a time-dependent CBO frequency.  The
straw tracker system measures this dependence directly in each subset
of \runone, and the substitution  
\begin{equation}
\omega_{\rm CBO}\cdot t \rightarrow \omega_{\rm CBO}\cdot t + A_{1}e^{-t/\tau_{1}} +  A_{2}e^{-t/\tau_{2}}
\label{eq:cbo_freq}
\end{equation}
from integration of the instantaneous frequency model
replaces the static frequency term in Eqs.~\ref{eq:nxt}--\ref{eq:phixt}. 
The parameter $\omega_{\rm CBO}$ floats freely in the fits, while the
time variations remain fixed.  The trackers provide the exponential
parameters of the time dependence, with short and long lifetimes of
order \mus{8} and \mus{80}, respectively. The integrated form captures
both the frequency shift and the accumulated phase shift.

In a weak-focusing storage ring, the vertical oscillation ($\omega_{y}$) 
and horizontal CBO frequencies satisfy the relationship
\begin{equation}
\omega_{y}(t) = \kappa_{y}\cdot \omega_{\text CBO}(t) 
\left(\frac{2\omega_{c}}{\kappa_{y}\cdot\omega_{\text CBO}(t)} -1\right)^{1/2}.
\end{equation}
For continuous ESQ plates generating a perfectly linear field around
the ring  $\kappa_{y}=1$, 
but the partial coverage and field non-linearities distort the relationship
between $\omega_{y}$ and $\omega_{\text CBO}$.
A shift in $\kappa_{y}$ at the $1\%$ level reflects these distortions.
 The correction parameter
$\kappa_{y}$ floats in the fit, and the best fit values agree with beam motion measurements 
with the straw tracking system.  The vertical oscillation frequency aliases down to the
vertical width frequency via
\begin{equation}
\omega_{\text VW}(t) = \omega_{c} - 2\omega_{y}(t).
\end{equation}

A similar function models the time series obtained with the integrated
energy analysis, though two additional effects require further
modeling.  These effects, described below, require a
multiplicative correction to the normalization in a manner analogous
to the muon loss correction $\Lambda(t)$. 

\subsubsection{Electronics ringing term}
\label{subsec:electronics ringing term}

As discussed in Sec.~\ref{subsec:qmethodrecon}, for the integrated
energy approach the average of the
time bins in the pedestal window provides an estimate for the pedestal
in the signal bin.  Consequently, any change in the slope of the
pedestal over the window introduces a bias. 

The dominant source of pedestal bias arises from electronics ringing,
with a period comparable to the pedestal window, following the $t = 0$
injection flash. 
The average difference between (a) the time samples with no pulse above
threshold, and (b) the pedestal estimates for that sample, provides an
estimate of the ringing as a function of time into the fill. 
This ringing term and an associated normalization parameter are then
incorporated in the fit function in the same manner as the muon loss
term.  
The anomalous precession frequency \oa changes
  by only $\mathcal{O}(10~\text{ppb})$ when including or excluding
  this term.

\subsubsection{Vertical drift term}
\label{subsec:vertical drift term}

As discussed in Ref.~\cite{\BD}, the vertical distribution of
stored muons for \runone changes slightly over the fill because of a
time dependence of the ESQ voltages on two of the 32 plates. 
Consequently, the positron
acceptance at the top and bottom of the calorimeters will change, and
introduce further time dependence of the fit normalization.  With the
low positron energy threshold for the integrated energy analysis, and
thus a correspondingly broader vertical distribution at the
calorimeter, this method becomes sensitive to the drift.  The time
distributions of the energy deposited in the three upper rows of
crystals in the calorimeter show a gradual decrease in deposited
energy as a function of time into the fill, while those in the lower
rows of crystals show an increase.  The magnitude of the effect varies
systematically with row -- maximal at the outermost rows and smallest
in the central rows. 

Tracking-based studies indicate that the drift and width changes occur
with the similar time dependences.  By carrying the measured dependence from
the crystal row studies through to the
normalization, we obtain a data-driven correction to the
normalization, analogous to the lost muon correction. 

In addition, we investigate the possible effects of vertical drift on
the asymmetry parameter. The measured asymmetry correlates with
vertical position through acceptance effects, and therefore a change
in vertical profile can also change the asymmetry as a function of the
fill time.  

Excluding versus including the vertical drift correction in the fit
shifts the extracted value of the
anomalous precession frequency \oa by $\mathcal{O}(100~\text{ppb})$.
The normalization term dominates this shift, with the effect from the
correction to the asymmetry parameter entering at least an order of magnitude
smaller. 

\subsection{Software blinding of \oa}
\label{subsec:analysis-blinding}

Each analysis group introduces an independent blinding of \oa at the software level within their fits, which prevents unconscious biasing towards the central value of any particular group.  This blinding proceeds through the introduction of an offset $\Delta R$, defined as
\begin{equation}
\oa(R) = \omega_{\rm ref}[1-(R-\Delta R)\times 10^{-6}],
\label{eq:Rblinding}
\end{equation}
where the reference frequency $\omega_{\rm ref} = 2\pi \times
0.2291~\rm{MHz}$.  This parameterization
expresses \oa in terms of the shift $R$ in parts per million (ppm)
from the reference frequency, and it introduces the blinded shift between
the value used in the fit model and the displayed results.  Each
analysis group chooses a blinding text phrase, which a standardized
package converts to a value of $\Delta R$, keeping the shift itself
unknown to the group.  An MD5 hash algorithm converts the blinding
phrase to four 32-bit seeds for a Mersenne Twister random number
generator.  Using this seeded generator, the package draws the
blinding factor $\Delta R$ from a flat $\pm 24$\,ppm  distribution
with 1\,ppm Gaussian tails.  This procedure always produces the same
blinded shift $\Delta R$ for a given blinding phrase. 

Unblinding at the software level proceeded in two stages.  The first relative unblinding occurred after each analysis group completed their analysis, including all cross checks and systematic uncertainty evaluation.  At that point, all groups adopted a common blinding offset to allow a direct comparison of results.  The final common software blinding and the hardware-level blinding were only removed after the final decision to proceed with publication.

\subsection{Parameter determination}
\label{subsec:par_extract}

All analyses determine the best fit parameters through minimization of the Neyman $\chi^2$
\begin{equation}
    \chi^2 = (\mathbf{N} - \mathbf{F})^{T}\mathbf{V}^{-1}(\mathbf{N} - \mathbf{F}),
\end{equation}
with the MINUIT numerical minimization package~\cite{James:2004xla}
either directly or through the \textsc{ROOT} software package
\cite{Brun:1997pa}.  The vectors $\mathbf{N} = \{N_{i}\}$ and $\mathbf{F} = \{F_{i}\}$
correspond to the measured data time series and corresponding model prediction, respectively,
while $V$ represents the data  covariance matrix.
When correlations may be neglected, analyses employ the simpler form
\begin{equation}
    \chi^2 = \sum_{i} \frac{\left[N_{i} - F(t_{i}, \vec{p})\right]^2}{\sigma_{i}^2}.
\end{equation}
The vector
$\vec{p}=(N_{0},\tau_{\mu},\oa,...)$ represents the free
parameters described in Sec.~\ref{subsec:fit_model}
together with the function $F(t_{i})$.  The number of
parameters floating in the fit varies with analysis method, the
details of the beam dynamics model, and the size of the dataset (which
determines sensitivity to the higher-order, lower-amplitude effects
from beam dynamics).  The number of free parameters ranges from 16
(ratio method) through 27 (integrated energy analysis), with 22 being
the typical number for the event-based analyses. 

A minimum of 30 to 100 positrons (depending on analysis group)
contribute to the weighted sums in even the least populated bins 
(149.2 ns wide) for
the event-based analyses, so a Gaussian approximation to the Poisson
distribution works well in estimating uncertainties. 
Standard error
propagation for the asymmetry weighting, and for the corrections for
pileup and muon loss also apply.  The event-based and ratio analyses
have about 4000 degrees of freedom in the fits, while the integrated
energy analyses have
about 1210.  We require that all fits contributing to this work have a
reduced $\chi^{2}$ consistent with unity within the expected standard
deviation of 0.02 (0.04) for the event-based (integrated energy)
analyses -- a necessary but not sufficient condition for an unbiased
determination of \oa. In addition, all fits had to exhibit a
structure-free residual distribution in both time and frequency
domains.

\begin{table*}[tb]
\centering
\begin{tabular}{crcr}\hline\hline
Parameter                     & Fit result                      & Parameter           & Fit result              \\
\hline
blinded $R$ (ppm)             & $  -16.01  \pm 0.68$            & $\tau_y$ (\si{\micro\second}) & $168       \pm 98$      \\
$N_0$                         & $ (7249.8  \pm 3.5)\times 10^3$ & $A_{N,y,2,2}$       & $0.00039   \pm 0.00022$ \\
$\gamma\tau_{\mu}$ (\si{\micro\second}) & $ 64.4478  \pm 0.0023$          & $\phi_{N,y,2,2}$    & $2.10      \pm 0.65$    \\
$A_0$                         & $0.355193  \pm 0.000021$        & $A_{N,x,2,2}$       & $0.000198  \pm 0.000059$\\
$\phi_0$                      & $2.07519   \pm 0.00013$         & $\phi_{N,x,2,2}$    & $-3.35     \pm 0.30$    \\
$\omega_\text{CBO}$ (s$^{-1}$)  & $2.33593   \pm 0.00030$         & $A_{A,x,1,1}$       & $0.00059   \pm 0.00014$ \\
$\tau_\text{CBO}$ (\si{\micro\second})  & $190       \pm 11$              & $\phi_{A,x,1,1}$    & $-0.38     \pm 0.24$    \\
$A_{N,x,1,1}$                 & $0.003237  \pm 0.000097$        & $A_{\phi,x,1,1}$    & $0.000108  \pm 0.000072$\\
$\phi_{N,x,1,1}$              & $-6.081    \pm 0.029$           & $\phi_{\phi,x,1,1}$ & $-3.19     \pm 0.66$    \\
$K_{\rm loss}$           & $0.00903   \pm 0.00036$         & $A_{N,y,1,1}$       & $-0.000082 \pm 0.000046$\\
$\kappa_y$                    & $1.01398   \pm 0.00063$         & $\phi_{N,y,1,1}$    & $-5.98     \pm 0.58$    \\\hline\hline
\end{tabular}
\caption{The (blinded) fit results for the asymmetry-weighted event analysis for
  the \runoned dataset.  The fit used the model and parameters
  described in Equations~\ref{eq:LM} through~\ref{eq:phixt} and
  Eq.~\ref{eq:Rblinding}.\label{tab:fit_asymm_r1d_50}} 
\end{table*}

\begin{figure*}[tb]
\centering
\includegraphics[width=.48\linewidth]{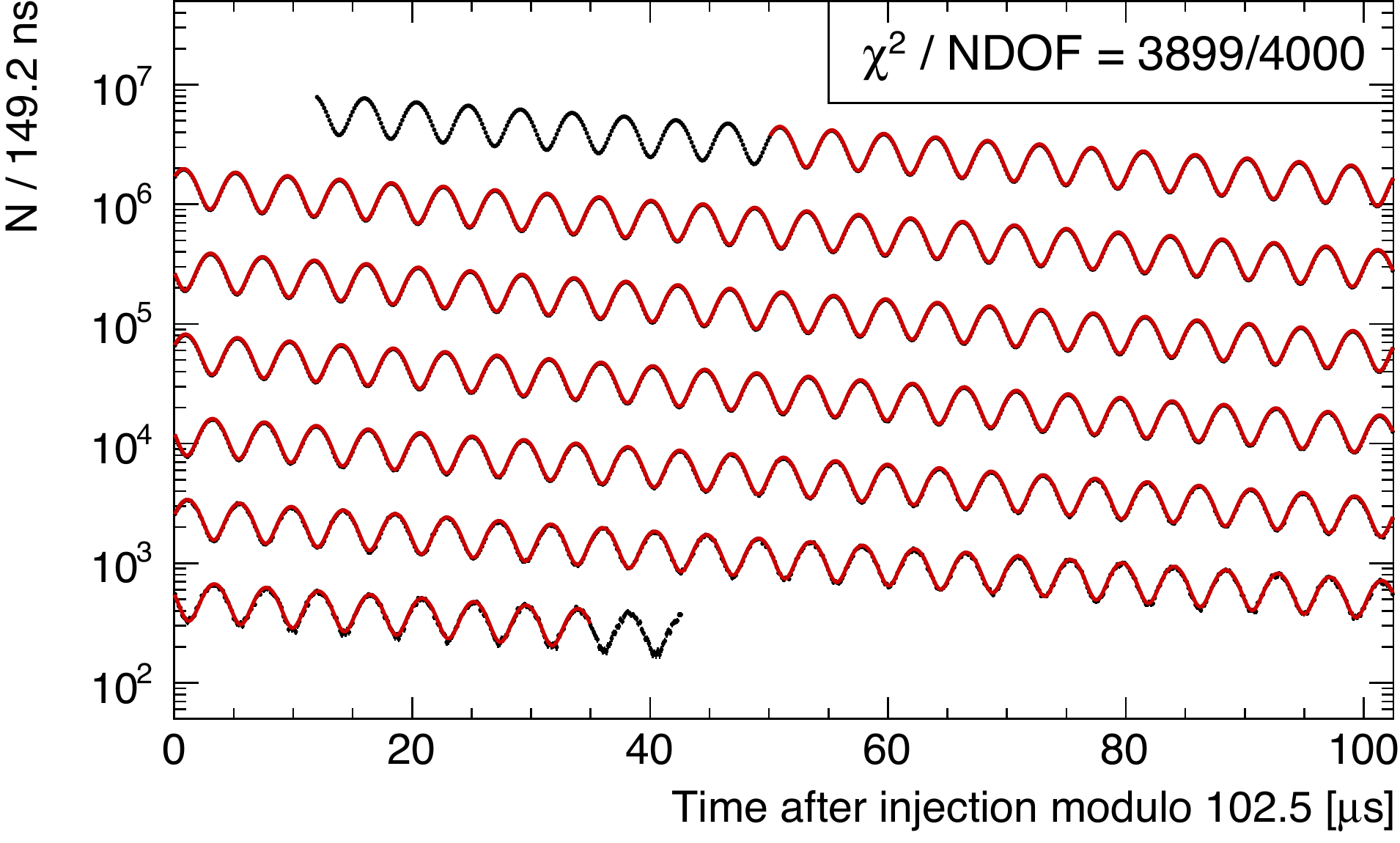}
\includegraphics[width=.48\linewidth]{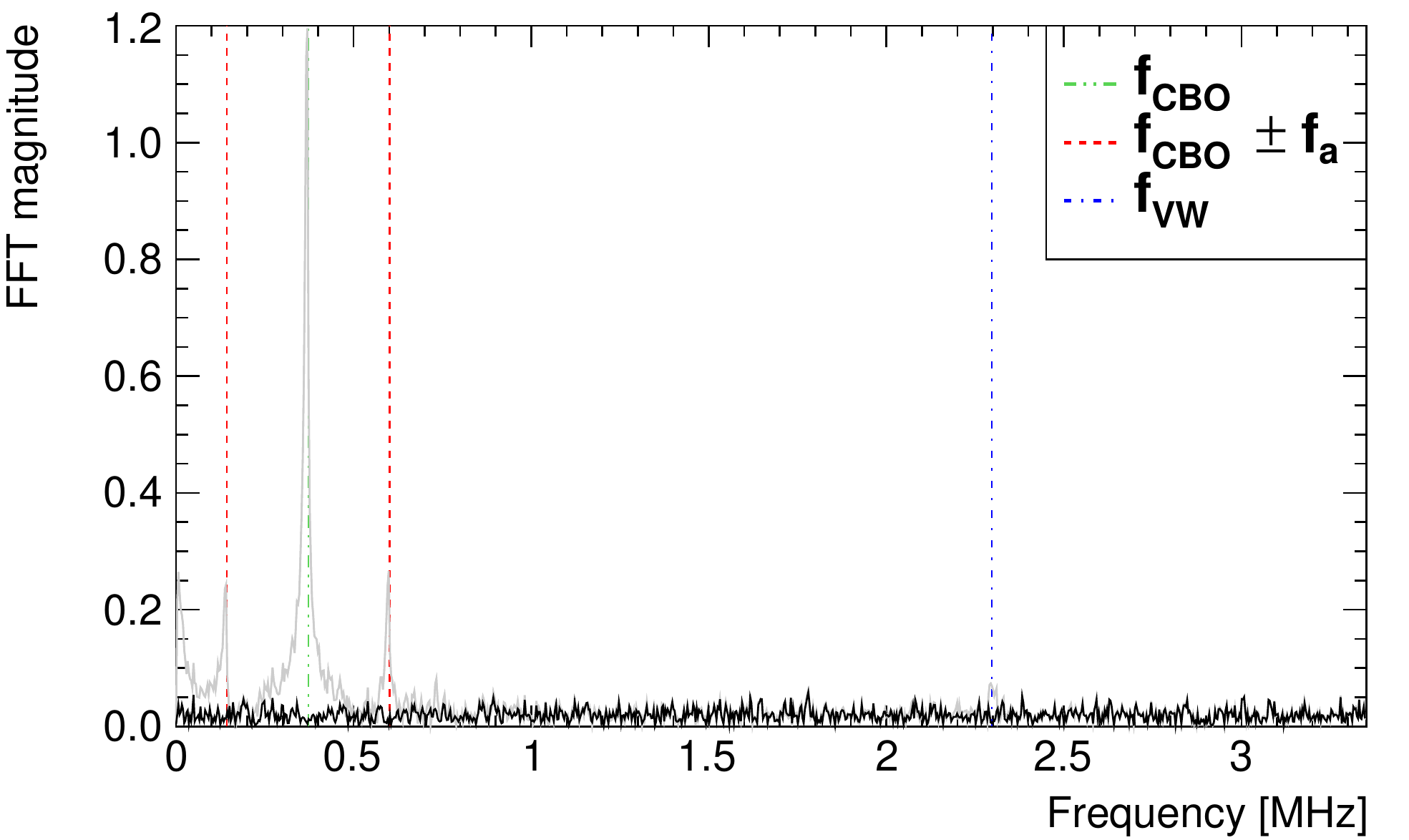}
\caption{Left: the overlay of the fit described in the text on the \runoned 
precession data.  Right: the FFT of the time distribution of
  residuals to that fit (black), which shows no remaining characteristic
  frequencies in the spectrum. For contrast, the residuals of the 5-parameter 
  fit with no beam modeling are also shown (light gray), which helps
  to highlight the 
  excellent performance of the fit including the modeling.}
\label{fig:Run1dFit_overlay}
\end{figure*}

Table~\ref{tab:fit_asymm_r1d_50}  presents the results of a fit to the
\runoned data, the subset with the largest statistics, for an analysis
using the model exactly as presented
above. Figure~\ref{fig:Run1dFit_overlay} 
shows both the result of the above fit overlaid on the precession data
and the FFT of the residual distribution.  With the full beam dynamics
model incorporated into the fit, this residual distribution no longer
exhibits any characteristic structure. 

Table~\ref{tab:correl_asymm_r1d_5} shows the correlation
coefficients for the fundamental five parameters of
Eq.~\ref{eq:5param} and the most significant beam dynamics
component, while Appendix~\ref{app:correlations} provides the full
correlation matrix. The strongest correlation of \oa ($R$) in the fits
occurs with the average initial precession phase, $\phi_0$, analogous
to the slope-intercept correlation in a linear fit. It has only small
correlations with all other parameters. While the correlations of \oa
with the CBO-related parameters are small, the strength of the leading
terms in the CBO model (reflected by the significance of the signal in
the fit) requires that we include these parameters in the fit.  If we
drop all CBO-related effects in the model, \oa shifts significantly
(of order 100 ppb).  Suppose we include the $N_{x,1,1}$ and $A_{x,1,1}$-related
terms in Eqs.~\ref{eq:nxt} and~\ref{eq:axt}, which correspond to the
main peak at the frequency $f_{\rm CBO}$ in the residuals to the
five-parameter fit (Fig.~\ref{fig:fiveParamTMeth}).  
The remaining terms in the CBO modeling affect \oa by 
at most 20\,ppb.

The correlation matrix also shows a strong correlation among the
overall normalization and the two parameters controlling a slow
variation over the time of the fill -- the lifetime parameter and the muon
loss normalization.  Increasing the muon lifetime, or the fraction of
lost muons, the overall normalization increases. 

Because of aliasing of the radial oscillations at positions 180$^{\circ}$ apart
in the ring, the effects of CBO in one calorimeter tend to 
compensate for the effects in the calorimeter directly across the ring.
To leading order, and neglecting decoherence, the sum of data from all
calorimeters 
provides a complete cancellation that is  
independent of variation in the radial betatron frequency.  Small differences
in the calorimeter acceptances result in a residual effect.
Nevertheless, summing the
data over all calorimeters significantly suppresses the effects of the
CBO in the fits.  
Excluding the CBO terms in the fit function in fits to individual calorimeters
results in shifts in \oa an order of magnitude larger than those observed
for fits to data summed over all calorimeters.

\begin{table*}[tb]
\centering
\begin{tabular}{rrrrrrrrrrr}\hline\hline
                    & $R$   & $N_0$ & $\gamma\tau_{\mu}$
                                            & $A_0$ & $\phi_0$
                                                            & $\omega_\text{CBO}$
                                                                    & $\tau_\text{CBO}$
                                                                            & $A_{N,x,1,1}$
                                                                                    & $\phi_{N,x,1,1}$
                                                                                            & $K_{\rm loss}$
                                                                                                     \\
\hline
$R$                 &  1.00 & -0.01 & -0.00 &  0.00 & -0.87 &  0.01 &  0.02 & -0.03 & -0.02 & -0.01 \\
$N_0$               &       &  1.00 &  0.86 & -0.03 &  0.01 & -0.00 & -0.03 &  0.05 &  0.00 &  1.00 \\
$\gamma\tau_{\mu}$  &       &       &  1.00 & -0.02 &  0.00 & -0.00 & -0.02 &  0.03 &  0.00 &  0.89 \\
$A_0$               &       &       &       &  1.00 & -0.01 &  0.01 & -0.01 &  0.01 & -0.02 & -0.04 \\
$\phi_0$            &       &       &       &       &  1.00 & -0.02 & -0.03 &  0.04 &  0.02 &  0.01 \\
$\omega_\text{CBO}$ &       &       &       &       &       &  1.00 & -0.03 &  0.03 & -0.92 & -0.00 \\
$\tau_\text{CBO}$   &       &       &       &       &       &       &  1.00 & -0.92 &  0.03 & -0.03 \\
$A_{N,x,1,1}$       &       &       &       &       &       &       &       &  1.00 & -0.03 &  0.04 \\
$\phi_{N,x,1,1}$    &       &       &       &       &       &       &       &       &  1.00 &  0.00 \\
$K_{\rm loss}$ &       &       &       &       &       &       &       &       &       &  1.00 \\ \hline\hline
\end{tabular}
\caption{The correlation matrix among the main parameters (full matrix
  in Appendix) from the fit
  whose results are presented in Table~\ref{tab:fit_asymm_r1d_50}. 
The parameters are defined in Equations~\ref{eq:LM} through~\ref{eq:phixt} and Eq.~\ref{eq:Rblinding}.
For purposes of display, the elements  below the diagonal for this
symmetric matrix have been not been
included. \label{tab:correl_asymm_r1d_5}} 
\end{table*}

Table~\ref{tab:allFits} presents the values for $R$ from each of the
11 fits to each of the four datasets.  Also provided are the simple
statistical weighted averages over the four datasets for a higher
precision comparison.  Note that the simple averages presented here do
not incorporate the small shifts in the magnetic field value and
changes in the beam dynamics corrections that vary set by set.
The averages are
only provided to allow assessment of the level of agreement among the
results from the different analysis methods. Reference~\cite{\PRL}
incorporates all necessary changes for a dataset by dataset comparison of the
anomalous magnetic moment. The values presented here also have the
hardware blinding and a common software blinding still applied.

\begin{table*}[tb]
\centering
\footnotesize
\begin{tabular}{rcrlllll}\hline\hline
 &  & \multicolumn{5}{c}{$R$ [ppm] for each dataset}
& \multicolumn{1}{c}{Naive $R$} \\ 
Recon. & Method & Pileup    &  \runonea        &  \runoneb        &  \runonec        & \runoned         & average  [ppm] \\ \hline
    global & A & empirical & -82.98 $\pm$ 1.21 & -81.70 $\pm$ 1.03 & -82.30 $\pm$ 0.82 & -82.34 $\pm$ 0.68 & -82.30 $\pm$ 0.43 \\
    local & A & shadow & -83.23 $\pm$ 1.20 & -81.77 $\pm$ 1.02 & -82.35 $\pm$ 0.82 & -82.48 $\pm$ 0.67 & -82.41 $\pm$ 0.43 \\
    local & A & shadow & -83.17 $\pm$ 1.21 & -81.84 $\pm$ 1.03 & -82.50 $\pm$ 0.83 & -82.45 $\pm$ 0.68 & -82.44 $\pm$ 0.44 \\
    local & A & pdf & -83.39 $\pm$ 1.22 & -81.72 $\pm$ 1.04 & -82.32 $\pm$ 0.83 & -82.42 $\pm$ 0.68 & -82.39 $\pm$ 0.44 \\
    local & T & shadow & -83.55 $\pm$ 1.36 & -81.80 $\pm$ 1.16 & -82.67 $\pm$ 0.93 & -82.45 $\pm$ 0.76 & -82.54 $\pm$ 0.49 \\
    global & T & empirical & -82.96 $\pm$ 1.34 & -81.96 $\pm$ 1.14 & -82.77 $\pm$ 0.91 & -82.47 $\pm$ 0.75 & -82.52 $\pm$ 0.48 \\
    local & T & shadow & -83.64 $\pm$ 1.33 & -81.83 $\pm$ 1.12 & -82.64 $\pm$ 0.91 & -82.63 $\pm$ 0.74 & -82.62 $\pm$ 0.48 \\
    local & T & shadow & -83.49 $\pm$ 1.34 & -81.75 $\pm$ 1.13 & -82.64 $\pm$ 0.91 & -82.42 $\pm$ 0.75 & -82.50 $\pm$ 0.48 \\
    local & T & pdf & -83.37 $\pm$ 1.33 & -81.76 $\pm$ 1.13 & -82.65 $\pm$ 0.91 & -82.47 $\pm$ 0.74 & -82.51 $\pm$ 0.48 \\
    local & R & shadow & -83.72 $\pm$ 1.36 & -81.96 $\pm$ 1.16 & -82.67 $\pm$ 0.93 & -82.52 $\pm$ 0.76 & -82.62 $\pm$ 0.49 \\
    n/a & Q & n/a & -83.96 $\pm$ 2.07 & -79.70 $\pm$ 1.76 & -81.03 $\pm$ 1.45 & -82.74 $\pm$ 1.29 & -81.82 $\pm$ 0.78 \\\hline\hline
\end{tabular}
\caption{The unblinded \oa fit results, in terms of the parameter
  $R$, from all analyses efforts for the four sets, as well as the
  naive weighted average of the results for a more stringent
  comparison among the different analyses.  The ``Recon.'' column
  indicates whether the local or global reconstruction methods (see
  Sec.~\ref{sec:positron_recon}) provided the positron candidates.
  Under the ``Method'' column, {\it T} corresponds to an event-based
  analysis with unit weighting (equivalent to a simple energy
  threshold), {\it A} corresponds to an asymmetry-weighted event-based
  analysis, {\rm R} corresponds to the ratio method applied to the
  unit-weighted event-based sample, and {\it Q} corresponds to the
  integrated energy (akin to a charge integration) analysis. 
\label{tab:allFits}}
\end{table*}

\subsection{Corrections to and comparisons of \oa}

Table~\ref{tab:analysisCorrelations} shows the expected level of
correlations among the different analysis and reconstruction
types. Statistically-allowed differences arise, for example, from
differences in the local and global reconstruction, in parameter
choices within the local reconstruction, in the weighting of positron
events in different analysis methods, in different positron energy
thresholds and fit start time choices, in binning differences, and in
different choices in the  lost muon selection algorithms, among other
effects.  We determined these correlations from 
$\sim 10^3$ Monte Carlo
simulation trials that incorporate the major reconstruction and
analysis differences that drive the range of allowed fluctuations.
Given these correlation coefficients, the expression
\begin{equation}
 \label{eq:allowedDiff}
\Delta\sigma_{12} = \sqrt{\sigma_{1}^2 + \sigma_{2}^2 - 2\rho\sigma_{1}\sigma_{2}}
\end{equation}
provides the allowed $1\sigma$ statistical deviation
$\Delta\sigma_{12}$  between fit values for \oa from two different
analyses.
The parameters $\sigma_{1}$ and $\sigma_{2}$ correspond to the
statistical uncertainties of the two measurements, while $\rho$
corresponds to the correlation between the two analyses.

The different analyses are strongly correlated and it is
known (\cite{Lyons:1988rp,Valassi:2013bga})
that, for two positively correlated results,
the variance of the combination has a maximum for 
\begin{equation}
\rho_{\text{crit}}=
\mbox{min}(\sigma_1/\sigma_2)/\mbox{max}(\sigma_1/\sigma_2),
\label{eq:rhocrit}
\end{equation}
while it drops to zero when the correlation moves from $\rho_{crit}$
to 1.
Because of this, particular care is required in combining the
different analyses, as described in Sec. \ref{sec:combination}.

The pulls of different R measurements (see Tab.~\ref{tab:allFits}) on the same dataset 
distribute approximately as a unit Gaussian. The integrated energy 
measurements show a moderate systematic shift with respect to the event based 
measurements, and these correlated shifts coupled with the O(200 ppb) difference
in the corrections for the damaged quad resistor largely explain the  
differences of stop times in these two categories of measurements.


\begin{table*}[tb]
\centering
\begin{tabular}{cllllll}\hline\hline
Recon./Method & global/T & global/A & local/T & local/A & local/R & Q \\ \hline
global/T      &  1.00    & 0.91   & 0.95   & 0.91   &  0.95  & 0.51 \\
global/A      &          & 1.00   & 0.90   & 0.99   &  0.90  & 0.58 \\
local/T       &          &        & 1.00   & 0.91   &  1.00  & 0.51 \\
local/A       &          &        &        & 1.00   &  0.90  & 0.57 \\
local/R       &          &        &        &        &  1.00  & 0.50 \\
Q             &          &        &        &        &        & 1.00 \\
\hline\hline
\end{tabular}
\caption{The statistical correlations found from Monte Carlo trials for the different types of \oa analyses and positron reconstruction methods.  The reconstruction and analysis shorthands are defined in Tab.~\ref{tab:allFits}.
\label{tab:analysisCorrelations}}
\end{table*}

\subsection{Internal consistency}
\label{subsec:internal_consistency}

\begin{figure*}[htbp]
\centering
\includegraphics[width=.95\linewidth]{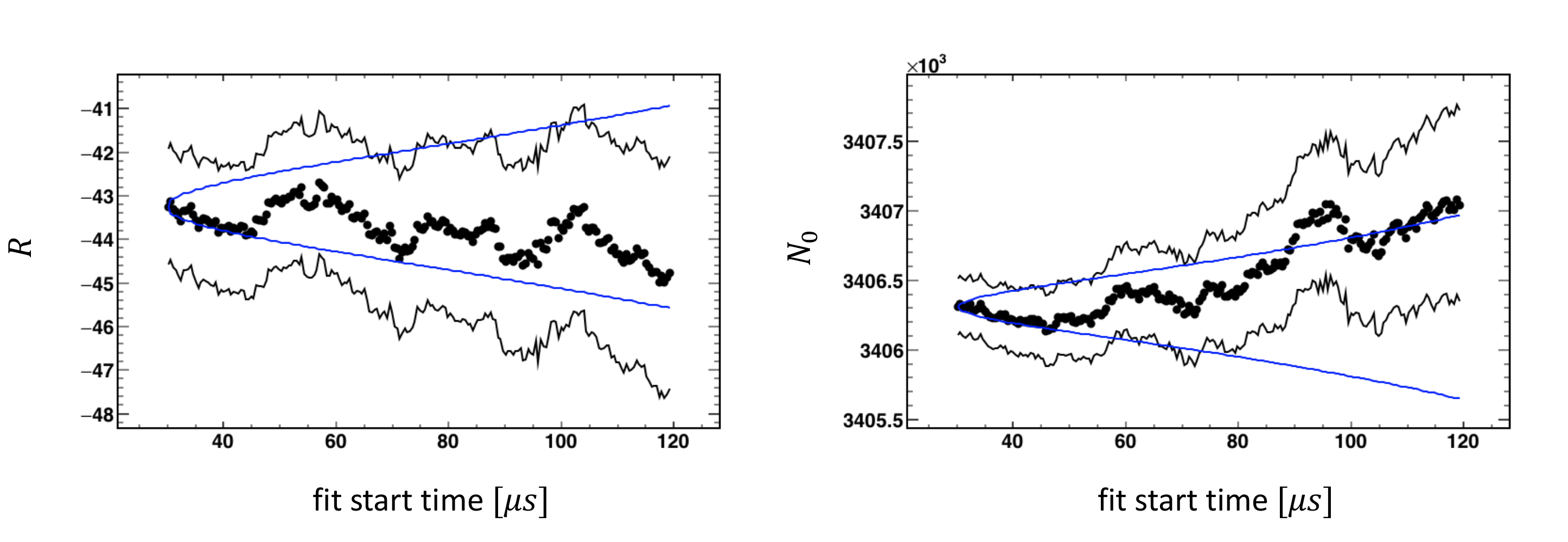}
\caption{The $R$ (left) and $N_0$ (right) parameters from a scan of
  the full fit to the \runonea data subset over the fit start time.
  The black curves above and below the data points indicate the full
  statistical error on $R$ or $N_{0}$ from the fits.  The one standard
  deviation bands (blue) show the allowed statistical variation of any
  given point relative to the nominal fit (starting point), and take
  into account the highly correlated statistics between those two
  points.  No scans show any systematic trends away from the
  statistically consistent region, nor any oscillation at the \oa
  period -- typical indicators of data mismodeling. The trend near the
  one standard deviation band simply indicates that data from the
  earliest fit times drive that statistically compatible shift.} 
\label{fig:rNVsStartTime}
\end{figure*}

To add further confidence in our data model and resulting fit, and to
probe for residual systematic effects, a number
of consistency checks have been performed.
The fit results should remain
stable with respect to the fit start time.  
Later start times reduce potential bias from residual effects that are
pronounced at early times, such as cyclotron motion, effects
from the dynamics of the stored beam, positron pileup and gain changes
related to the 
injection process.  Improper modeling of slow effects, such as those
due to gain stability or muon loss, would appear as an oscillation of
the extracted value of \oa at the period of the anomalous precession
itself.  Stability of the fitted \oa as a function of start time
indicates that these effects are controlled to within the allowed
statistical variation given the small change in statistics relative to
the nominal start time.  Figure~\ref{fig:rNVsStartTime} shows the two parameters
$R$ (see Eq.~\ref{eq:Rblinding}) and $N_{0}$ 
from a fitting start time
scan for one analysis.  
Both these combined scans and the individual subset
scans show excellent \oa stability.  Most of the data remains
common to each point in the start time scan, significantly correlating
the parameter values for each point in the start time scan.  The scans
therefore reveal trends, as opposed to exhibiting the statistical scatter
of statistically independent samples.  The maximum excursion in $N_{0}$
at a start time of $\sim$\mus{90} means that $N_{0}$ from that fit
agrees with $N_{0}$ from the nominal start time at $\sim 1.5$ standard deviations
given the change in statistics.

\begin{figure}[bt]
\centering
\includegraphics[width=.95\linewidth]{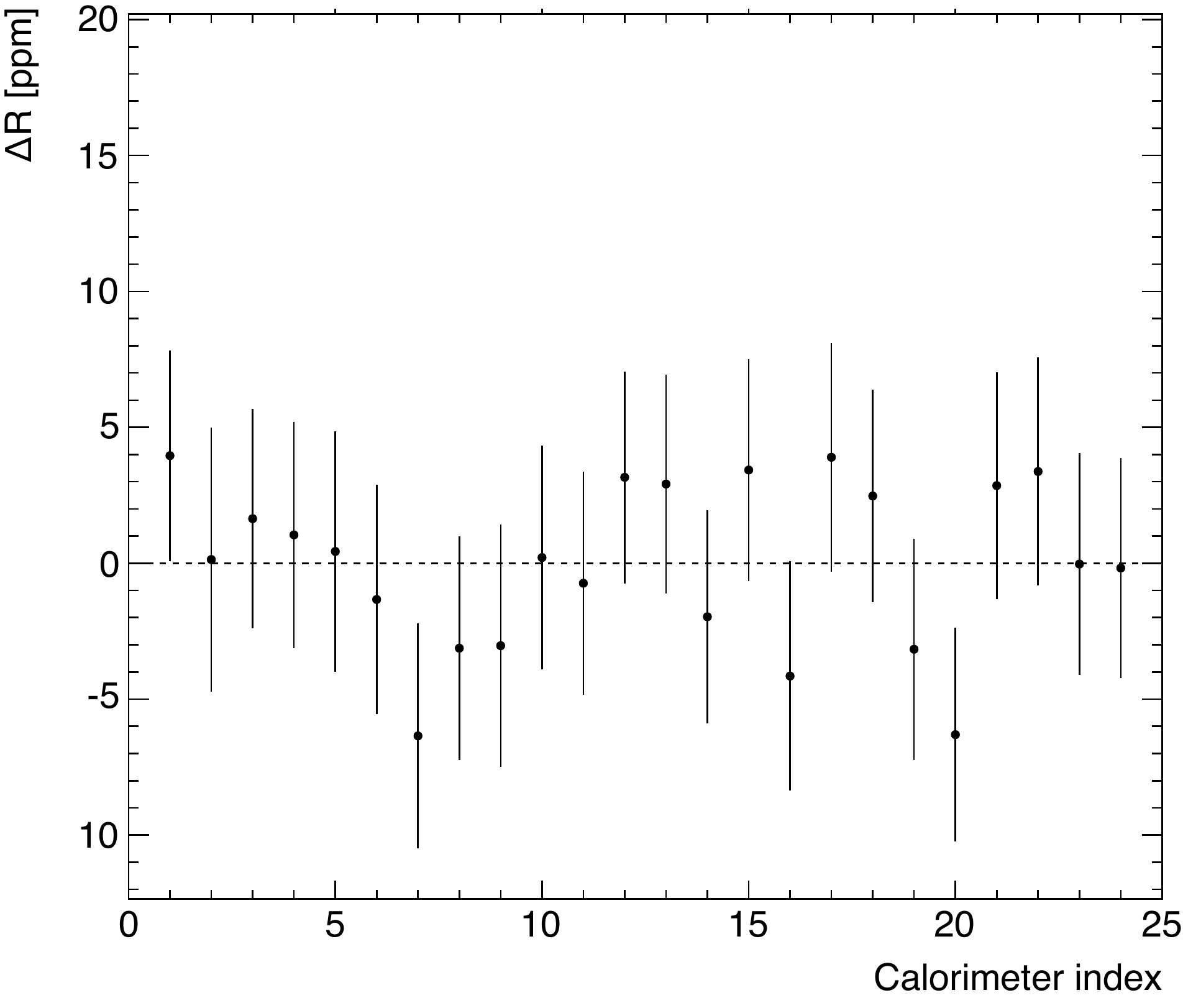}
\caption{The $R$ parameters from fits to the \runonec
  dataset by individual calorimeter, relative to their average.} 
\label{fig:calofits}
\end{figure}

We have also fit for \oa using the data in each of the 24 individual
calorimeter stations (Fig.~\ref{fig:calofits}).  As noted earlier, the data from an individual
station have a significantly more pronounced CBO motion than the
combined data. Thus, we can use the individual fits as sensitive
probes to evaluate our beam dynamics model.
Residual effects from the cyclotron motion can also induce a bias of
\oa as a function of position around the storage ring.  The value of
\oa remains stable as a function of calorimeter station, indicating
proper accounting for these effects.

Extracting \oa as a function of positron energy probes systematic
effects that depend on positron energy, such as positron pileup and
instability in the energy scale.  The energy scans show no systematic
dependence of \oa on energy.
The energy scans do show an unphysical variation of the muon loss
normalization parameter $K_{\rm loss}$.
 A number of sources can
contribute to such an effect, such as a residual gain miscalibration
on the order of a few parts per $10^{4}$, an overall drift in positron
or lost muon acceptance as a function of time into the fill, or
residual issues with the pileup correction.  The pileup correction,
for example, becomes more pronounced
at larger positron energies.  The different sources can shift \oa in different
directions, so we do not apply an overall correction to the central
value of \oa.  The systematic uncertainty receives a contribution from
this residual early-to-late effect, as discussed in the next section.

Other cross checks include fits for \oa versus run number, time of day, the
bunch number within the supercycle of 16 fills, calorimeter column and row number,
none of which show any systematic trend.

\section{Systematic uncertainties}
\label{sec:systematics}

The known  potential systematic effects and their possible biasing of the
extracted \oa value were evaluated for each analysis. For all datasets
and analyses, the statistical uncertainties exceeded the systematic uncertainties 
by one to two orders of
magnitude. The dominant systematic uncertainties arise from
uncertainties in the calorimeter gain corrections 
(Sec.~\ref{subsec:gain_fluctuation}), in multi-positron pileup
(Sec.~\ref{subsec:multi_positron_pileup}), in the beam dynamics model
(Sec.~\ref{subsec:fit_model}), and from the unknown source of the 
unphysical energy dependence
of the lost muon normalization parameter. This section will discuss
the methods used to estimate these uncertainties.  
While we have investigated many other 
sources of potential bias, 
 the estimated systematic uncertainty on \oa
fell below 10 ppb and has negligible effect on the result.  
Table~\ref{tab:averages}
summarizes the systematic uncertainty on the extracted \oa value for
each dataset and for each source of systematic uncertainty. The 
following section describes the method used to combine the
different analyses and thus to arrive at this summary table.


\subsection{Detector gain corrections}
\label{subsec:gainsys}

Short term and in-fill gain
corrections (Sec.~\ref{subsec:gain_fluctuation}) remove the energy
scale variation in each calorimeter channel as a function of time into
the muon fill.  
The statistical uncertainties of the gain
functions' best-fit amplitudes and characteristic time constants,
which are both typically between 10 and 20\%, introduce a systematic
uncertainty on the extracted \oa value. The  long-term gain correction,
on the other hand, does not pose a systematic bias to the extracted \oa
value because it is constant across each muon fill. 

\begin{figure}[htbp]
\centering
\includegraphics[width=.99\linewidth]{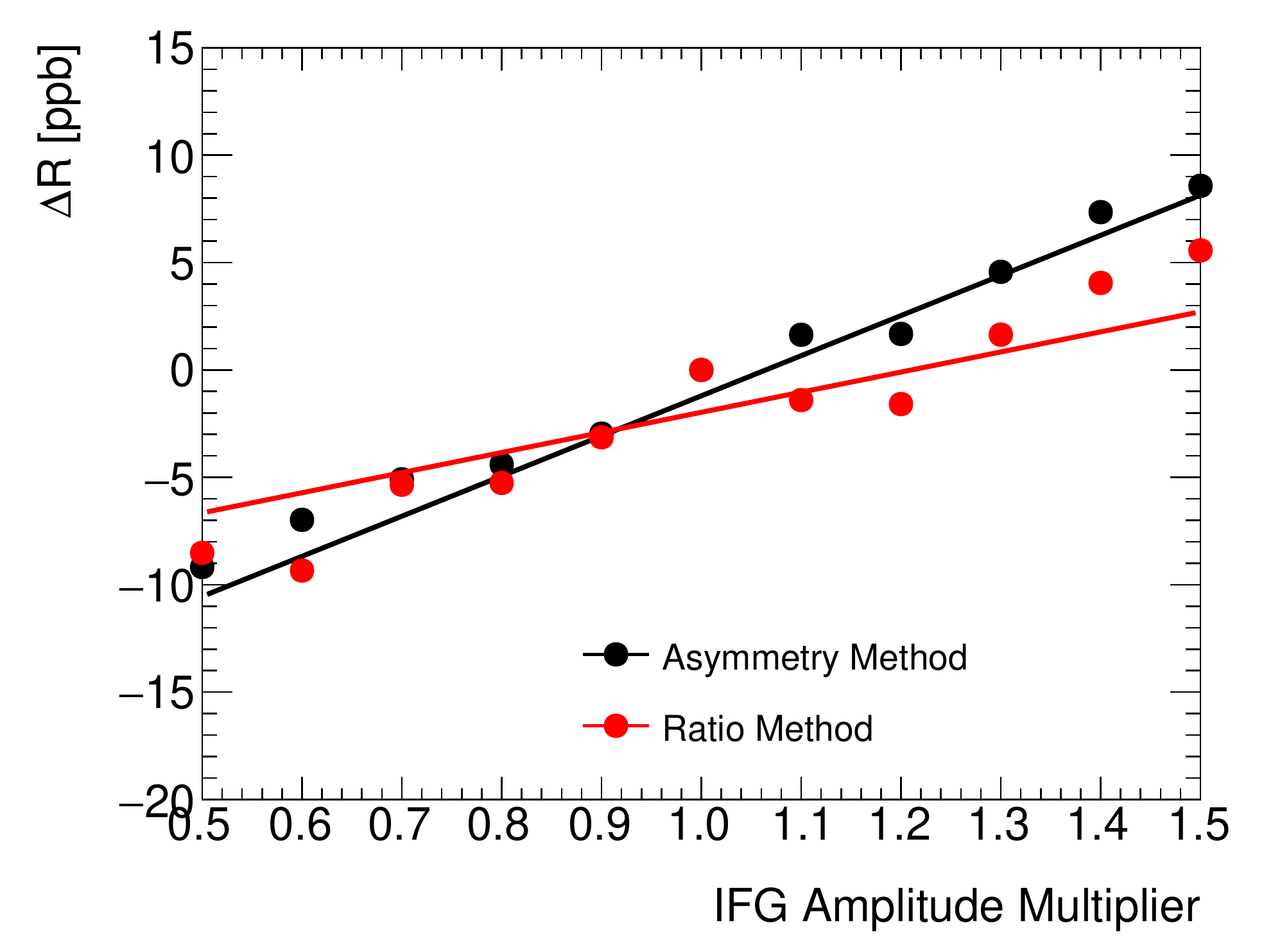}
\caption{Sensitivity of \oa to the amplitude of the in-fill gain
  correction for one of the Asymmetry-weighted analyses (black) and
  the Ratio method analysis (red).  The precession frequency changes by
  $18.8\,$ ppb and $9.4\,$ ppb, respectively, for a variation of the amplitude
  scaling factor that corresponds to one standard deviation in its
  average uncertainty.
As described in
  Sec.~\ref{subsec:r_method}, the ratio method is less sensitive to these
  ``slow effects''.}
\label{fig:IFG_sys_MP}
\end{figure}

A sweep of the amplitude of the exponential gain correction function
through a common multiplicative scaling applied to all calorimeter
channels provides an assessment of the collective sensitivity of
\oa to the in-fill correction and to the short term correction.
Figure~\ref{fig:IFG_sys_MP} illustrates the sensitivity obtained for
different methods from two analysis groups. The average uncertainty
of the amplitudes for all crystal corrections provides the range
that determines the uncertainty estimate for \oa given the measured
sensitivity.  The determination of the uncertainty from the time
constants in the exponential form employed an analogous procedure.

We find systematic uncertainties on \oa from the in-fill and
short term gain correction of order of 10 and 1 ppb,
respectively, across all data subsets.

\subsection{Multi-positron pileup}

The sources of systematic uncertainty on \oa related to multi-positron
pileup depend on the reconstruction (Sec.~\ref{sec:positron_recon})
and correction (Sec.~\ref{subsec:multi_positron_pileup}) approaches
used. For instance, the global-fitting approach to reconstruction
significantly reduces the amount of pileup, leading to a smaller
correction and in turn a smaller systematic uncertainty on \oa. We estimate
an uncertainty due to pileup under 5 ppb
across the datasets for the analysis that used this reconstruction
approach. This subsection  therefore focuses on the remaining
analyses that used the local-fitting approach to reconstruction, along
with either the shadow-window or probability-density-function
approaches to the pileup correction. 

In these analyses, the dominant systematic uncertainties on \oa arise
from uncertainties in the pileup correction's amplitude and phase. 
A scaling procedure, like the one used to assess the gain correction
amplitudes, provides 
the sensitivity of the extracted value of \oa to the amplitude.

To determine the uncertainty in the amplitude itself, the analysis groups
use one of two methods.  The first method tabulates the $\chi^{2}$
from the full  fit as a function of the scaled pileup amplitude.  
A quadratic interpolation to the 
$\chi^{2}$ distribution near its minimum then provides the amplitude uncertainty via $\Delta\chi^{2} = \pm 1$.
The second method uses agreement between the estimated pileup distribution and
the data in the energy spectrum in the non-physical region above 3.5
GeV (Fig.~\ref{fig:eDepPileup}) 
to estimate the uncertainty.

To estimate the systematic uncertainty from the phase, some analyses
shift the correction in time to evaluate the
sensitivity of \oa to the phase. We combine this sensitivity with
the ambiguity in the pileup time estimate in each method to obtain the
systematic uncertainty. Other analyses that use the
shadow-window approach vary the time (and energy) models in
Eqs.~\ref{eq:Edoublet} and \ref{eq:tdoublet}. The resulting change in the
extracted \oa value yields the uncertainty estimate.

For analyses using the local-fitting approach to reconstruction, the
total systematic uncertainty on \oa from the pileup correction ranges
from about 30 to 40 ppb across the
datasets. 

\subsection{Beam dynamics}
\label{subsec:BeamDynamics}

The fit function accounts for the imprint of beam dynamics on the
calorimeter data  
through the terms in Eqs.~\ref{eq:fit_function}--\ref{eq:cbo_freq}, and
the uncertainty in that modeling leads to a systematic uncertainty on
the extracted \oa value.  The dominant modeling uncertainties come
from the time-dependent CBO frequency ($\omega_\text{CBO} (t)$) and
the CBO decoherence envelope ($e^{-t/\tau_\text{CBO}}$).  Information
provided by the tracking system determines the time dependence of the 
CBO frequency caused by the damaged ESQ resistor.  The difference
in the parameterization obtained  separately from the two tracker
stations
provides the estimate of the uncertainty on $\omega_\text{CBO}$. 
We find an
uncertainty of order 10 ppb across all 
datasets and analyses.

The tracker data also constrain the uncertainty from the modeling of
time dependence of the CBO envelope.  The data show consistency
with an exponential behavior for the decoherence of the betatron
oscillations at the current level of precision.  However, beam dynamics
simulations of the \gm and other storage rings 
indicate that the betatron oscillations within the stored beam can
re-cohere.  Fits using the two alternate CBO envelope models
\begin{align}
&e^{-t / \tau_\text{CBO}} + B, \\
&e^{-t / \tau_\text{CBO}} \left[1 + C \cdot \cos \left(\omega_C t + \phi_C \right) \right],
\label{eq:cbo_model2}
\end{align}
where $B$, $C$, $\omega_{C}$, and $\phi_{C}$ are additional fit
parameters that we either float freely in the fit or fix to the values determined from the
tracker data, bound the sensitivity of \oa to the envelope. 
Note that the model in Eq.~\ref{eq:cbo_model2} is itself
motivated by beam-dynamics simulations.  

Additionally,  each envelope model assumes a common  $\tau_\text{CBO}$
for the CBO modulation of the normalization, asymmetry,
and phase terms in Eq.~\ref{eq:fit_function}.  However, simulations
suggest that these time constants could vary by as much as 50~\%.
The largest shift in \oa observed under
variation of each of these time constants by up to 50\% provided an
additional contribution to the systematic uncertainty.  Depending upon
the analysis technique and data subset, this uncertainty ranged from 10 to
50 ppb.

The interplay between the three classes of beam dynamics parameters discussed in this section
likely correlates them.  We therefore conservatively combine these three contributions
linearly to arrive at the total systematic uncertainty on \oa from the beam dynamics
modeling. For the average of the asymmetry-weighted event-based analyses presented
below, this uncertainty ranges from 30 to 50 ppb across the
data subsets. 

%


\subsection{Residual early-to-late effect}
The introduction of an {\it ad hoc}  time-dependent
correction to the energy scale can eliminate the unphysical
positron energy dependence of the muon loss rate (see
Sec.~\ref{subsec:internal_consistency}).  The scale of the 
required correction typically amounts to a few $\times 10^{-4}$, depending upon
the reconstruction method and dataset.  A
small time-dependent acceptance variation can similarly ameliorate
this effect, but with a shift of \oa in the opposite direction. The
source of the effect remains under investigation.  For this analysis,
we do not apply an overall correction, but we assign a systematic
uncertainty on \oa of $\simeq 20$ ppb based on the shift of its central value
upon application of one of the corrections.

\subsection{Additional systematic effects}

We have evaluated many other potential sources of bias on \oa, and find their
effects to be under 10 ppb on the final \runone \amu average,
and therefore negligible for the result from this \runone dataset.
Two of these effects of note that have been considered include
muon loss 
(see Sec.~\ref{subsec:muloss}) and time randomization (see 
Sec.~\ref{subsec:beam_bunch_structure}).

The contribution of the muon loss correction has been evaluated by
modifying the shape of the lost muons function $L(t)$ in
Eq.~\ref{eq:LM} according to different selection criteria.

Tests of the stability of the time randomization procedure include variation of the
binning size in time, incorporating the spread of cyclotron periods (from the spread
of stored muon energies) into the time randomization process, and by comparing the
time randomization for a cluster-by-cluster versus a fill-by-fill basis.  Variation of
\oa in these tests remained well below 10 ppb.  To minimize the statistical fluctuations
introduced by the minimization procedure, each analysis effort reanalyzed and refit the
data using many random number seeds.  The quoted uncertainty reflects the residual
uncertainty from the finite number of seeds employed.


Both these contributions have an effect of less than 10 ppb on the \amu average.

Other items investigated but below threshold for significant discussion include studies of
$\chi^{2}$ vs likelihood fitting, the extracted cluster time of the reconstructed positron
candidates, the short time gain correction parameters, biases in the reconstructed time and
energy in the empirical pileup estimation method, the lost muon selection criteria and the 
master clock stabilility.  In all, the full list of investigated uncertainty categories included
thirty seven separate categories.  Some of these were likely highly correlated and were combined
linearly to obtain the final categories above, or, like those listed here, were found to be negligible.

\section{Analysis combination}
\label{sec:combination}
Ideally, combination of the \oa results for each of the four data
subsets from each of the eleven analyses would proceed through a
best linear unbiased estimator (BLUE) averaging procedure.  For
example, one could minimize 
\begin{equation}
\chi^{2} = \bm{\Delta_{\omega}}^{T}\bm{C}^{-1}\bm{\Delta_{\omega}},
\label{eq:blue}
\end{equation}
where $\Delta_{\omega} = \omega_{a,i} - \bar\omega_{a}$ is the
difference between the i$^{th}$ measurement and the average
$\bar\omega_{a}$, and the covariance matrix $\bm{C}$ includes the
correlations, statistical and systematic, between the eleven
determinations of \oa.  When combined with the magnetic field
measurements for each subset, this approach would expand to $11\times
4$ determinations to be averaged.  

For the \runone sample presented
here, the statistical uncertainties dominate  the
covariance matrix for a given data subset, and the significant
statistical correlations among the eleven results for a given dataset
pose practical impediments to a well-behaved
procedure~\cite{Lyons:1988rp, Valassi:2013bga}.  
In particular, the correlation between different analyses often
reaches the ``critical value'' defined in Equation~\ref{eq:rhocrit}.

Therefore, to correctly compute an average, 
accurate estimates of the
statistical correlations are required. 
These have been estimated with
toy Monte Carlo simulations and have been shown in 
Tab.~\ref{tab:analysisCorrelations}.
Additional systematic uncertainties, due to imperfections of the
simulation, have not been estimated and are assumed to be subdominant.
The simulation confirms that measurements on the same dataset 
are all consistent with being ``critically correlated''.

As documented in the literature \cite{Valassi:2013bga, cowan1998},
correlations beyond the critical values cause the weights of the less
statistically precise measurements to become negative and reduce the
uncertainty of the BLUE combination average. We have found that in
our conditions the finite precisions of the estimated uncertainties
and correlations of the $11 \times 4$ 
measurements of \oa make the BLUE
procedure highly unstable.

When averaging two
 measurements that are exactly critically correlated, the BLUE
 combination has weight = 1 for the most precise 
result and weight = 0 for the
 least precise one (see \cite{cowan1998}).
 In the limiting case when two measurements have exactly
 the same uncertainty and are critically correlated,
 the two weights are 1/2 each. 
In our case, it
 is convenient and reasonable to set all statistical correlations to
 the critical values, and to set the measurement uncertainties
to be exactly the same when using the same method.
Under
 these assumptions, the most precise method, which is the
 asymmetry-weighted method, gets all the weight, while the other ones
get no weight in the combination.
This is justified  as long as uncorrelated
systematic uncertainties
are much smaller than the quadratic
 difference of the total uncertainties between the different methods.
In these
 conditions, there is a negligible benefit in including the other
 methods' measurements in the average with the goal of reducing the
 systematic part of the total uncertainty. 

The analyses that use different
reconstruction algorithms (local vs. global) 
are less correlated than the ones using the same reconstruction
program (see Tab.~\ref{tab:analysisCorrelations}). Thus we perform a
“staged” average of the asymmetry-weighted results for
\oa by first combining with equal weights all analyses that use the
local reconstruction and all analyses that use the global
reconstruction and then combining with equal weights the two
\oa averages of the first stage.
The \oa results of this simplified procedure have
been compared with several other different more complex procedures,
all designed to address the issue of the instability of the
combination average in case of highly correlated results.
Within the context of the BLUE approach, the covariance
matrix calculation either caps the correlation coefficients at
$\rho_{ij} < \rho^{\rm crit}_{ij}$ or uses Tikhonov
regularization~\cite{tikhonov1977}, which effectively rescales all
correlation coefficients down.  These calculations assume fully
correlated systematic uncertainties across the analyses within each
category: gain, muon loss estimation, etc.

For all these approaches, the average of the individual subsets varied
by up to 10 ppb in all cases, except one outlier, which varied by 30
ppb. 
 In summary, our results show very
good stability over all reasonable average approaches that we have
investigated. 

Here we present results from the staged averaging approach using only
the asymmetry-weighted analyses.  This method both makes optimal use
of the statistical information and shows the smallest sensitivity to
the phase-related correction from the damaged electrostatic
quadruples.   
The statistical uncertainties across the different datasets are
uncorrelated, while the systematic uncertainties are strongly
correlated, as shown in Table~\ref{tab:systcorrel}.  

Table~\ref{tab:averages} presents the resulting average value of \oa
foreach  the four data subsets.  When combining these values, along
with their associated magnetic field measurements, to obtain the final
Run-1 determination of $a_\mu$, these results contribute a total
statistical uncertainty of 434 ppb, while their systematic
contribution amounts to 56 ppb.

\begin{table}[htb]
\begin{tabular}{crrrr}
\hline
Correlation & 1a & 1b & 1c & 1d\\
\hline
1a & 1.0000 & 0.9935 & 0.9884 & 0.9812\\
1b &  & 1.0000 & 0.9820 & 0.9935\\
1c &  &  & 1.0000 & 0.9669\\
1d &  & & & 1.0000\\
\hline
\end{tabular}
\caption{Correlation matrix among different data sets for systematic uncertainties.} 
\label{tab:systcorrel}
\end{table}

\begin{table*}[tb]
\centering
\footnotesize
\begin{tabular}{rcccc} \hline\hline
\runone data set & 1a & 1b & 1c & 1d \\ \hline
\oa/2$\pi$ (s$^{-1}$) &  229080.957 & 229081.274 & 229081.134 &
229081.123\\
$\Delta$(\oa/2$\pi$) (s$^{-1}$) & 0.277 & 0.235 & 0.189 & 0.155\\
\hline
statistical uncertainty (ppb) & 1207 & 1022 & 823 & 675 \\
\hline
Gain changes (ppb)       & 12 &  9 &  9 &  5 \\
Pileup (ppb)             & 39 & 42 & 35 & 31 \\
CBO (ppb)                & 42 & 49 & 32 & 35 \\
Time randomization (ppb) & 15 & 12 &  9 &  7 \\
Early-to-late effect (ppb) & 21 & 21 & 22 & 10 \\
\hline
total systematic uncertainty (ppb) & 64 & 70 & 54 & 49 \\
\hline
total uncertainty (ppb) & 1209 & 1025 & 825 & 676 \\
\hline\hline
\end{tabular}
\caption{The combination result for each data set when using a staged
  approach. } 
\label{tab:averages}
\end{table*}

\section{Discussion and conclusion}
\label{sec:discussion}

In this article, we have described the full procedure for the
extraction of the muon precession frequency \oa for the four datasets
collected in 2018.
As described in Section \ref{subsec:subsets}, the ESQ
and kicker settings were modified over the course of \runone, in
order to optimize the quality of the stored beam.
To optimize the determination of \amu, in Ref.~\cite{\PRL} we combine the 
four \oa values presented here with corrections and field measurements
determined individually for the four datasets.  The final value corresponds
to the average of those four combined values.

Six analysis  groups produced measurements of
\oa by using two independent reconstruction algorithms,
four different histogramming methods and many variants of the
procedures used to evaluate
the correction factors and
to fit the final spectrum. 
Each analysis was carried out
with a different blinding offset. The 
relative unblinding was
performed during a Collaboration Meeting,
after all analyses were completed and shown to have an
overall agreement. 

All analyses show that the error on \oa, for \runone,
is dominated by the statistical contribution.
The systematic
uncertainties described in section \ref{sec:systematics}
have reached or approached the goal that has been set
in the Technical Design Report~\cite{Grange:2015fou} for the full statistics.

\appendix
\section{Important frequencies and full correlation matrix}
\label{app:correlations}

Table~\ref{tab:frequencies} summarizes the relevant frequencies which
characterize the \gm storage ring. The beam related 
frequencies are evaluated
according to the formulae and coincide with the measured values at the
$1\%$ level, the difference being due to decoherence effects discussed
in Sec.~\ref{subsec:BeamDynamics}.
 
Table~\ref{tab:correl_asymm_r1d_50} provides the full set of
correlation coefficients for the fit to the \runoned dataset described
in Section~\ref{sec:fitting}.  

\begin{table*}[tbp]
\centering
\footnotesize
\begin{tabular}{ccccccc}\hline\hline
& & & \multicolumn{2}{c}{n=0.108} & \multicolumn{2}{c}{n=0.120}\\
Physical frequency & Variable &Expression &
Frequency&Period&Frequency&Period\\
& & & (MHz) & ($\mu$s)  & (MHz) & ($\mu$s) \\
\hline
Anomalous precession& $f_{\rm a}$& $\frac{e}{2\pi m} \amu B$&0.229&4.37&0.229&4.37\\
Cyclotron& $f_{\rm c}$& $\frac{v}{2\pi R_0}$&6.71&0.149&6.71&0.149\\
Horizontal betatron& $f_{\rm x}$& $\sqrt{1-n}\,f_{\rm c}$&6.34&0.158&6.29&0.159\\
Vertical betatron& $f_{\rm y}$& $\sqrt{n}\,f_{\rm c}$&2.20&0.453&2.32&0.430\\
Horizontal CBO& $f_{\rm CBO}$& $f_{\rm c} - f_{\rm x}$&0.37&2.68&0.42&2.41\\
Vertical waist& $f_{\rm VW}$& $f_{\rm c} - 2f_{\rm y}$&2.31&0.433&2.07&0.484\\
\hline \hline
\end{tabular}
\caption{Frequencies and periods which characterize the \gm storage ring.}
\label{tab:frequencies}
\end{table*}

\begin{table*}[tbp]
\centering
\footnotesize
\begin{tabular}{rrrrrrrrrrrr}\hline\hline
                    & $R$   & $N_0$ & $\gamma\tau_{\mu}$ 
                                            & $A_0$ & $\phi_0$
                                                            & $\omega_\text{CBO}$
                                                                    & $\tau_\text{CBO}$
                                                                            & $A_{N,x,1,1}$
                                                                                    & $\phi_{N,x,1,1}$
                                                                                            & $K_{\rm loss}$
                                                                                                    & $\kappa_y$ \\ 
\hline
$R$                 &  1.00 & -0.01 & -0.00 &  0.00 & -0.87 &  0.01 &  0.02 & -0.03 & -0.02 & -0.01 &  0.00  \\
$N_0$               &       &  1.00 &  0.86 & -0.03 &  0.01 & -0.00 & -0.03 &  0.05 &  0.00 &  1.00 & -0.01  \\
$\gamma\tau_{\mu}$  &       &       &  1.00 & -0.02 &  0.00 & -0.00 & -0.02 &  0.03 &  0.00 &  0.89 & -0.01  \\
$A_0$               &       &       &       &  1.00 & -0.01 &  0.01 & -0.01 &  0.01 & -0.02 & -0.04 & -0.00  \\
$\phi_0$            &       &       &       &       &  1.00 & -0.02 & -0.03 &  0.04 &  0.02 &  0.01 & -0.00  \\
$\omega_\text{CBO}$ &       &       &       &       &       &  1.00 & -0.03 &  0.03 & -0.92 & -0.00 & -0.21  \\
$\tau_\text{CBO}$   &       &       &       &       &       &       &  1.00 & -0.92 &  0.03 & -0.03 &  0.01  \\
$A_{N,x,1,1}$       &       &       &       &       &       &       &       &  1.00 & -0.03 &  0.04 & -0.01  \\
$\phi_{N,x,1,1}$    &       &       &       &       &       &       &       &       &  1.00 &  0.00 &  0.20  \\
$K_{\rm loss}$ &       &       &       &       &       &       &       &       &       &  1.00 & -0.01  \\
$\kappa_y$          &       &       &       &       &       &       &       &       &       &       &  1.00  \\
                    & $\tau_y$
                            & $A_{N,y,2,2}$
                                    & $\phi_{N,y,2,2}$
                                            & $A_{N,x,2,2}$
                                                    & $\phi_{N,x,2,2}$
                                                            & $A_{A,x,1,1}$
                                                                    & $\phi_{A,x,1,1}$
                                                                            & $A_{\phi,x,1,1}$
                                                                                    & $\phi_{\phi,x,1,1}$
                                                                                            & $A_{N,y,1,1}$
                                                                                                    & $\phi_{N,y,1,1}$ \\
$R$                 & -0.00 &  0.00 &  0.01 &  0.01 & -0.00 &  0.02 & -0.01 & -0.00 & -0.01 & -0.00 & -0.01 \\
$N_0$               &  0.00 & -0.00 & -0.01 & -0.01 &  0.01 &  0.05 & -0.02 & -0.05 & -0.05 &  0.00 &  0.01 \\
$\gamma\tau_{\mu}$  &  0.00 & -0.00 & -0.01 & -0.01 &  0.00 &  0.03 & -0.01 & -0.03 & -0.03 &  0.00 &  0.01 \\
$A_0$               &  0.00 & -0.00 & -0.00 &  0.00 &  0.00 &  0.00 &  0.03 &  0.02 & -0.01 & -0.00 &  0.00 \\
$\phi_0$            &  0.00 & -0.00 & -0.01 & -0.01 &  0.01 & -0.03 &  0.01 &  0.00 &  0.01 &  0.00 &  0.01 \\
$\omega_\text{CBO}$ &  0.00 & -0.00 & -0.01 & -0.03 & -0.16 & -0.00 & -0.11 & -0.01 & -0.06 &  0.00 &  0.01 \\
$\tau_\text{CBO}$   & -0.00 &  0.00 &  0.00 & -0.14 & -0.01 & -0.08 &  0.03 & -0.01 & -0.00 & -0.00 & -0.00 \\
$A_{N,x,1,1}$       &  0.00 & -0.00 & -0.01 &  0.12 &  0.00 &  0.09 & -0.02 &  0.01 &  0.00 &  0.00 &  0.01 \\
$\phi_{N,x,1,1}$    & -0.00 &  0.00 &  0.01 &  0.03 &  0.14 & -0.01 &  0.12 &  0.01 &  0.05 & -0.00 & -0.01 \\
$K_{\rm loss}$ &  0.00 & -0.00 & -0.01 & -0.01 &  0.01 &  0.04 & -0.02 & -0.05 & -0.04 &  0.00 &  0.01 \\
$\kappa_y$          & -0.47 &  0.45 &  0.95 &  0.00 &  0.02 & -0.00 &  0.03 &  0.01 &  0.00 & -0.10 & -0.51 \\
$\tau_y$            &  1.00 & -0.95 & -0.47 &  0.00 &  0.00 &  0.00 & -0.00 & -0.00 &  0.00 &  0.47 &  0.14 \\
$A_{N,y,2,2}$       &       &  1.00 &  0.45 & -0.00 & -0.00 & -0.00 &  0.00 &  0.00 & -0.00 & -0.45 & -0.13 \\
$\phi_{N,y,2,2}$    &       &       &  1.00  & -0.00 & -0.01 & -0.00 &  0.00 &  0.01 & -0.01 & -0.09 & -0.50 \\
$A_{N,x,2,2}$       &       &       &       &  1.00 & -0.00 & -0.00 & -0.01 &  0.03 & -0.03 & -0.00 &  0.00 \\
$\phi_{N,x,2,2}$    &       &       &       &       &  1.00 &  0.02 &  0.01 &  0.02 &  0.03 &  0.00 &  0.01 \\
$A_{A,x,1,1}$       &       &       &       &       &       &  1.00 & -0.02 & -0.01 & -0.01 &  0.00 &  0.00 \\
$\phi_{A,x,1,1}$    &       &       &       &       &       &       &  1.00 &  0.00 &  0.03 & -0.00 & -0.00 \\
$A_{\phi,x,1,1}$    &       &       &       &       &       &       &       &  1.00 &  0.00 & -0.00 & -0.01 \\
$\phi_{\phi,x,1,1}$ &       &       &       &       &       &       &       &       &  1.00 &  0.00 &  0.01 \\
$A_{N,y,1,1}$       &       &       &       &       &       &       &       &       &       &  1.00 & -0.00 \\
$\phi_{N,y,1,1}$    &       &       &       &       &       &       &       &       &       &       &  1.00 \\
\hline\hline
\end{tabular}

\caption{The correlation matrix from the fit whose results are presented in Table~\ref{tab:fit_asymm_r1d_50}. The parameters are defined in Equations~\ref{eq:LM} through~\ref{eq:phixt} and Eq.~\ref{eq:Rblinding}. For clarity,
only the above-diagonal elements of the symmetric matrix have been displayed.
 \label{tab:correl_asymm_r1d_50}}
\end{table*}
\section*{Acknowledgments}
We thank the Fermilab management and staff for their strong support of
this experiment, as well as the tremendous support from our university
and national laboratory engineers, technicians, and workshops.


The Muon \gm Experiment was performed at the Fermi National
Accelerator Laboratory, a U.S. Department of Energy, Office of
Science, HEP User Facility. Fermilab is managed by Fermi Research
Alliance, LLC (FRA), acting under Contract No. DE-AC02-07CH11359.
Additional support for the experiment was provided by the Department
of Energy offices of HEP and NP (USA), the National Science Foundation
(USA), the Istituto Nazionale di Fisica Nucleare (Italy), the Science
and Technology Facilities Council (UK), the Royal Society (UK), the
European Union's Horizon 2020 research and innovation programme under
the Marie Sk\l{}odowska-Curie grant agreements No. 690835,
No. 734303, the National Natural Science Foundation of China 
(Grant No. 11975153, 12075151), MSIP, NRF and IBS-R017-D1 (Republic of Korea), 
the German Research Foundation (DFG) through the Cluster of 
Excellence PRISMA+ (EXC 2118/1, Project ID 39083149).

\end{document}